\documentclass{aa}  

\usepackage{graphicx}
\usepackage{txfonts}
\usepackage{xcolor}
\usepackage{soul}
\usepackage{ulem}
\usepackage{float}
\usepackage{placeins}
\usepackage[version=4,arrows=pgf-filled,
textfontname=sffamily,
mathfontname=mathsf]{mhchem}
\newcommand{\half}{\frac{1}{2}}

\definecolor{darkorange}{RGB}{210, 70, 0}  % DarkOrange RGB values

\begin{document} 

\title{Exoplanet climate characterization with transit asymmetries }

\subtitle{A comprehensive population study from the optical to the infrared}

   \author{L. Carone
          \inst{1},
          Ch. Helling
          \inst{1,2}
         S. Gernjak \inst{1,2},
           H. Leitner \inst{1,2}
         \and
          T. Janz \inst{1,2} 
          }

   \institute{
$^1$Space Research Institute, Austrian Academy of Sciences, Schmiedlstrasse 6, A-8042 Graz, Austria\\
 \email{ludmila.carone@oeaw.ac.at}\\
$^2$Institute for Theoretical Physics and Computational Physics, Graz University of Technology, Petersgasse 16, 8010 Graz
             }

   \date{Received TBD; TBD}

  \abstract
 {Space missions (e.g., CHEOPS, JWST, PLATO)
   facilitate detailed characterization of exoplanets, in particular for close-in gas giants with their wide range of global temperatures and climate regimes.}
   {The aim of this work is to provide a  
   framework to characterize cloud and climate properties of close-in gas giants via transit depth asymmetries from the optical to the infrared ($\lambda = 0.33\,\ldots\,10~\mu$m). }
   {The AFGKM ExoRad 3D GCM grid 
   provides 3D  gas temperature profiles, assuming a chemical equilibrium gas-phase, for an ensemble of 50 tidally locked gaseous planets orbiting A, F, G, K and M-type host stars. It is combined with a kinetic cloud formation model (with nucleation, surface growth/evaporation, gravitational settling, mixing, element conservation). The end result is a set of synthetic transit spectra and evening-to-morning transit asymmetries that span all relevant climate regimes: warm ($T_{\rm global}=800$~K $\ldots$  1000~K), intermediately hot ($T_{\rm global}=1200$~K\,$\ldots 2000$~K) and ultrahot ($T_{\rm global}=2200$~K\,$\ldots 2600$~K). }
   {WASP-39b observations suggest that clouds are iron-free, and cloud condensation nuclei are
   less abundant than previously expected. The ensemble study shows that clouds increase transit limb differences due to asymmetries in cloud coverage and by enhancing horizontal differences in the gas temperatures. 
   
   For planets in the hot climate regime, 
   evening-to-morning differences of up to 150~ppm are suggested in the optical and 100~ppm in the infrared (2-8~$\mu$m).
   For ultra-hot Jupiters, the observed evening-to-morning transit differences are dominated by the morning cloud for a cloud-free evening limb: They are strongly negative in the PLATO band (0.5-1~$\mu$m, -500 ppm), moderately negative in the near-infrared (1-1.5~$\mu$m, -200 ppm) and moderately positive (+100 ppm) for $\lambda > 2\mu$m. For a partly cloudy evening terminator, the evening-to-morning transit asymmetry is moderately positive in the whole wavelength range. Warm Jupiter planets exhibit negligible transit asymmetries.
   The reported transit asymmetries are conservative: Planets 30\% larger than 1~$R_{\rm Jup}$ increase the transit asymmetry signal by a factor of two.
   }
   {PLATO may
   help to characterize climate regimes as well as cloud properties.
   The combination of precise PLATO and JWST transit asymmetry observations between 1-2~$\mu$m is optimal to characterize cloudy planetary atmospheres around K -A stars.  JWST/MIRI observations are most effective for planets around M stars with transit differences $\geq +500$~ppm for 8-10~$\mu$m.}


   \maketitle
%
%-------------------------------------------------------------------

\section{Introduction}

Tidally locked hot Jupiters are a treasure trove for understanding planetary climates in 
context compared to Solar system planets, thus widely expanding the boundary of our knowledge. These planets are tidally locked (with a permanent day side) and orbit very closely to their host stars with orbital periods of $P < 20$ days. Tidal locking leads to highly asymmetric irradiation conditions and to exoplanet climates dominated by a strong  equatorial wind jet ($1-10$~km/s), the superrotating jet.

Such an efficient transport of cold gas from the night side to the dayside results in a cold morning and a warm evening terminator with all its consequences for the local gas  composition and cloud formation (\citealt{2019A&A...626A.133H, 2019A&A...631A..79H, 2020A&A...641A.178H, Helling2021}).
Cloud asymmetries between the limbs 
will, hence, be a consequence as well as an indicator for climate regime dominated by superrotation. \citet{Esteves2015} showed that Kepler-7b with $T_{\rm global}=1630$~K (\footnote{The global temperature is defined with respect to the planet's interior temperature $T_{\rm int}$ and the planet's equilibrium temperature $T_{\rm eq}$ assuming averaging over the whole planet. The latter depends on the stellar radius $R_{\star}$ and temperature $T_{\star}$ and planetary semi major axis $a$, which yields: $T_{\rm global}^4= T_{eq}^4+T_{\rm int}^4$, where $T_{\rm eq}=T_{{\star}} \sqrt{R_{\star}/2a}$}) has a morning terminator with a higher cloud opacity than the evening side by analysing the contribution of reflected stellar light in the planetary optical phase curve.  Already, \citet{Fortney2010} explored the challenges of probing the limbs on a 3D planet and suggested that spatial variations across the limbs may be important and potentially observable. \citet{VonParis2016,Line2016} suggested that cloud asymmetries may be detectable by performing transmission spectroscopy over the transit ingress and egress part of the transit lightcurve separately. Two effects may lead to detectable differences between the morning and evening terminator: Gas temperature differences across the limbs may lead to a more extended atmosphere, 
and differences in the location of the cloud top block stellar light at different altitudes. \citet{2019A&A...626A.133H,2019A&A...631A..79H} and \citet{Powell2019} 
highlighted the need of a microphysical cloud model to realize the potential of transit asymmetry observations.

The first detection of morning and evening asymmetries by transmission spectroscopy for the hot Jupiter WASP-39b  ($T_{\rm global}\sim$1100~K) with JWST/NIRSpec PRISM proved that detailed transit asymmetry observations are possible \citep{Espinoza2024}. The investigation of transit depth asymmetries with modern space telescopes has, hence,  the potential 
to derive spatially resolved atmospheric properties of the evening and morning terminators. For example, the transit ingress and egress observations of WASP-39b \citep{Espinoza2024} suggest that the morning terminator has a higher cloud opacity compared to the evening limb as already  qualitatively predicted by atmosphere simulations (\citealt{Carone2023}). 
\cite{Murphy2024} observed  a morning and evening limb asymmetry in the relatively cool  transiting gas giant WASP-107b ($T_{\rm global}\sim$770~K). This  suggests that even in relatively cool planets terminator asymmetries are detectable, where advection is efficient enough to homogenize the temperature field across the planetary limbs \citep[e.g.][]{Komacek2016,Helling2023}.  The use of transmission photometry for the study of exoplanet gas giants with PLATO was explored in \citet{Grenfell2020}
for cloud-free planets orbiting F, G, K and M host stars. It was suggested the PLATO transmission photometry may be used to explore bulk atmospheric compositions and geometric albedos. Previous work by, e.g., \cite{Roth2024,2024MNRAS.528.1016T} explored an ensemble of cloud-free planets in the warm to ultra-hot Jupiter regime to understand horizontal differences due to atmosphere dynamics effects. \citet{Roman2021,Helling2023,Kenneth2025} performed ensemble studies to explore horizontal cloud coverage changes for similar ranges of temperatures. \citet{Kenneth2025}  
explored the impact of cloud opacity on JWST transit asymmetries assuming equilibrium condensation (S=1) and uniform cloud properties.

This paper conducts an ensemble study of exoplanet gaseous planets of solar metallically and uniform size ($1R_{\rm Jup}$) with $T_{\rm global}=800\,\ldots\,2600$K to analyze the underlying physical reason for transit asymmetries between the morning and evening terminator in diverse superrotating climates and to identify particularly interesting climate and cloud scenarios. The aim is to use terminator asymmetries 
to characterize exoplanet climate regimes utilizing observations from CHEOPS, TESS, JWST and also from PLATO 
that cover the wavelength range $\lambda=0.33\,\ldots\,10~\mu$m. 

For this purpose, a grid of 3D GCM atmosphere models for tidally locked planets that orbit A,  F, G, K and M-type host stars \citep{Plaschzug2025} is utilized in combination with a kinetic cloud formation model to derive local cloud properties (e.g., mean particle size, cloud particle material composition) for the calculation of transmission spectra for 50 grid 3D model atmospheres. This ensemble of planets covers the three climate regimes previously  introduced \citep{Helling2023}: warm ($T_{\rm global}=800\,\ldots\,1000$K), intermediately hot ($T_{\rm global}=1200\,\ldots\,1800$K) and ultra-hot ($T_{\rm global}=~2000\,\ldots\,2600$K) gas planets.

The paper addresses the following research questions:
\begin{itemize}
    \item  For which wavelengths are transit asymmetries the largest?
    \item Which physical mechanism cause 
    observable asymmetries?
    \item Which wavelength ranges yield complementary insights about cloud properties like cloud coverage and composition?

\end{itemize}

Section~\ref{sec: approach} presents the approach of this paper where  the 3D GCM \texttt{ExoRad}, the kinetic cloud modelling and radiative transfer modelling  are described.
Section~\ref{sec: WASP39b}
presents the case study of the JWST early release science (ERS) target WASP-39b and demonstrates how kinetic cloud modelling 
reproduces the available combined spectra from  JWST (NIRCAM, NIRSpec, NIRISS), HST and VLT.
Section~\ref{sec: Three cases} presents transmission spectra and focuses on three examples cases, $T_{\rm global}=800, 1200, 2400$K for planets orbiting a  G-type host star. These are chosen to represent the three major climate scenarios (warm, intermediately hot, ultra-hot; \citealt{Helling2023}). Different cloud scenarios are explored in order to study the impact of the uppermost atmosphere where cloud condensation nuclei (CCN) determine the cloud formation.
Section~\ref{sec: PLATO} presents the transit depth asymmetry results for PLATO's white band (normal cameras) and the fast cameras' red and blue band for $T_{\rm global}=800\,\ldots\,2600$K, focusing first on G-type host stars. It is shown that PLATO may help to identify cloud scenarios with efficient CCN formation. Sections~\ref{sec: NIR} and \ref{sec: MIRI} explore the complementary power of observations from the optical (HST, CHEOPS, TESS, PLATO) and the near-IR (1-2~$\mu$m) and mid-IR (5-8~$\mu$m and 8-10~$\mu$m) respectively. The study of transit depth differences in different wavelength ranges allow to coherently constrain horizontal gas phase differences across the limbs, latitudinal cloud coverage at the evening terminator as well as to provide constraints for of the cloud mass load in the upper-most  atmosphere that complement optical measurements.
Section~\ref{sec: all} present the transit depth analysis for the whole ensemble of the 50 3D AFGKM Exorad GCM grid atmosphere models. Detailed cloud properties for gas planets orbiting M dwarf stars can still be identified with MIRI, even if the low luminosity of M dwarf stars in the PLATO band makes detailed characterization difficult. 
In Sect.~\ref{sec: discus} insights from WASP-39b observations that indicate a reduction of 
of the cloud mass load of sub-micron particles 
as well as a potential locking of iron into the interior are explored. Further, we present the range of transit asymmetries for different climate regimes and analyze which properties of the cloud lead to the amplification of  transit depth limb asymmetries. Section~\ref{sec: Conclusion} concludes that this grid study has the potential to provide a coherent guidance to plan transit asymmetry observations. Transit limb asymmetries can provide unique information about climate regimes, detailed cloud properties as well as cloud-climate structure by coherently tying together observations from the optical to the infrared for planets around diverse host stars.

%--------------------------------------------------------------------
\section{Approach}
\label{sec: approach}

Transmission spectra and transit photometry covering the PLATO, CHEOPS, TESS and JWST wavelength ranges based on results from 3D GCM simulations (Sect.~\ref{sec:ExoRad}) are explored in combination with detailed kinetic cloud formation  from a gas in chemical equilibrium (Sect.~\ref{sec:Cloudmodelling}) and radiative transfer calculations (Sect.~\ref{sec:Transmission}). 
The 3D GCM provides the thermodynamic input for the cloud model from which the necessary gas and cloud opacities are then derived. 
The hierarchical modelling approach aims to enable the best performance of each component (\citealt{Helling2021,Helling2023}) and was also applied to individual JWST hot Jupiter targets \citep{Espinoza2024,Carone2023,Chubb2024,Samra2023}.

\subsection{The 3D AFGKM \texttt{ExoRad} GCM grid}
\label{sec:ExoRad}

The latest version of the \texttt{Exorad} 3D climate modelling framework \citep{Carone2020} that employs full LTE radiative transfer (\texttt{expeRT/MITgcm}) is used. For a full description of the set-up using the \texttt{MITgcm} dynamical core \citep{Adcroft2004} with an Arakawa C type cubed sphere grid (C32 $\approx 2.8^{\circ}\times 2.8^{\circ}$ horizontal resolution) and 47 vertical levels between 700~bar and $10^{-5}$~bar, see \citet{Carone2020} and \citet{Schneider2022a}. 
%We note, however, that a 
The simulations used here have adopted a
dynamical timescale of $\Delta t =25~$s. The gas pressure levels between $10^{-4}-10^{-5}$ bar comprise the upper boundary sponge layers, in which the zonal horizontal velocity $u$ is damped by a Rayleigh friction
term towards its longitudinal mean $u$ to avoid unphysical gravity reflection. Since vertical velocities are calculated via the mass continuity equation from the horizontal velocities, the sponge layer also results in a damping of the vertical wind. The lower boundary is stabilized with basal drag \citep{LiuShowman2013ApJ} between 500 and 700~bar \citep{Carone2020}.

In this work, the following gas-phase opacity species were considered: H$_2$O (from ExoMol\footnote{\url{https://www.exomol.com}} -- \citealt{Tennyson2016_ExoMol, TennysonEtal2020jqsrtExomol2020}), Na \citep{Allard19_Na_K}, K \citep{Allard19_Na_K}, the latter two including pressure broadening, CO$_2$, CH$_4$, NH$_3$, CO, H$_2$S, HCN, SiO, PH$_3$ and FeH, as well as H$^-$ scattering suitable for an ionised atmosphere as listed in \citet[][Tab. 1]{Schneider2022a} excluding TiO and VO to provide a homogeneous framework for transit depth asymmetry comparison and to avoid the uncertainties in TiO/VO condensation \citep[e.g.][]{Sing2019, Parmentier2013,Schneider2022a}. 
Collision-induced absorption for H$_2$--H$_2$ \citep{BorysowEtal2001jqsrtH2H2highT, Borysow2002jqsrtH2H2lowT, Richard2012} and H$_2$--He \citep{BorysowEtal1988apjH2HeRT, BorysowFrommhold1989apjH2HeOvertones, BorysowEtal1989apjH2HeRVRT}, Rayleigh scattering for H$_2$ \citep{Dalgarno1962}, He \citep{Chan1965}, and H$^-$ free-free and bound-free opacities\citep{Gray2008} are also included. The the correlated-k tabulated gas opacities are combined in 11 spectral bins, spanning 0.26 -300~$\mu$m \citep{Schneider2022a}. The gas is assumed to be in local thermal equilibrium (LTE) such that the number densities of the opacity species are calculated in chemical equilibrium.
A constant mean molecular weight of 
$\mu=2.3$ is assumed for the 3D atmospheric modelling due to the current numerical limitations of many GCMs \citep[see also][]{Roth2024,Kenneth2025}. We note that this is a valid assumption for planets with global temperatures smaller than 1600~K \citep{Helling2023}, but fails at the dayside for hotter planets due to thermal dissociation of \ce{H2} \citep{Helling2023,Tan2019}. A reduction to $\mu=1.8$ is predicted in the upper atmosphere at the dayside for the hottest planets in the grid (2400-2600~K). For planets with global temperatures between 1800~K - 2200~K, a more localized reduction at the hottest point at the dayside is expected.

\texttt{expeRT/MITgcm} is used to generate a grid of 50 simulated tidally locked $1 R_{\rm Jup}$-sized gas planets ($g = 10$~m/s$^{2}$) with solar metallicity ([Fe/H]=0), solar C/O ratio (C/O=0.55) and global temperatures $T_{\rm global}=800\,\ldots\,2600$K , orbiting M5V, KV5, GV5, F5V and A5V host stars.  Stellar parameters from \citet{Pecaut2013} and interior temperatures were derived with the parametric fit of \citet{Thorngren2019}. The 3D AFGKM \texttt{Exorad} GCM grid is an update from the previous \texttt{ExoRad} grid that used Newtonian cooling \citet{Baeyens2021} and successfully applied to gain more insights about disequilibrium chemistry \citep{Baeyens2022} and cloud formation~\citep{Helling2021}. The new \texttt{ExoRad} grid also comprises short-period planets ($<1$~days) around K and M stars for which the closest counterparts are brown dwarf-white dwarf pairs like WD 0137-349B \citep{Lee2022} and SDSS 1557 \citep{Amaro2024}. The 3D AFGKM \texttt{Exorad} GCM grid is expanded to include A-type host star to facilitate also studies for the impact of extreme stellar environments in the scope of the NewAthena space mission \citep{Cruise2025}. The full grid is introduced in \citet{Plaschzug2025}.
 The simulations are run for $1000\,{\rm days}$ and time averaged over the last 100 days.

The output of the \texttt{ExoRad} simulations are down-sampled, that is, the model is mapped from the original cubed-grid sphere to a latitude-longitude grid of lower resolution ($22.5^{\circ}$ in latitude ($\theta$) and $15^{\circ}$ in longitude ($\phi$) ) before hand-over to the cloud model (see \citet{Ronchi1996} for the mathematical description of transformation between latitude-longitude and cubed sphere grids). Here, the open-source package \texttt{gcm-toolkit}\footnote{https://gcm-toolkit.readthedocs.io} was used to transform the original C32 grid from \texttt{Exorad} to the latitude-longitude grid used by \texttt{IWF Graz} cloud model. 

In this work, we thus provide a coherent baseline in climate and cloud interplay in the superrotating climate regime over a large temperature range, building upon experience from previous large grid studies \citet{Helling2023,Baeyens2022,Komacek2019,Roman2021}. However, the impact of magnetic drag and TiO that may affect the dayside of  ultra-hot Jupiters \citep{Deline2025} was neglected. It is still debated how much TiO remains in gas phase at the cooler limbs as cold trapping may remove the molecules from the gas phase \citep{Parmentier2013,Roth2024}. 
Since magnetic coupling only occurs at the partly ionized dayside, superrotation persists on the colder nightside, that is, for the majority of the cloud forming regions \citep{Beltz2022}.
Table~\ref{tab: stellar_params} lists the %most important 
stellar parameters, including stellar radius $R_*$ that is important to interpret the observability of transit depth differences that are scaled with $R_P^2/R^2_*$, where $R_P$ is the planetary radius.

\subsection{The \texttt{IWF Graz} 3D cloud model}
\label{sec:Cloudmodelling}

For each planet in the 3D AFGKM \texttt{Exorad} GCM grid, $120$ 1D 
 ($T_{\rm gas}$(z), $p_{\rm gas}$(z), $v_{\rm z}(z)$)-profiles are extracted. $T_{\rm gas}$(z) is the local gas temperature [K], $p_{\rm gas}$(z) is the local gas pressure [bar], and $v_{\rm z}$(z) the vertical velocity component. Each \texttt{ExoRad} 3D model is sampled every 15$^{\circ}$ in longitude ($\phi$) and at five latitudes ($\theta = 0^{\circ}, 22.5^{\circ}, 45^{\circ}, 67.5^{\circ}, 90^{\circ}$). The north and south hemisphere are assumed to be identical.  
These 1D ($T_{\rm gas}$(z), $p_{\rm gas}$(z), $v_{\rm z}(z)$) profiles are used as input for the kinetic, non-equilibrium cloud formation model, which calculates the formation of cloud condensation nuclei, the growth and evaporation by surface reactions,  gravitational settling, mixing and element conservation (\citealt{Woitke2003,Woitke2004,2004A&A...423..657H,Helling2006,2008A&A...485..547H}; see also  \citealt{helling2013RSPTA,2022arXiv220500454H}).

To calculate the formation of cloud particles, the local gas composition needs to be known. The local gas-phase composition is calculated assuming chemical equilibrium by applying {\sc GGChem} (\citealt{2018A&A...614A...1W}) which is part of our cloud formation code. The gas phase is assumed to be in chemical equilibrium for all planets considered here. Deviation from LTE in the cloud forming regions only affect cloud formation marginally (\citealt{2020A&A...635A..31M}). Out of the total set of elements considered for the gas-phase composition, 11 elements  (Mg, Si, Ti, O, Fe, Al, Ca, S, K, Cl, Na) are considered for the bulk growth of the cloud particles and 6 (Ti, Si, O, K, Cl, Na) participate in the formation of cloud condensation nuclei. The formation of 4 nucleation species are considered and the total nucleation rate is $J_* =\sum_{\rm i}J_{\rm i}$ [cm$^{-3}$ s$^{-1}$] (i=TiO$_2$, SiO, NaCl, KCl). KCl and NaCl appear in negligible amounts for the planets considered here.  To calculate $J_*$, the modified classical nucleation theory (e.g., see \citealt{helling2013RSPTA})  is used which previously was compared  to a Monte Carlo approach treating individual cluster collisions for for TiO$_2$ (\citealt{2021A&A...654A.120K}).

The total nucleation rate, $J_*$ determines the number of cloud particles that  form locally. These cloud condensation seeds grow through 132 gas-surface growth reactions (see Table~\ref{tab:chemreak}) to macroscopic particles. The growth (or evaporation) 
of 16 materials (sorted in 4 groups) is considered:
metal oxides (SiO[s], SiO$_{2}$[s], MgO[s], FeO[s], Fe$_{2}$O$_{3}$[s]), silicates (MgSiO$_{3}$[s], Mg$_{2}$SiO$_{4}$[s], CaSiO$_{3}$[s], Fe$_{2}$SiO$_{4}$[s]), high temperature condensates (TiO$_{2}$[s], Fe[s], Al$_{2}$O$_{3}$[s], CaTiO$_{3}$[s], FeS[s]) and salts (KCl[s], NaCl[s]).

In total, 31 ODEs  are solved in order to model the formation of cloud particles as a sequence of nucleation, surface growth/evaporation, gravitational settling, element replenishment and element conservation. The undepleted gas is assumed to be of solar composition. The modelling of the vertical mixing to facilitate element replenishment follows \citealt{Helling2023} (Sect. 2.3) where the mixing time scale is derived from the hydrodynamic velocity field (Eq. B 26). This approach uses the information from the local velocity field with which the hydrodynamic fluid is transported across a number of computational cells. 5 cells are used here and only the vertical velocity component is considered. It is therefor different from the large-scale convection focused approach outline in the comparison paper \cite{2008MNRAS.391.1854H} (their Sect. 2.2.4.4). Guided by our previous study (\citealt{2023A&A...669A.142S}), a 100x reduced vertical mixing is used in this paper. Horizontal mixing as transport mechanism for cloud particles is neglected, but its effect on the local thermo- and hydrodynamic is taken into account as it determines the local state variables (T$_{\rm gas}$, p$_{\rm gas}$, v$_{\rm gas}$) which are the input for the cloud model. The 3D nature of any non-vertical velocity component on the hydrodynamics is therefore taken into account. As pointed out in \citealt{Helling2023} (Sect. 4.1), horizontal advection of cloud particles could  smear out the
boundaries of the partial cloud coverage. However, such a smearing will be
limited by the thermal stability of the cloud particle in particular
towards all high-temperature regions (e.g. on the dayside). \citet{Powell2024} re-iterate the effect of thermal stability:  In their ensembles of homogeneous material  cloud particles,  the materials with the highest thermal stability (Fe[s], Al2O3[s], TiO2[s], Cr[s]) survive a homogeneous velocity shift into regions of higher temperature (e.g. the day side of hot Jupiters) better than low-temperature condensates (Mg2SiO4[s]). For the very same reason are cloud particles in the inner, deeper atmosphere regions predominantly composed of high-temperature condensates (e.g., \citealt{Helling2021}). In the hierarchical modelling framework applied here, a higher resolution in longitude may shift the cloud coverage somewhat to the dayside. If the thermal stability argument is used to determine how far into the dayside a cloud may be present, the vertically changing sizes and their respective frictional forces need to be considered.  A process that may actually cause clouds to extend into higher-temperature regions is hydrodynamic turbulence because it increases the chemically active volume where seed formation can act (\citealt{2001A&A...376..194H,2004A&A...423..657H}, see also Sect. 3.1.3. in \citealt{2019AREPS..47..583H} for a review).

In order to calculate the cloud opacity, the particle number densities and material compositions are required which is expressed in terms of material volume fractions ($V_{\rm s}/V_{\rm tot}$ with $s$ being the 16 different materials). 
The cloud opacity is derived from the cloud dust-to-gas mass ratio, $\rho_{\rm d}/\rho_{\rm g}$ (see Eqs. 14 and 15 in \citealt{Kiefer2024iron}).

\subsection{Transmission spectrum} 
\label{sec:Transmission}

Transmission spectra are calculated including the output from our cloud formation model described in Sect.\ref{sec:Cloudmodelling}. 
Following  \citet{Brown2001},  a transmission spectrum or wavelength dependent transit depth $T_{\rm Depth}(\lambda)$ is  the wavelength dependent ratio of the flux observed during transit and the flux outside of transit:
\begin{equation}
T_{\rm Depth}(\lambda)=\frac{F_{\rm transit}(\lambda)}{F_0 (\lambda)}.
\end{equation}

The flux outside of transit $F_0(\lambda)$ comprises the stellar flux $F_\star(\lambda)$ and the planetary flux $F_{Pl}(\lambda)$:
\begin{equation}
F_0(\lambda) \left[\rm \frac{ erg} {cm^2 s}\right]= F_\star(\lambda) + F_{Pl}(\lambda)\approx F_\star(\lambda),   
\end{equation}
where the observer receives 
thermal planetary flux from the cold nightside just before and after transit, which is negligible in comparison to the stellar flux. The same holds for the scattered light contribution. The dominant contribution from the transmission therefore stems from rays of stellar light passing through the planetary atmosphere. Here, also refraction of light, that is, deviations from the initial ray path due to the density gradient of the atmosphere is neglected. Transmission spectra are calculated for the optical and infrared $0.33-10\mu$m for low resolution JWST and PLATO observations that typically have a resolution $R\leq 300$.

For simplicity it is assumed that the planet is passing through the center of stellar disk during transit. Further, the stellar limb darkening and the so called stellar light source effect \citep{Rackham2018}, that is, stellar surface inhomogeneities and stellar molecular bands for M dwarfs are neglected for this study. \citet{Kostogryz2024} showed that both effects, 3D planetary atmospheres and 3D stellar surface impacts need a more detailed understanding to fully realize the next generation of transmission spectra observations.  The present  work aims to provide improved insights in the impact of an extended cloudy 3D planetary atmosphere on transmission spectra. The focus lies here to bring to the fore the potential of combining complex models with detailed and high precision observations with current and upcoming mission.

With $F_\star$ being the the disk integrated stellar flux \citep[][Eq.~11]{Brown2001}, the relative diminution of the stellar flux by the occultation of a stellar disk of radius, $R_\star$, by a planetary disk of radius, $R_{\rm Pl,0}$, with an extended atmosphere of vertical extent $z$ is
\begin{equation}
\label{eq:tautransit}
T_{\rm Depth}(\lambda)  = \frac{1}{\pi R^2_\star} \int_0 ^{z_{\rm max}} 2\pi (R_{Pl,0}+z)\left[1-\exp(-\tau(z,\lambda))\right]dz.
\end{equation}
\

The optical ray passing tangentially to the planetary atmosphere during transit has to be mapped to the radial extent of the occulting planetary disk until which the planetary atmosphere is thick at wavelength $\lambda$ such that $R_{Pl}(\lambda)=R_{Pl,0}+z_{\rm max}=R_{Pl}(\lambda)$. 
Using \texttt{petitRADTRANS} for transit depth calculations, we follow the formalism as outlined in \citep{Molliere2019} who derive $R_{Pl}(\lambda)$ by calculating the optical depth of a ray of light, grazing the atmosphere above the planetary center through a series of concentric circles with $r_1$ to $r_N$, where $r_1$ is the radius of the outermost and $r_N$ the radius of the innermost circle, assumed to be optical thick for all wavelengths. In our case, this corresponds to $p_{\rm bottom}=700$~bar and the outermost circle corresponds to $p=10^{-5}$~bar, where hydrostatic equilibrium and the ideal gas law is assumed to transform between pressure and radial extent of the atmosphere. 

For each circle segment that the ray traverses, the local gas density and total opacity is calculated to derive the local optical depth $\tau_i(\lambda)$ between to concentric circles of radii $r_i, r_i+1$. The path of the grazing ray, for which the deepest ring with radius $r_N$ fulfills the condition that the total optical depth, that is, the sum of optical depths in each concentric ring 
$\mathcal{T}
=\sum_1^N \tau_i(\lambda),...\tau_N(\lambda)$ is equal to one, is mapped to the radial direction via the path integral

\begin{equation}
\pi R^2_{Pl}(\lambda)=2 \pi \int_0^{R_{Pl,0}} r \left(1-\mathcal{T}(\lambda)\right) \, dr = \pi R^2_{P,0}\left(1-\mathcal{T}(\lambda)\right),   
\end{equation}
$R_{Pl,0}$ is the reference planetary radius, here $R_{Pl,0}= 1 R_{\rm Jup}$.

For each vertical level at gas pressure, $p_{\rm gas}$, the necessary cloud properties from the kinetic cloud models are used: cloud species volume fractions ($V_s/V$, where $V_s$ is the volume of material $s$ in an individual cloud particle of volume $V$), cloud particle number density, $n_{\rm d}$, and mean cloud particle size $\langle a \rangle$ at height $z$. The cloud opacity $\kappa_{\rm cloud}$ is then

\begin{equation}
\kappa_{\rm cloud} (\lambda, z) \left[\frac{1}{\rm cm}\right] = \pi \langle a \rangle^2(z) Q_{\rm ext} \left(\lambda,\langle a \rangle,z,V_{\rm s}/V_{\rm tot} \right) n_{\rm d}(z),
\end{equation}

where $Q_{ext}$ is the quantum extinction efficiency of the cloud particles, assuming well-mixed heterogeneous cloud composition (effective medium theory) and applying Mie theory for spherical particles of radius $\langle a \rangle$ (\citet{Kiefer2024iron} for the full set of equations and the relation (see their Eqs. 14 and 15) between $n_{\rm d}$ and cloud dust-to-gas mass ratio $\rho_{\rm d}/\rho_{\rm g}$.

At each p$_{\rm gas}(z)$
%pressure 
layer, %the corresponding 
gas and cloud opacities are added up %to the gas phase opacities
\footnote{Using the same opacities as in \texttt{ExoRad}.}, where equilibrium chemistry is assumed to calculate molecular abundances with  \texttt{GGchem} \citep{GGchem}.  The above transit depth formalism is applied to a series of %temperature-pressure 
(T$_{\rm gas}$, p$_{\rm gas}$)
-columns that cover a segment of the planet extracted from the \texttt{ExoRad} simulations as described below. 

\paragraph{One vs. many latitudes:}
Two cases are explored to understand potential uncertainties in terminator asymmetry calculations: a) only equatorial information (latitude $\theta=0$) are used, b) all planetary latitudes $\theta$ are taken into account, 
hence,
$T_{\rm Depth} \left(\theta\right)$ is calculated and averaged for each limb. The evening and morning transmission spectra and transit depths are calculated for both cases. The (planetary) average  transmission spectra and transit depths  are  the mean of evening and morning transmission spectrum or transit depths. The planetary average transit depths are also calculated for both cases, a and b.

Case a) (equator only)  uses the local  ($T_{\rm gas}$(z), $p_{\rm gas}$(z)) profiles  centered at longitudes $\phi=+90^{\circ}$ (evening terminator) and $\phi=-90^{\circ}$ (morning terminator) and at the equator, that is, latitude $\theta=0^{\circ}$. Each equatorial cell covers $\Delta\phi = 15^{\circ}$ in longitude and $\Delta\theta=\pm 22.5^{\circ}$ in latitude. They are assumed to be representative for all other latitudes $\theta$ at each limb. This case was applied e.g. by \citet{Carone2023} hence,  assumes that the equatorial morning and evening limb each is representative for the whole terminator.

Case b) (all latitudes) derives a transmission spectrum that includes all latitudes, hence, sampling the 3D~cloud model centered at latitudes $\theta=0^{\circ},\pm 22.5^{\circ}, \pm 45^{\circ},  \pm 67.5^{\circ}, \pm 90^{\circ}$. The selected locations encompass each a circular ring segment of $\Delta\theta=22.5^{\circ}$ in latitude at the morning $\phi=-90^{\circ}$ and evening $\phi=+90^{\circ}$ terminator, respectively, except the polar cells that cover $\Delta\theta=11.25^{\circ}$ in latitude at each terminator. Each segment further covers $ \Delta\phi=15^{\circ}$ in longitude as in case a). This case was applied e.g. by \citet{Espinoza2024}.

\paragraph{The (planetary) average transmission spectrum:} For the whole planet this is the combination of the morning and evening transit depth, 
$T_{\rm Depth}(\lambda) = T_{\rm Evening}(\lambda) + T_{\rm Morning} (\lambda)$ as derived from Eq.~\ref{eq:tautransit}.
It is thus assumed that the morning transit depth is representative of the situation where the morning half of the planetary disk is obscuring the stellar disk during transit ingress. By analogy, the evening half of the planetary disk will obscure the stellar disk during transit egress.
Strictly speaking  is this only true if the stellar disk is uniformly bright, which is not the case at the limbs. \citet{Espinoza2024}, however, showed that this contribution can be corrected for by accounting for  limb darkening  during data extraction. For WASP-39b, three different methods of limb darkening treatment were tested to ensure the robustness of correction and thus to enable the comparison of the NIRSpec PRISM data with theoretical evening and morning transit depth calculations from atmosphere models like those presented here.

\paragraph{Terminator asymmetries:} For a  3D planetary atmosphere they may be derived by measuring the transit depth during transit ingress and egress, respectively. When the planet transits centrally across the stellar disk with no positional offset between the planet's projected path and the star's centrum ,i.e., with an impact parameter of zero, then the morning terminator covers the star during ingress and the evening terminator during egress. The ingress transit depth therefor captures the morning limb, $T_{\rm Morning}(\lambda)$,  and the egress transit depth the evening limb, $T_{\rm Evening}(\lambda)$.

In this work, the evening/morning transit depth is calculated by assuming as in \texttt{catwoman} \citep{Jones2020,Espinoza2021} that the morning and evening limb are each represented by half circles, whereas the planetary average transit depth ($T_{\rm Morning}(\lambda)$+$T_{\rm Evening}(\lambda)$) is representative of a full circle. Henceforth, transit depth asymmetries are always expressed as evening minus morning terminator transit depths, $\Delta T_{\rm Depth} = T_{\rm Evening} -T_{\rm Morning}$. Positive transit depth asymmetries indicate a more extended evening limb compared to the morning limb for a given wavelength. Conversely, negative transit depth differences indicate a more extended morning limb for a given wavelength.

\paragraph{Transit photometry:} To derive photometric information for the PLATO, CHEOPS and TESS pass bands from model atmospheres, the instrument transmission function $B^{\rm Instr}$ (Appendix~\ref{sec: Pass}) needs to be considered.
This is facilitated by integrating for a band pass covering the wavelengths $\lambda_1$ and $\lambda_2$ as
\begin{equation}
T_{\rm Depth}^{\rm Instr}(\Delta\lambda) = \int_{\lambda_1}^{\lambda_2} T_{\rm Depth}(\lambda)\, B^{\rm Instr}(\lambda) d\lambda.  
\end{equation}

\section{ The WASP-39b test case}

\label{sec: WASP39b}

The aim of this paper is to explore how terminator asymmetries may be used to characterize exoplanet climate regimes and possibly exoplanets atmospheres in general with observations from  CHEOPS, TESS, JWST and PLATO. The well-know  JWST early release science (ERS) target WASP-39b is %therefore
used to test the hierarchical modelling chain employed in this paper in application to %a large 
the wavelength range covering  JWST (NIRCAM, NIRSpec, NIRISS), HST and VLT. The aim is to arrive at  an as-consistent-as-possible fit to the complete wavelength range $\lambda=0.3\,\ldots\,5.25\mu$m with one model. This exercise is conducted to also gain understanding for potential adaption needs of the kinetic cloud model utilizes here.

\begin{figure}
    \centering
     \includegraphics[width=0.24\textwidth]{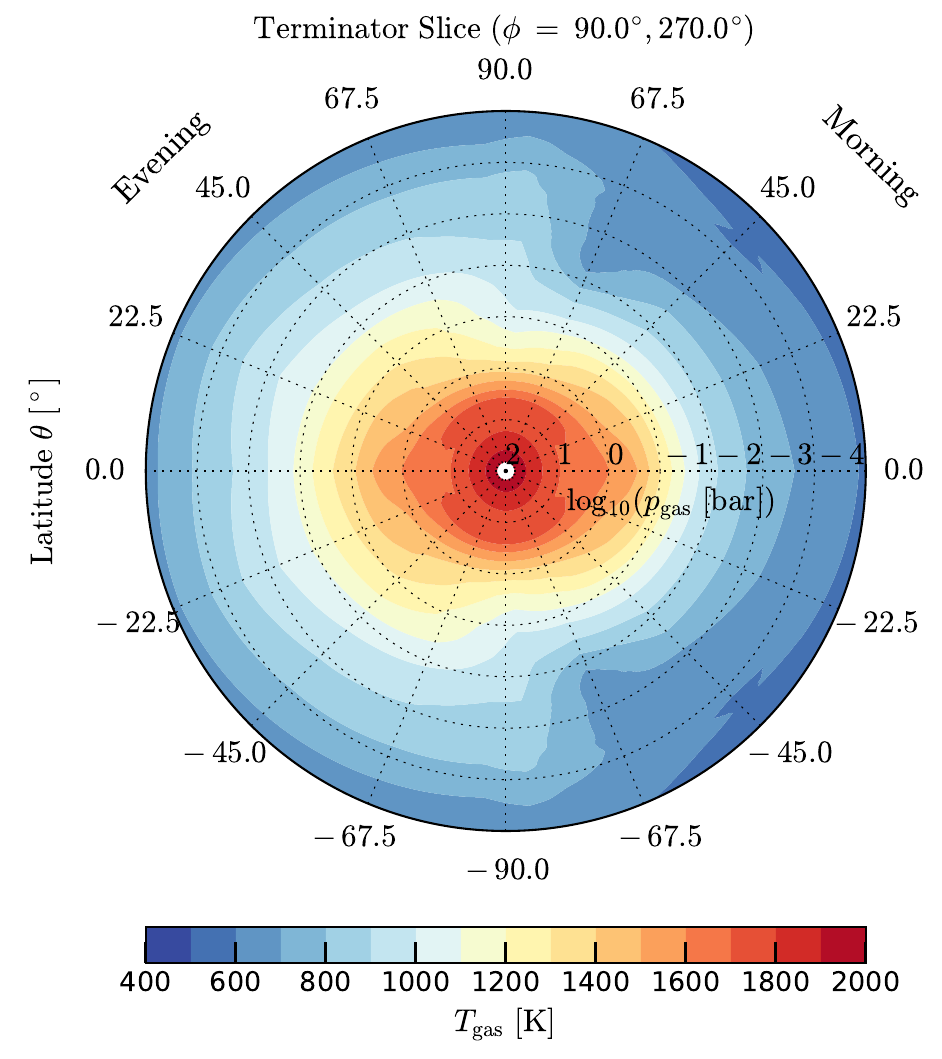}
    \includegraphics[width=0.24\textwidth]{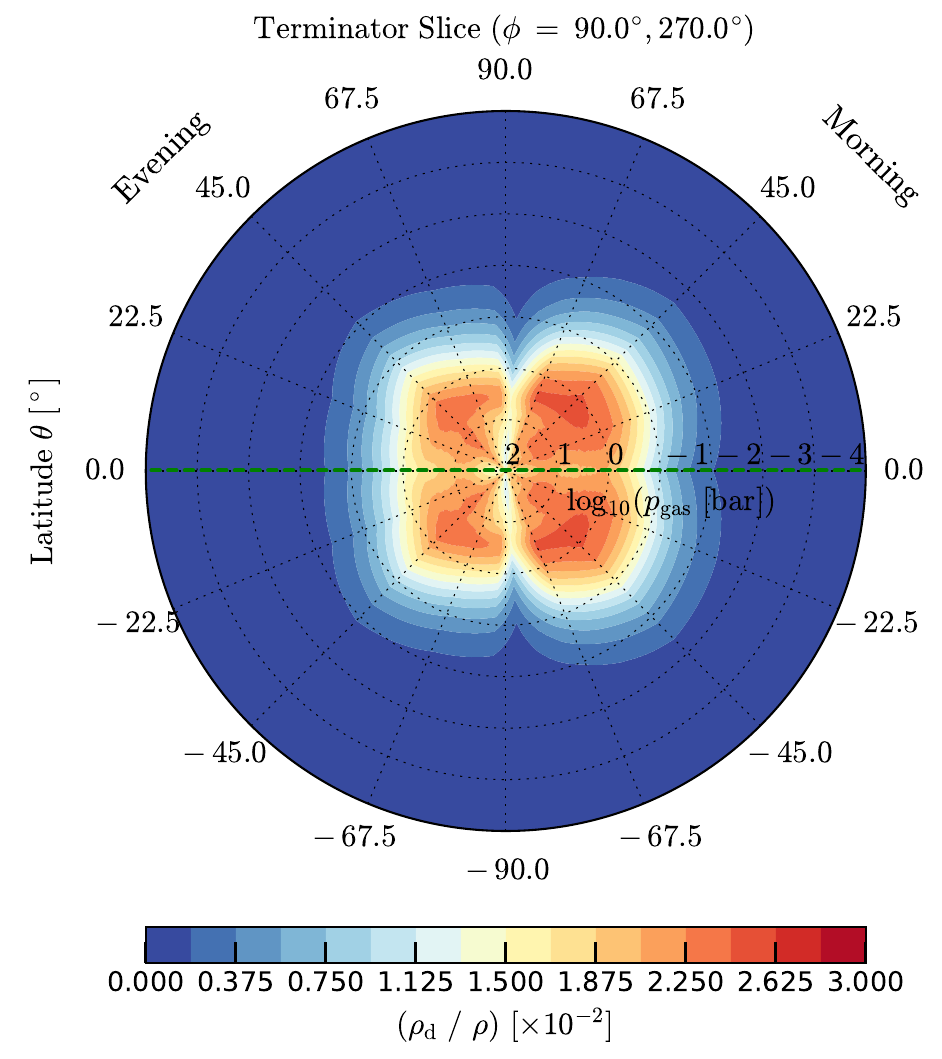}
    \caption{WASP-39b 3D model cross sections. Shown are the local gas temperature, T$_{\rm gas}$ [K] (left)  and cloud dust-to-gas mass ratio, $\rho_{\rm dust}/\rho_{\rm gas}$  (right). The values are derived from the 3D GCM results and shown as cross section across the evening (left hemisphere) and morning terminator (right hemisphere).}
    \label{fig:W39b_Terminator_slices}
\end{figure}

\smallskip
The cloudy hot Jupiter WASP-39b provides with precise observations from the optical (VLT/FORS2), near-infrared (HST/WC3/IR) to the infrared (JWST/NIRSpec, NIRISS, NIRCam and MIRI) a prime data set to gain insights about detailed cloud properties \citep{Ahrer2023,Feinstein2023,Espinoza2024,Nikolov2016,Wakeford2018,Alderson2023,Rustamkulov2023,Powell2024b}. Already the first analysis suggested 'patchy' clouds and limb asymmetries that may be indicative of asymmetric cloud coverage in WASP-39b \citep{Espinoza2024}. We thus use the complete WASP-39b data set for a comprehensive re-analysis of cloud chemistry and physics of the detailed kinetic cloud model that underlies the \texttt{IWF Graz} 3D cloud model (Sect.~\ref{sec:Cloudmodelling}).

A similar approach led in \citet{Carone2023} to a re-evaluation of the vertical mixing for WASP-39b based on the previous version of the cloudy \texttt{ExoRad} grid \citet{Helling2023}. 
{In the present paper, the opacities arising from cloud properties are 
directly implemented in the radiative transfer in post-processing (Sect.~\ref{sec:Transmission}) based on the 3D temperature structure from the latest Exorad 3D climate model (Sect.~\ref{sec:ExoRad}) in combination with the kinetic cloud formation (Sect.~\ref{sec:Cloudmodelling}) and applied to WASP-39b\footnote{Here, an internal temperature of $T_{\rm in} =350$~K, radius of $R_P=1.27~R_{\rm Jup}$, solar C/O ratio of 0.55 is   10$\times$ solar metallicity was assumed. Further, a host star of radius $R_*=0.9 R_{\rm Sun}$ and $T_{\rm eff}=5400$~K is adopted, which gives for a semi major axis $a=0.0486$~AU a  global temperature $T_{\rm global}=1120$~K.}. 

The terminator cross section maps for the  gas temperature and cloud dust-to-gas mass ratio, $\rho_{\rm dust}/\rho_{\rm gas}$  for WASP-39b (Fig.~\ref{fig:W39b_Terminator_slices}) already provide first insights about cloud coverage asymmetries. There are only small differences between the equatorial and higher latitude regions for each terminator. The cloud distribution shows, however, a larger difference between the morning and evening limb with a higher cloud mass (by a factor of 2 at $p=10^{-3}$~bar) at the morning terminator. In any case, the majority of the cloud mass is located below $p=10^{-3}$~bar, which is in agreement with retrieved cloud top estimates based on e.g. JWST/NIRCam observations \citep{Ahrer2023} and predictions based on optical depth $\tau=1$ estimates \citep{Carone2023}.
The equatorial synthetic transmission spectrum of WASP-39b derived from these results (Fig.~\ref{fig:W39b_Fe_remove}) does, 
however,  suggest an atmosphere that is opaque also at atmosphere layers above $p=10^{-3}$~bar. 

How  should the cloud properties derived from a detailed, kinetic cloud model be adapted in order to reproduce the observed transmission spectra? Assuming that the cloud model is sufficiently complete regarding the formation and destruction processes, which properties need to change to reduce the cloud opacity in upper atmosphere? 

In the following, the optical properties of the cloud particles 
that affect the atmosphere regime probed by transmission spectroscopy ($p_{\rm gas} = 10^{-3}\,\dots\,10^{-5}$~bar) are analyzed and compared to the available WASP-39b data from  VLT/FORS2, HST/WFC3, NIRSpec PRISM and NIRSpec/G395H. For the grid calculations, equilibrium chemistry is assumed that also includes \ce{CH4}. For the special case of WASP-39b, NIRSpec PRISM observations \citep{Espinoza2024} suggest the absence of \ce{CH4} at the morning and the evening limb due to disequilibrium effects. Thus, here - and only here -  \ce{CH4} is omitted as a opacity source to mimic to first order the reduction of \ce{CH4} abundances to facilitate comparison between theoretical calculations and observational data.
Section~\ref{ss:Vs} explores the impact of iron-bearing materials on the cloud opacity, and Sect~\ref{ss:nd} explores the role of small particles in the cloud nucleation regions above the main cloud deck. Section~\ref{ss:oooo} show that both, iron content and small particles in the upper atmosphere, affect the synthetic evening and morning transmission spectra as well as averaged transmission spectra considerably.

\begin{figure}
    \centering
    \includegraphics[width=0.5\textwidth]{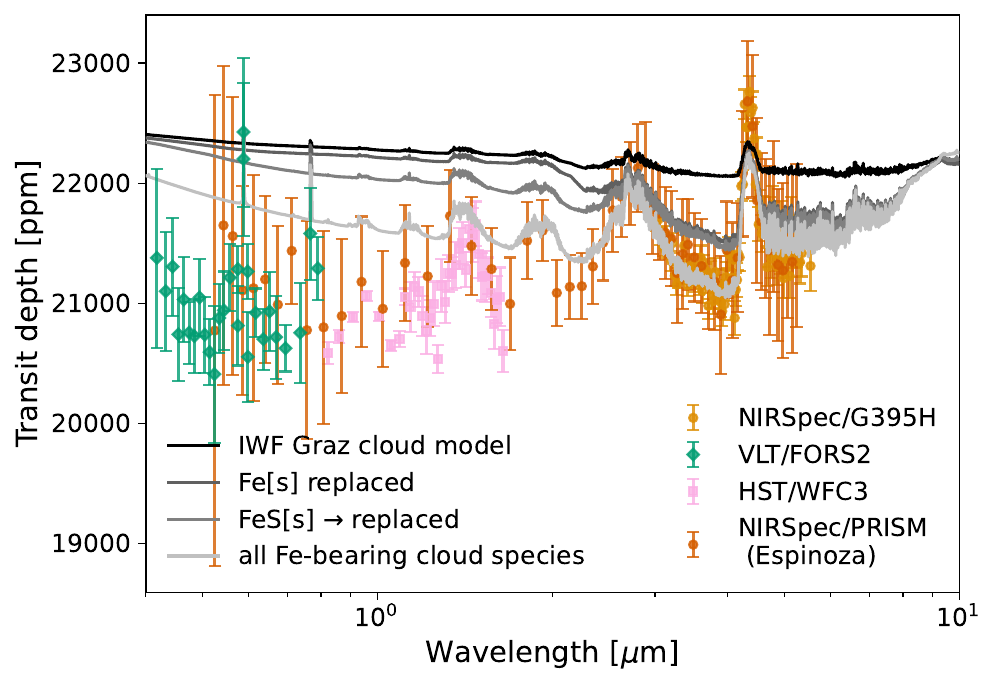}
    \caption{WASP-39b transmission spectra and cloud scenarios. Spectra are observed with JWST (orange), HST (magenta
    ) and VLT (green) in comparison to  
    models with different Fe content: full \texttt{IWF Graz} cloud model (black line), 
Fe[s] replaced (dark gray), FeS[s] replaced  (middle grey), all Fe-binding materials replaced  (light gray). Fe-free, mixed-material cloud compositions provides a good fit in the near-IR but not in the optical.} 

    \label{fig:W39b_Fe_remove}
\end{figure}

\begin{figure}
    \centering
    \includegraphics[width=0.5\textwidth]{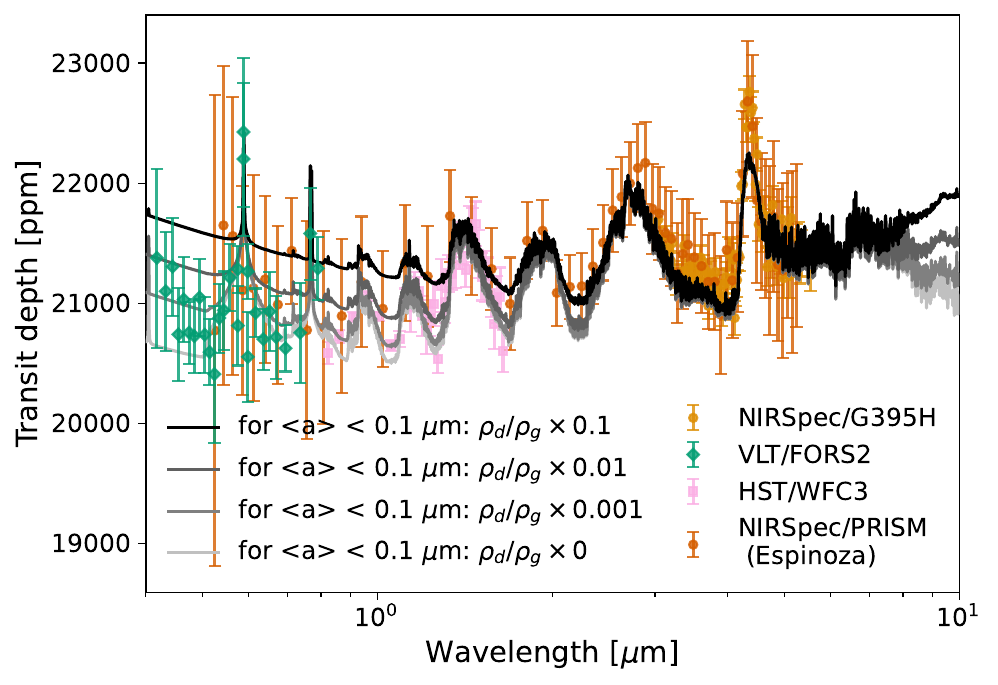}
    \caption{WASP-39b transmission spectra and calculations with different cloud mass load. Data are from  JWST (orange), HST (magenta) and VLT (green) shown in comparison to Fe-free cloud scenarios with  different cloud mass loads, $\rho_g/\rho_d(z)$, in atmospheric layers where $\langle a \rangle <  0.1~\mu$m (middle gray: $10^{-3}\rho_g/\rho_d(z)$, dark gray:  $10^{-2}\rho_g/\rho_d(z)$, black: $0.1\rho_g/\rho_d(z)$). The light gray line depicts the cloud-free case in the upper atmosphere. The models with Fe-free cloud materials and a reduced cloud mass load fit the whole wavelength range well.}  \label{fig:W39b_Fe_red_submic_detailed}
\end{figure}
    
\subsection{The iron conundrum}
\label{ss:Vs}

 WASP-39b is suggested to exhibit a two-layered cloud composition with silicate bearing condensate particles (e.g.\ce{SiO2}[s],\ce{Fe2SiO3}[s],\ce{MgSiO3}[s],\ce{Mg2SiO4}[s]) dominating in the upper atmosphere ($p_{\rm gas}<1$~bar) and with high temperature condensates, including Fe[s], dominating at deeper layers \citep[][Fig.~9]{Carone2023}.
 A model of similar complexity, \texttt{ExoLyn} \citep{ExoLyn2024}, yields a very comparable cloud structure with a transition in cloud material from silicate to iron dominated in deeper atmosphere layers between local gas temperatures, $T_{\rm gas}=1460\,\dots\,1600$K that occurs in our model at 1~bar for WASP-39b. Both cloud models also infer that while the upper silicate cloud is depleted in Fe[s] compared to the lower cloud, still a volume fraction of a few percent of Fe[s] persists even at $p_{\rm gas}=10^{-4}$~bar. Thus, the general cloud structure and major cloud components agree between those two kinetic cloud models. In contrast to our model, \texttt{ExoLyn} does not consider, however, a mixture of iron and silicate cloud composition in the form of \ce{Fe2SiO3}[s] that can result in up to 16\% in the cloud volume fraction \citep[][Fig.7]{Carone2023}.

It has been further shown that even small fractions of Fe[s] as well as \ce{Fe2SiO3}[s] have large refractive indices and exert a noticeable impact on the overall cloud opacity \citep{2006A&A...460L...9W,Chubb2024,Kiefer2024iron}. Observations by \citep{Grant2023} and \citep{Dyrek2024} indicate silicate (\ce{MgSiO3}[s],\ce{Mg2SiO4}[s]) and silica cloud (\ce{SiO2}[s]) composition for warm (WASP-107b, $T_{\rm global}=770$~K) to ultra-hot tidally Jupiters (WASP-17b, $T_{\rm global}=1800$~K), in agreement with our general cloud structure from two kinetic clouds models (\texttt{DRIFT} and \texttt{ExoLyn}). The same observations indicate, however, a dearth of Fe[s] and iron-bearing cloud species like \ce{Fe2SiO4}[s]. The complete lack of these species is puzzling, because iron makes up at least 6\% of the total solar elemental abundances \citep{Asplund2009}. Further, high resolution observations detected Fe in the gas phase of several ultra-hot Jupiters like WASP-121b \citep{Sing2019} and WASP-189b \citep{Sreejith2023} and even indications of Fe condensation in the ultra-hot Jupiter WASP-76b have been found \citep{Ehrenreich2020}.

To provide further guidance on the iron conundrum and to gain insights which physical mechanisms and chemical reaction rates may need to be revised in the complex kinetic clouds model, an adjusted cloud opacity formalism is introduced. It does not change the kinetic cloud model itself but instead adjusts the optical properties for transmission spectrum calculations (Sec.~\ref{sec:Transmission}). 
For this, the opacity of iron bearing condensates is replaced  with those of their closest chemical structure and/or monomer size. Fe[s] is the only atomic monomer material such that it is replaced by the only material observed so far (SiO$_2$[s]).
This approach maintains to first order the total cloud mass density and distribution in the \texttt{IWF Graz} model. 

The following volume fraction of cloud particle species in the Mie scattering calculations are replaced: Fe[s] $\rightarrow$ SiO$_2$[s], Fe$_2$SiO$_4$[s]  $\rightarrow$ Mg$_2$SiO$_4$[s], FeO[s] $\rightarrow$ MgO[s], Fe$_2$O$_3$[s] $\rightarrow$ MgO[s], FeS[s] $\rightarrow$ MgO[s]. Replacing successively 
the iron material content as
opacity source improves agreement  between  the synthetic spectra and the available WASP-39b observations  (Fig. \ref{fig:W39b_Fe_remove}). Figure \ref{fig:W39b_Fe_remove} also demonstrates  that even 3\% Fe[s] (dark gray line) would have an large impact on the transmission spectrum between $2\,\ldots\,5~\mu$m. FeS[s] (second most dark gray line) that makes up 7\% in cloud volume fraction mostly affects the transmission spectrum in the optical and NIR wavelength regions ($0.33\,\ldots\,2~\mu$m). Substituting \ce{FeO[s],Fe2O3[s]}, and \ce{Fe2SiO4[s]}, reduces the cloud opacity in the whole wavelength range.  The complete adjustment to iron-free silicate clouds yields a theoretical transmission spectrum that agrees within the error bars with observations for the $2\,\ldots\,5~\mu$m wavelength region\footnote{Since this model does not contain photochemistry, \ce{SO2} can't be reproduced.}. 

\subsection{Upper atmosphere cloud
particles}
\label{ss:nd}

Section~\ref{ss:Vs} may be considered to have addressed the discrepancy between model and observations in the near-IR spectral range. In the optical wavelength range, however, a large discrepancy remains between the WASP-39b observations and the theoretical transmission spectrum  
also for the iron-free calculations. Most notably, the pressure broadened \ce{Na}I lines centered at 0.59~$\mu$m and the \ce{H2O} absorption bands between 0.8-2~$\mu$m are strongly muted in contrast to the VLT/FORS2 data and HST/WFC3 data. 

Transmission spectra in the optical wavelengths regions are dominated by scattering of submicron particles in the planetary atmosphere. Further, the \texttt{IWF Graz} model predicts mean cloud particles sizes $\langle a \rangle \leq 0.1$~$\mu$m to dominate the entire upper atmosphere (Fig.~\ref{fig:W39b_Terminator_slices}, \citet[][Fig.~9]{Carone2023}). 
To test the impact of the amount of the small cloud particles
on the optical part of the transmission spectra,  the radiative transfer calculations are adjusted accordingly. As a first case, a cloud-free low-pressure atmosphere is tested for regions where the mean cloud particle sizes $\langle a \rangle <0.1~\mu$m: 
The cloud mass load, $\rho_{\rm dust}/\rho_{\rm gas}$,  is set to zero in each column of extent $z$ and local gas pressure $p_{\rm gas}(z)$ if the mean cloud particle size  $\langle a \rangle<0.1~\mu$m, hence for  
$p_{\rm gas} < p_{\rm gas}(\langle a\rangle (z)< 0.1~\mu m)$.

A yet better agreement with the data from the optical to the near infrared 
(Fig.~\ref{fig:W39b_Fe_red_submic_detailed})
was achieved. The transmission spectrum without any sub-micron particles in the upper most atmosphere layers, however, misses to reproduce the scattering slope observed with VLT/FORS2 between 0.3 and 0.4~$\mu$m.
A more detailed investigation indicates how high sensitivity of transmission observations between 0.3 and 1 $\mu$m can inform us about the density of submicron particles (Fig.~\ref{fig:W39b_Fe_red_submic_detailed}):
A reduction  of the cloud dust-to-gas mass ratio in atmospheric layers, for which submicron particles dominate ($\langle a \rangle <  0.1~\mu$m), by scaling $\rho_{\rm dust}/\rho_{\rm gas}$ with the constant factor $0.001$ provides a better fit to the optical data of WASP-39b than a model without any submicron particles. This investigation thus elucidates that the clouds in 3D hot Jupiters do not only consist of a thick cloud dominated by micron-size particles. Above these, the atmosphere is dominated by submicron particles, indicative of atmosphere layers
where cloud condensation seeds form.
Transmission spectra covering the optical ($<1 \mu$m) or infrared ($>8 \mu$m) wavelengths are the most informative for the diagnostics of 
the small particle density in the upper-most atmospheric regions.

\subsection{ 
Morning and evening transit asymmetry data}
\label{ss:oooo}

So far,  the simulated cloud properties for WASP-39b were inferred based on an average transmission spectrum, that is, averaged over the morning and evening limb. NIRSpec measurements give in addition also access to details of the morning and evening terminator \citep{Espinoza2024}. 

Applying the \texttt{IWF Graz 3D cloud} model without iron and sub-micron particles adjustment already yields qualitatively a morning and evening asymmetry, albeit with very muted spectral \ce{H2O} and \ce{CO2} features (Fig.~\ref{fig:W39b_asym}). The compositional adjusted cloud model, in which iron-bearing cloud particles are replaced, provides a better agreement with the observational spectra, in particular with the evening terminator spectrum. However, the amplitude of the \ce{CO2} band at 4.5~$\mu$m is under-predicted. The observed morning terminator spectrum of WASP-39b is also flatter, even when different submicron cloud particle adjustments are applied. The relative flatness of the observed morning transmission spectrum compared to the synthetic spectrum thus suggests that the location of the optically thick cloud top is higher than predicted.
Consequently, the quantitative differences between the simulated evening and morning transmission spectra are generally about 200~ppm, that is, smaller by a factor of 2 compared to observations (Figure~\ref{fig:W39b_asym}, bottom). Qualitatively, we agree with observations in that we also find the strongest asymmetry in the \ce{CO2} absorption band at 4.5~$\mu$m.

\begin{figure}
    \centering
    \includegraphics[width=0.5\textwidth]{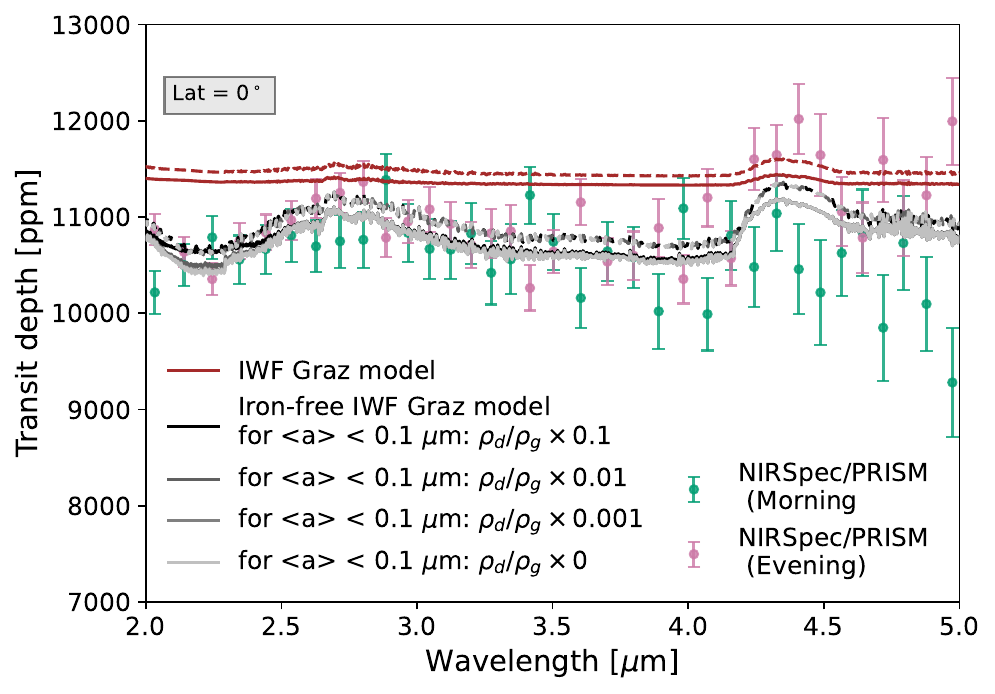}
       \includegraphics[width=0.5\textwidth]{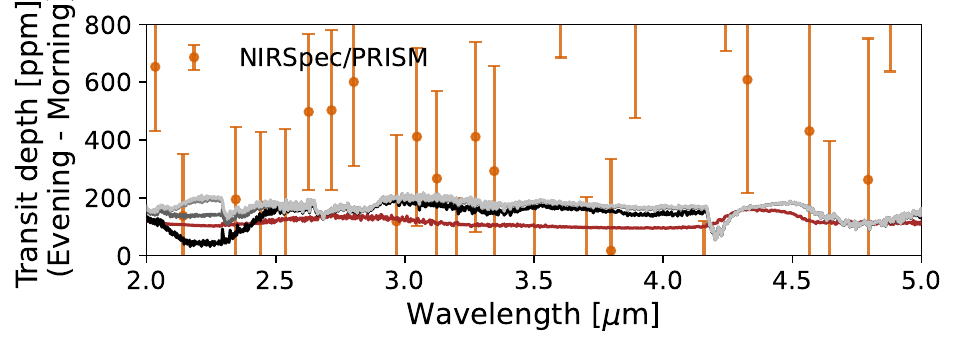}
    \caption{Equatorial transit asymmetry calculations in comparison with WASP-39b data. Top: JWST/NIRSpec PRISM transmission evening (pink) and morning transmission spectrum (green) for WASP-39b. Solid/Dashed  lines depict equatorial morning/evening transmission spectra calculated with the \texttt{IWF Graz} cloud model (red: with Fe and full cloud mass load, $\rho_g/\rho_d(z)$, that are offset to the data and other scenarios for clarity) and for Fe-free cloud scenarios with  different cloud mass loads, $\rho_g/\rho_d(z)$, in atmospheric layers where $\langle a \rangle <  0.1~\mu$m (middle gray: $10^{-3}\rho_g/\rho_d(z)$, dark gray:  $10^{-2}\rho_g/\rho_d(z)$, black: $0.1\rho_g/\rho_d(z)$).
    The light gray line depicts the cloud-free case 
    in the upper atmosphere.
      Bottom: Differences between the evening  and morning  terminator transmission spectra.}
    \label{fig:W39b_asym}
\end{figure}

So far, only the equatorial region was considered in the calculation of the transmission spectrum (Sec.~\ref{sec:Transmission}) in the previous sections. The quality of the JWST data and the improvement of the compositional adjusted cloud model allows to investigate in how far the results may change, when higher latitudes are considered.
We find that evening and morning transmission spectra, for which all latitudes are considered, generally yield similar results compared to the equatorial transmission spectra. The simulated transit asymmetries, however, can be locally reduced by about 50~ppm (Fig.~\ref{fig:W39b_asym_all_Lats}). Still, the differences between equatorial and full latitudes transmission spectra are within the uncertainty of the available data and thus not significant.

Interestingly, for both transmission spectra calculations, differences for morning and evening terminator transmission spectra occur for diverse submicron cloud particle scenarios for $\lambda\!=\!2\,\ldots\,2.5~\mu$m. The precision of the NIRSpec PRISM data at its current state is, however, not conclusive to rule out any of the scenarios.

\begin{figure}
    \centering
    \includegraphics[width=0.5\textwidth]{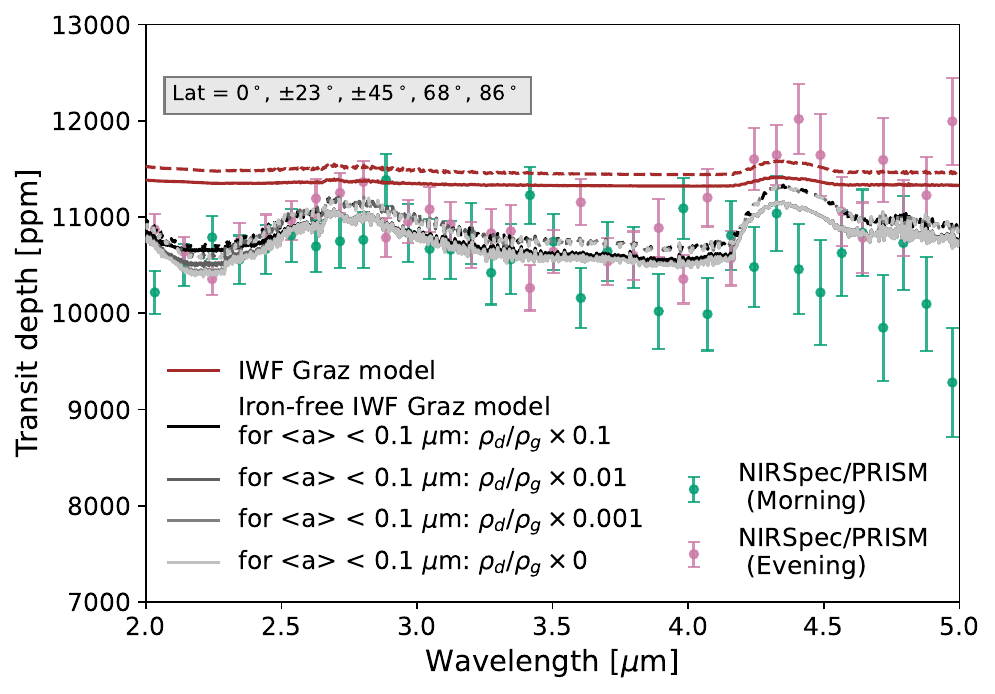}
       \includegraphics[width=0.5\textwidth]{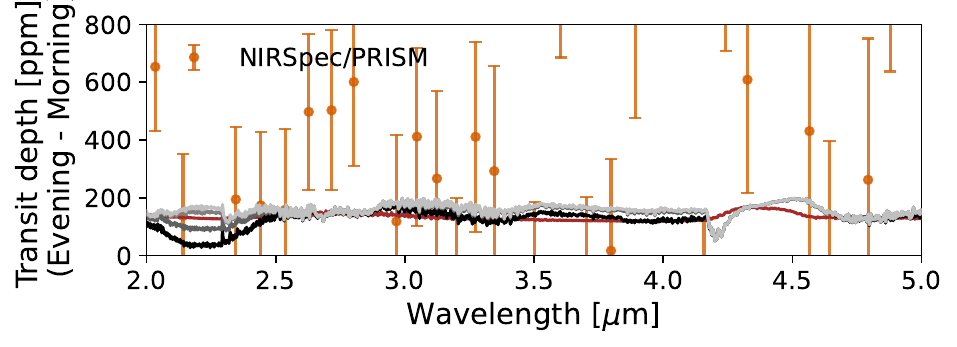}
    \caption{Latitudinally averaged transit asymmetry calculations in comparison with WASP-39b data. Top: JWST/NIRSpec PRISM transmission evening (pink) and morning transmission spectrum (green) for WASP-39b. Solid/Dashed  lines depict morning/evening transmission spectra considering all latitudes ($\theta = 0^{\circ}, \pm 23^{\circ},\pm 45^{\circ},\pm 68^{\circ},\pm 86^{\circ}$) calculated with the \texttt{IWF Graz} cloud model (red: Results with Fe and full cloud mass load, $\rho_g/\rho_d(z)$, that are offset to the data and other scenarios for clarity) and for Fe-free cloud scenarios with  different cloud mass loads, $\rho_g/\rho_d(z)$, in atmospheric layers where $\langle a \rangle <  0.1~\mu$m (middle gray: $10^{-3}\rho_g/\rho_d(z)$, dark gray:  $10^{-2}\rho_g/\rho_d(z)$, black: $0.1\rho_g/\rho_d(z)$). The light gray line depicts the cloud-free case 
    in the upper atmosphere.  Bottom: Differences between the evening  and morning  terminator transmission spectra. }
    \label{fig:W39b_asym_all_Lats}
\end{figure}

\subsection{Summary for the WASP-39b test case} 

The comparison between the \texttt{IWF Graz} 3D cloud model and low-resolution observations suggest the presence of iron-free
cloud particles and constrains the cloud particle mass load
of submicron cloud  particles in the upper atmosphere where p$_{\rm gas}<10^{-3}$bar. The adjusted hierarchical cloud model qualitatively agrees with the observations from the optical to the infrared, including the NIRSpec PRISM transit depth asymmetry data. Quantitatively, the transit asymmetries are underpredicted in particular at the morning terminator and can thus be seen as a conservative model.

The addition of higher latitudes does not lead to significant changes in the synthetic spectrum compared to an equatorial transmission  spectrum. The inspection of T$_{\rm gas}$ for latitudinal differences at each limb (Fig.~\ref{fig:W39b_Terminator_slices}) confirms that WASP-39b (T$_{\rm global}\sim$1120~K) shows little latitudinal variation and only moderate differences between the cloudy morning and evening limbs.

The WASP-39b test case highlights that combining data across a large wavelength range, from the optical to the infrared, allows to derive a coherent cloud picture that includes differences between the morning and evening clouds. The hierarchical cloud model captures important trends that can be used to inform observational strategies and improvement in cloud modelling.

\section{Evening-morning transmission spectra for different climate states}
\label{sec: Three cases}

The JWST target WASP-39b  was used to test the hierarchical modelling chain employed in this paper in application to a large wavelength range covering  JWST (NIRCAM, NIRSpec, NIRISS), HST and VLT. The complete wavelength range $\lambda=0.33\,\ldots\,5.25~\mu$m has been well reproduced by the \texttt{Exorad} 3D climate modelling in combination with a Fe-free kinetic cloud model with a reduced cloud mass load of small cloud particles in the uppermost atmosphere. With this insight, the iron-free \texttt{IWF Graz} 3D cloud model is applied to the 3D AFGKM \texttt{ExcoRad} grid to investigate the potential of limb asymmetry observations from the optical to the infrared for various climate states and host stars for a coherent ensemble of atmosphere models. 

Section~\ref{ss:3cs} explores the transmission spectra for the three climate regimes introduced in \cite{Helling2023}.  Section~\ref{ss:transdepth} %provides the transits depth to 
addresses the question of observability of terminator asymmetries across different wavelengths, where evening and morning limb differences in transit depths are explored (Sect.~\ref{sec:Transmission}). Section~\ref{sec:cloudimpact}   presents the results for the whole model grid ensemble with focus on observability with PLATO in the context with other space missions. To also facilitate a better understanding of the underlying physical mechanism behind the simulated transit asymmetry, results will also presented in comparison to cloud-free spectra. In the hierarchical cloud modelling used here, such a comparison is very helpful because both, the cloud-free and cloudy spectra are based on the same 3D temperature structure. Effects due to the local gas temperature differences and due to different cloud properties can, hence,  be better disentangle compared to e.g., \citet{Kenneth2025}. The results will also be presented for equatorial transmission spectra  and spectra, for which all latitudes are taken into account to assess the potential impact of latitudinal variations.

\subsection{Transmission spectra for gas giants in three climate states}
\label{ss:3cs}

\begin{figure*}%[ht]
%    \centering
    \includegraphics[width=0.275\textwidth]{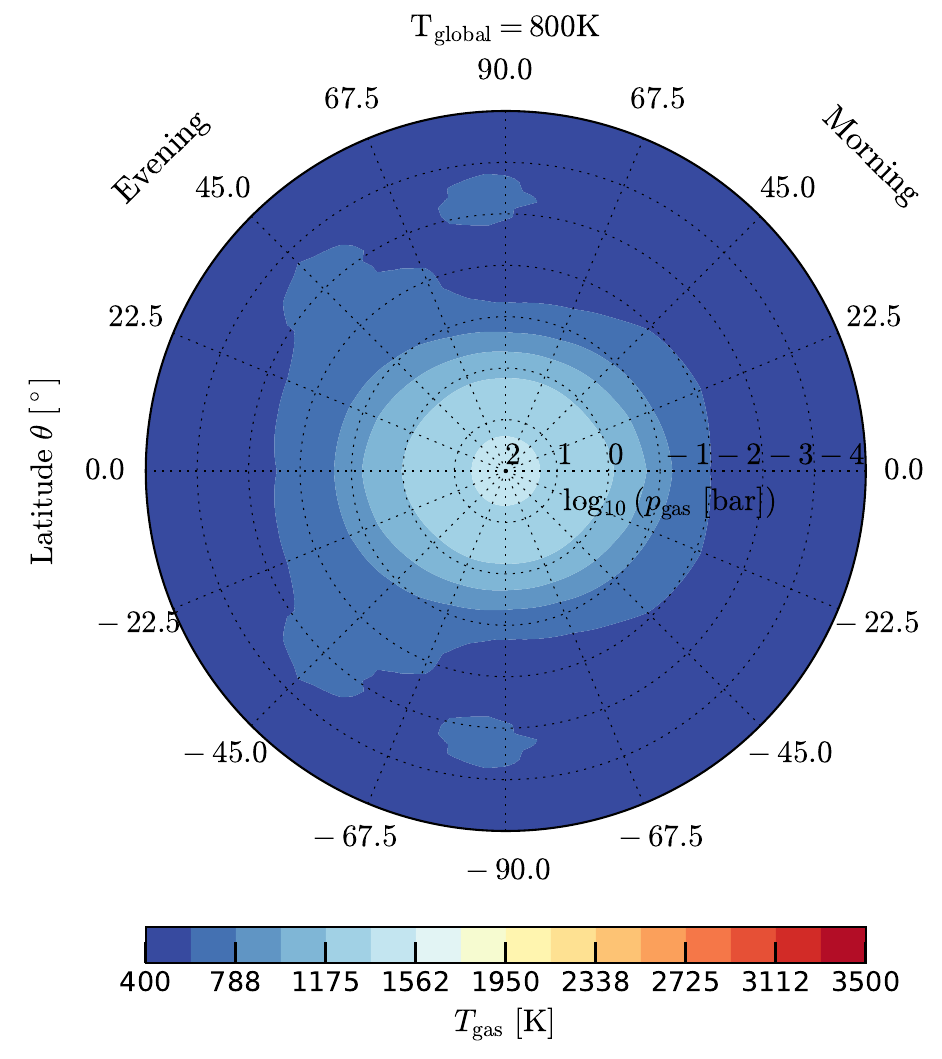}
        \includegraphics[width=0.275\textwidth]{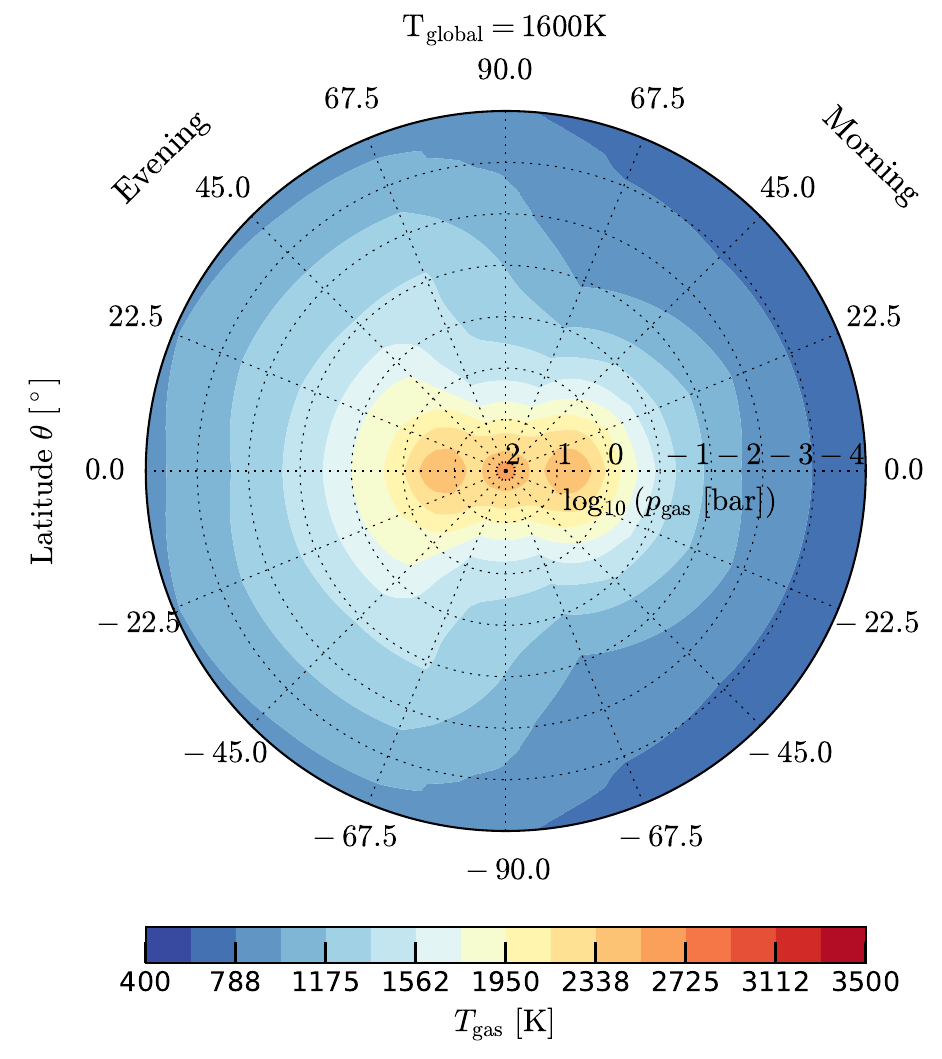}
        \includegraphics[width=0.275\textwidth]{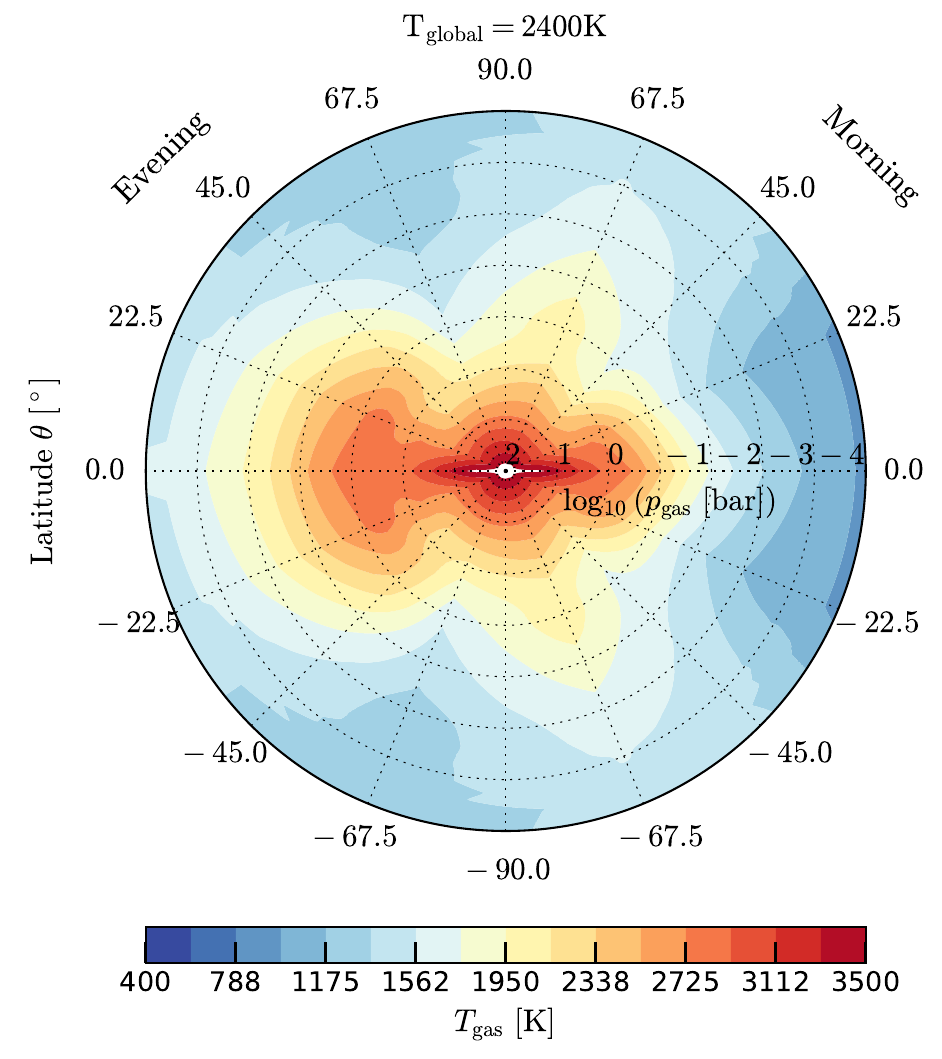}
            \includegraphics[width=0.275\textwidth]{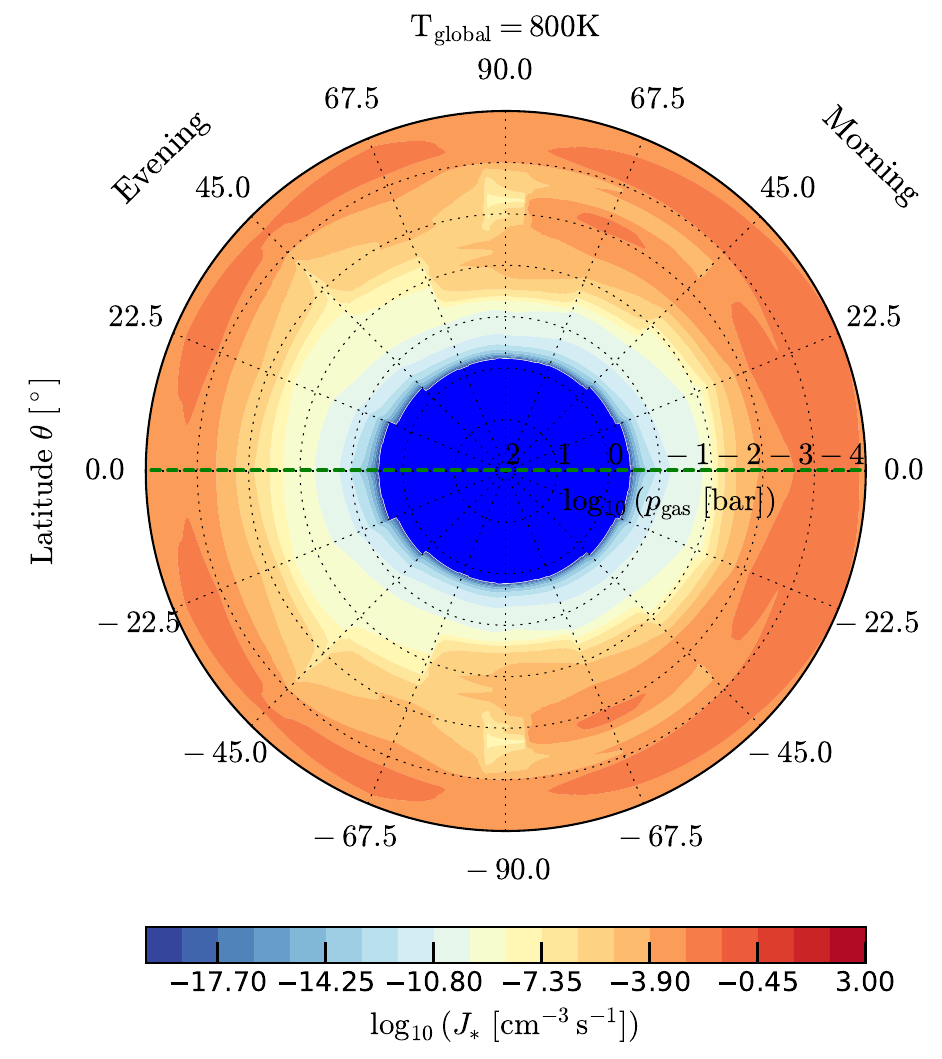}
        \includegraphics[width=0.275\textwidth]{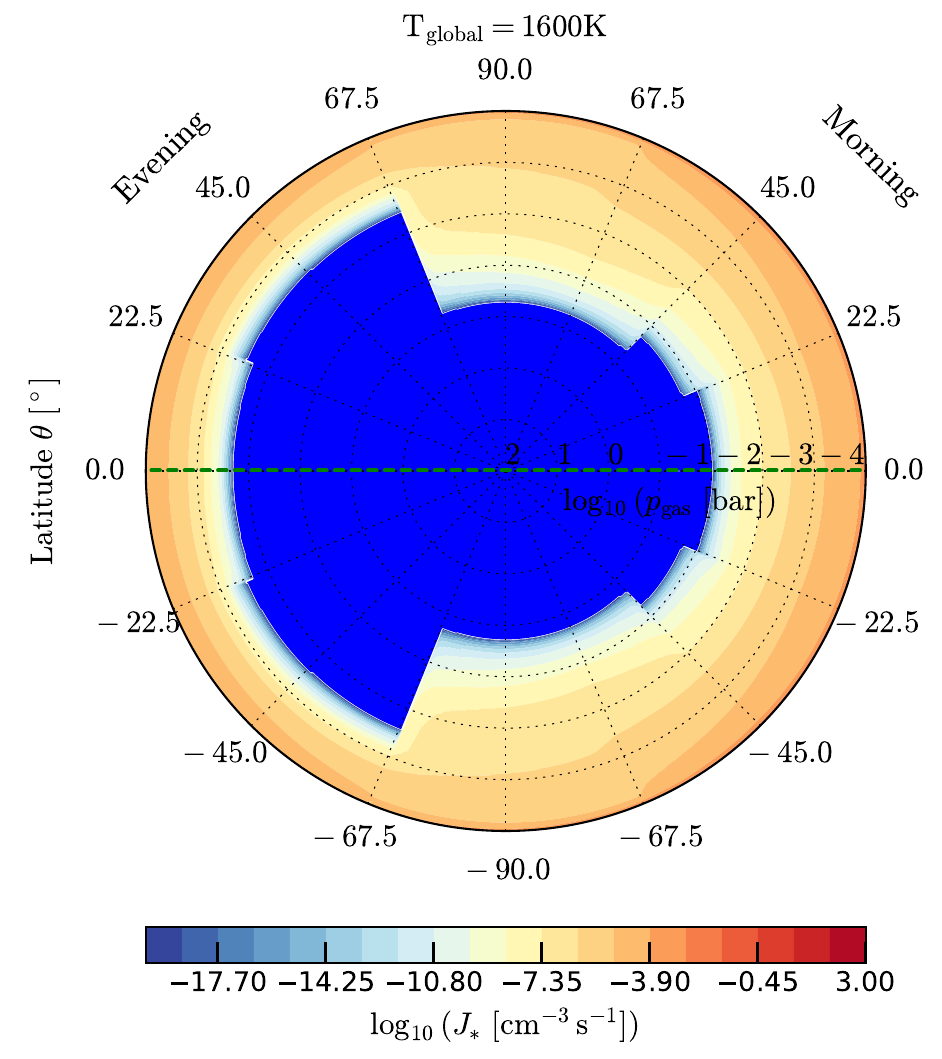}
        \includegraphics[width=0.275\textwidth]{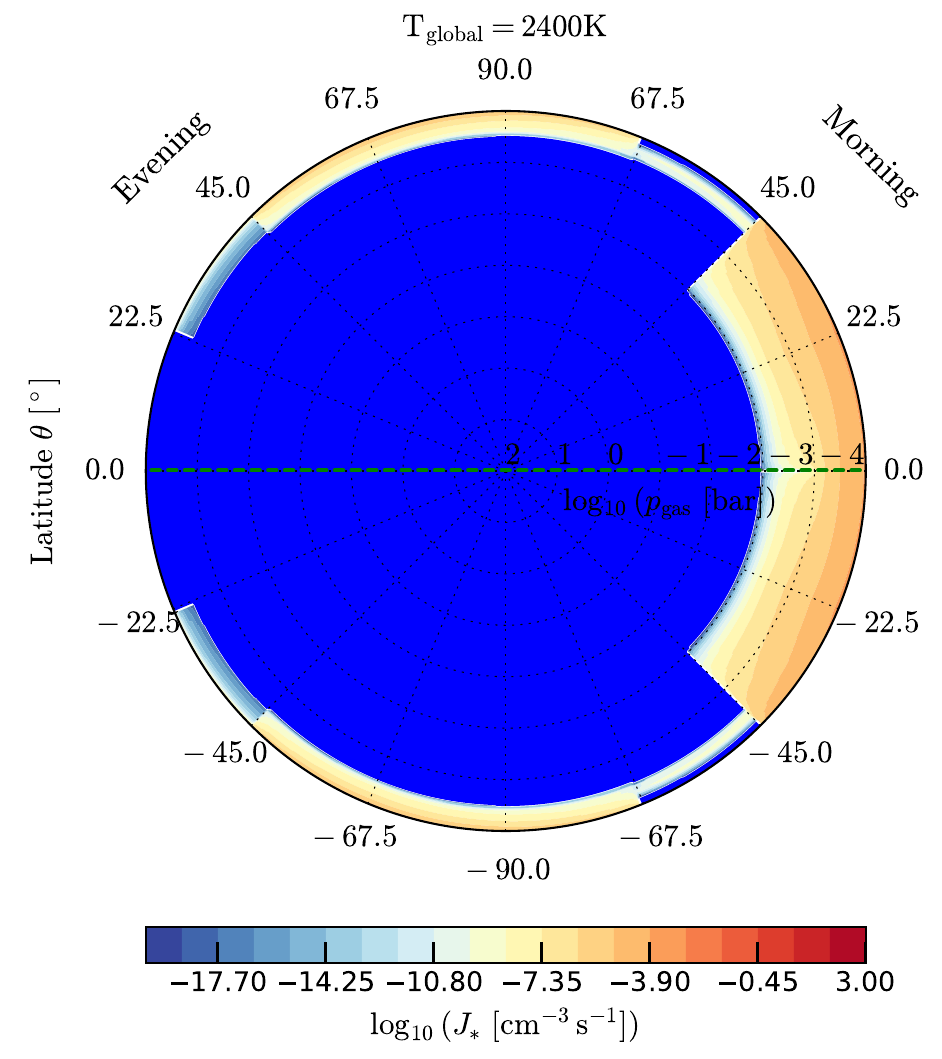}
        \includegraphics[width=0.275\textwidth]{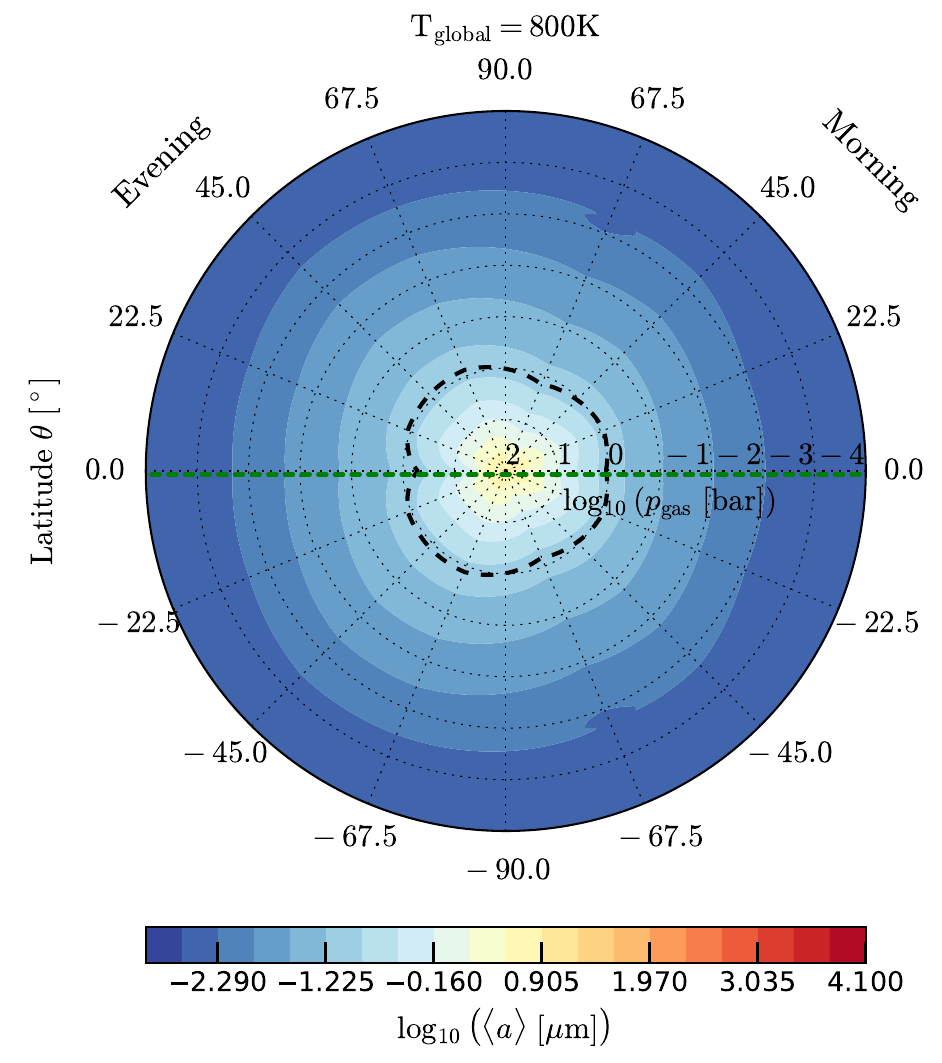}
        \includegraphics[width=0.275\textwidth]{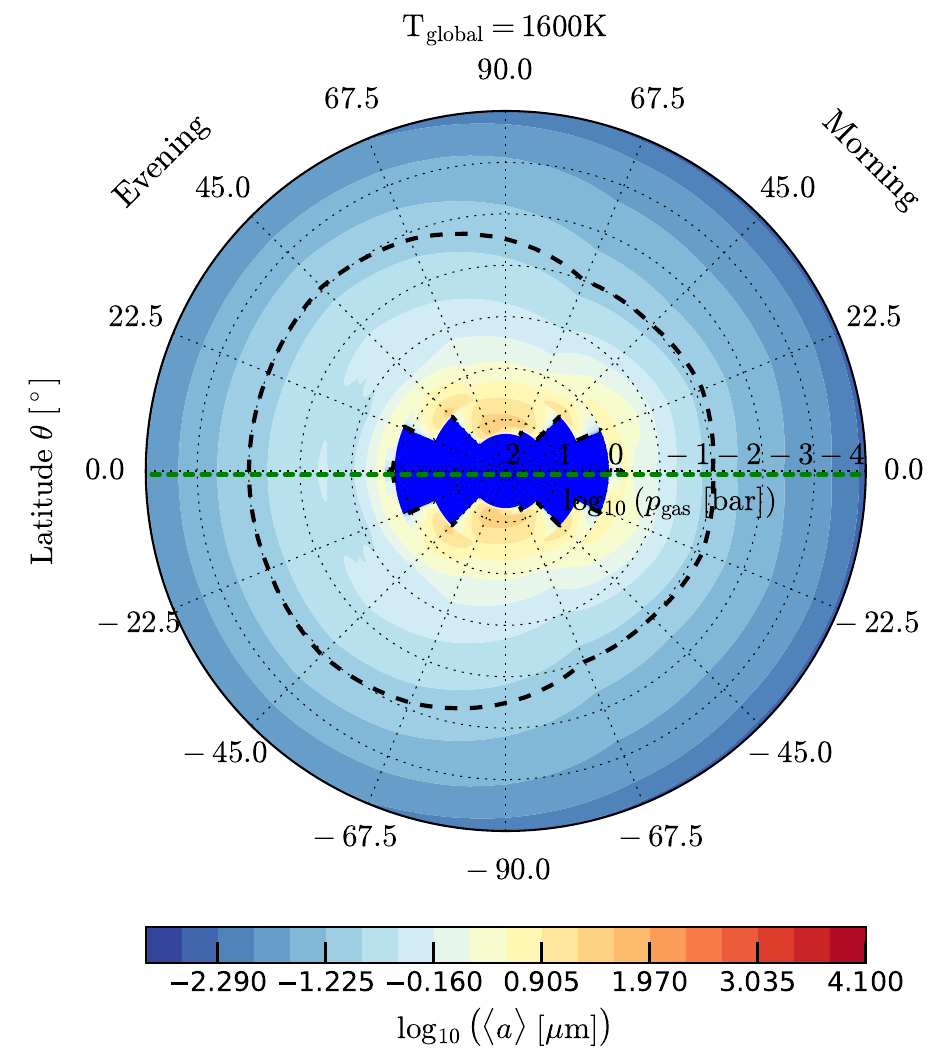}
        \includegraphics[width=0.275\textwidth]{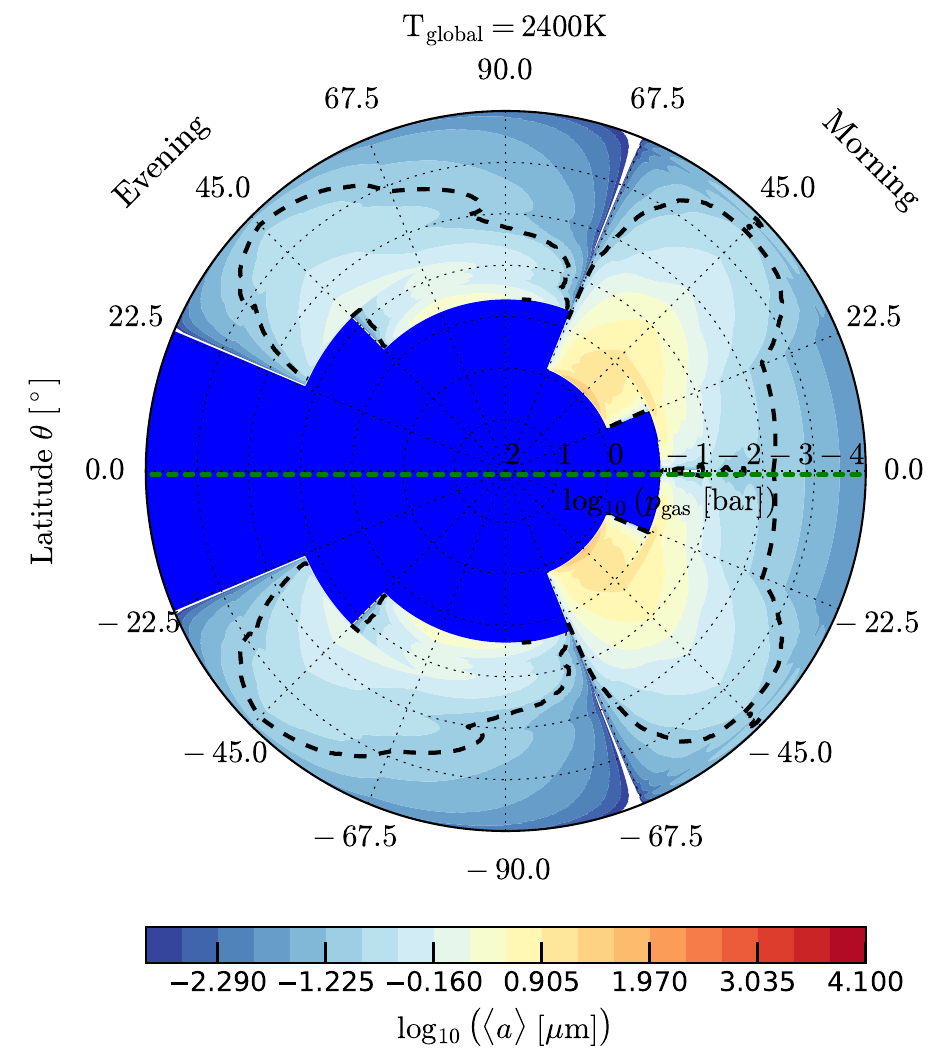}
        \includegraphics[width=0.275\textwidth]{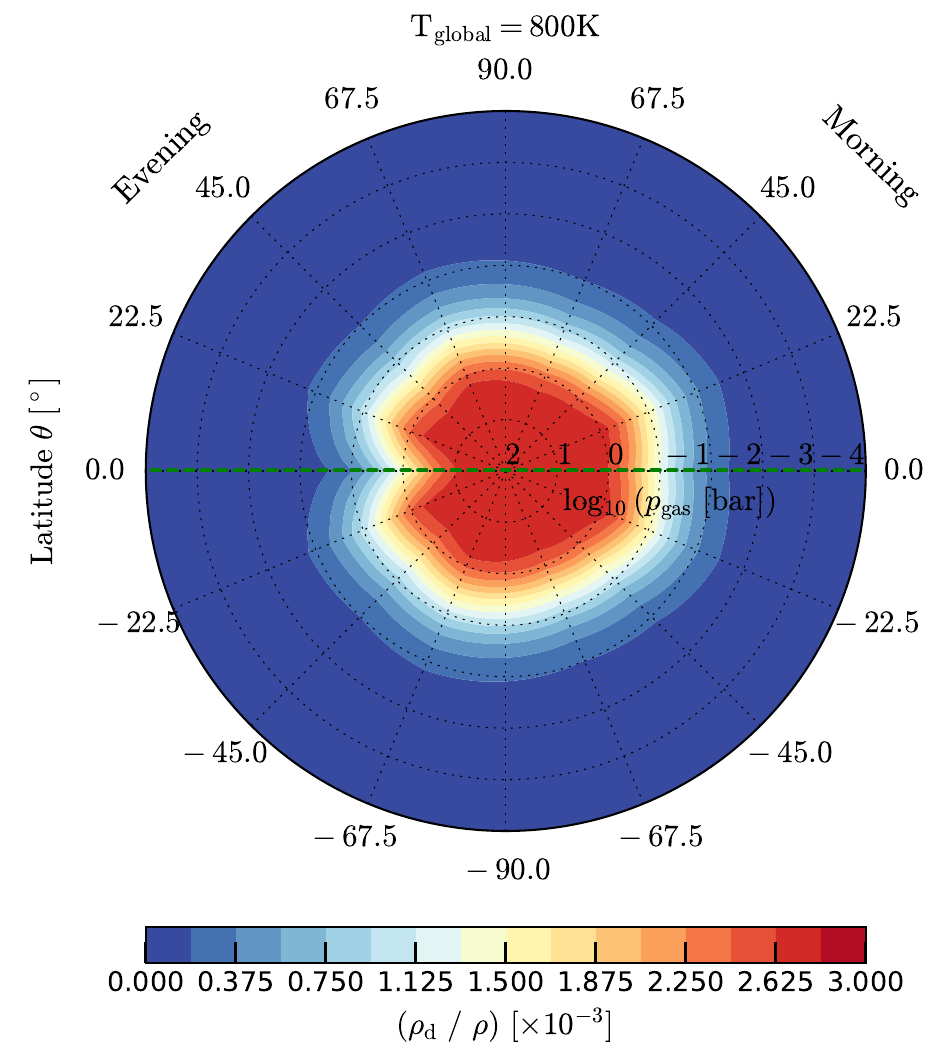}
        \hspace{1.5cm}
        \includegraphics[width=0.275\textwidth]{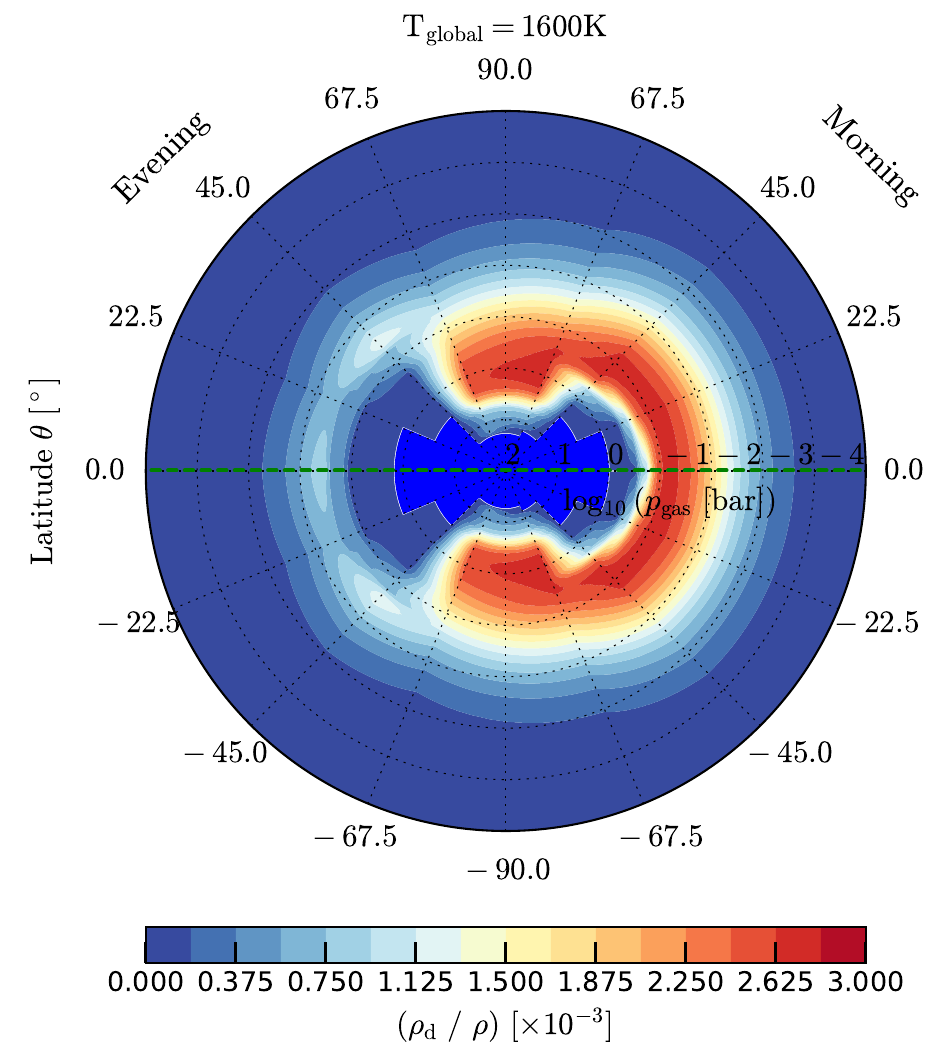}
        \hspace{1.5cm}
        \includegraphics[width=0.275\textwidth]{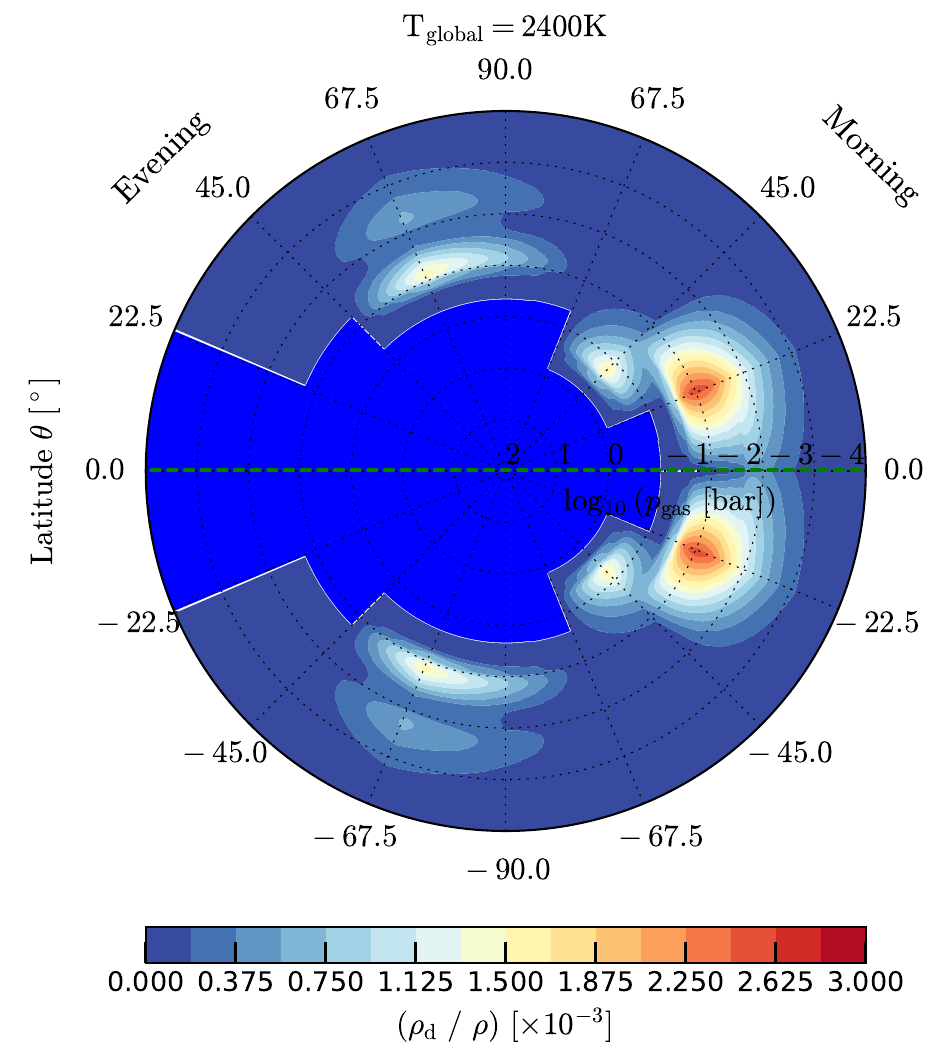}

    \caption{2D terminator slice plots for the three climate regimes. Shown are from top to bottom local gas temperatures, $T_{\rm gas}$ [K], total nucleation rate, [$\rm log_{10} (J_*/cm^{-3}s^{-1})$], mean cloud particle sizes, [$\rm log_{10} (\langle a \rangle/\mu m)$] (with contour line at $\langle a \rangle = 0.1 \mu$m ) and  cloud dust-to-gas mass ratio, $\rho_{\rm dust}/\rho_{\rm gas}$,  for tidally locked gas planet with global temperatures of T$_{\rm global}=800$~K, 1600~K, 2400~K (left to right). A G main sequence host star with $T_{\rm eff}=5660$~K, 0.98 $R_{\rm Sun}$, and 0.98 $M_{\rm Sun}$ is assumed. }
    \label{fig: 3DClimates}
\end{figure*}

We first introduce three simulated planets with global temperatures of 800~K, 1600~K and 2400~K orbiting a G type star as representative cases for three diverse 3D cloudy climate cases (Fig.~\ref{fig: 3DClimates}; \citealt{Helling2021}). The warm Jupiter regime exhibits a globally uniform cloud coverage. 
In the intermediately hot Jupiter regime, the 3D temperature structure at the limbs also resembles the simplified assumption used by \citet{Carone2023} that assumes a uniform evening/morning limb to analyze evening-morning transit asymmetries. That is, the temperature between limbs differ but the latitudinal changes in temperature are smaller. For this regime, both limbs are cloud covered, albeit with a higher cloud top at the evening terminator \citep[see also][]{Carone2023,Powell2019}.  The ultra-hot Jupiter 3D temperature structure is more complex. Here not only do the gas temperatures between the evening and morning limb differ, there are also large latitudinal differences in particular at the evening terminator up to 2500K. This climate regime thus results in very asymmetric cloud coverage, where the morning limb  and higher latitudes at the evening limb ($\pm (22.5^{\circ}-90^{\circ})$) are cloud covered, whereas the equatorial evening terminator is predicted to be cloud free.

The 800~K warm Jupiter shows relatively small cloud particles with  mean particle sizes  $\langle a \rangle\lesssim0.1~\mu$m  dominate throughout the atmosphere. This model outcome generally agrees with observations of \citet{Brande2024} who note very flat \ce{H2O} absorption features in similarly warm Exo-Neptunes that can be explained with a very extended atmosphere filled with small particles. Thus,  photochemistry is not absolutely necessary to create such a atmosphere configuration. For the warmer climates, generally larger cloud particles are predicted ($\langle a \rangle=0.1\,\ldots\,1~\mu$m).

\subsection{Equatorial transit depth differences}
\label{ss:transdepth}

\begin{figure*}[ht]
    \centering
    \includegraphics[width=0.3\textwidth]{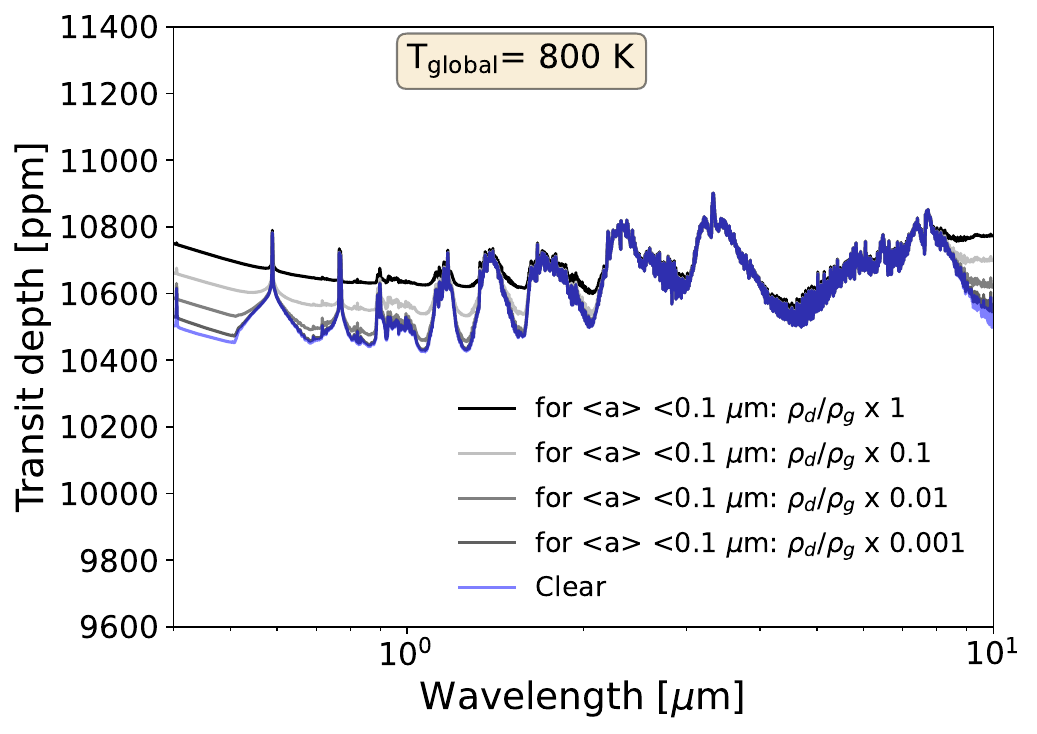}
        \includegraphics[width=0.3\textwidth]{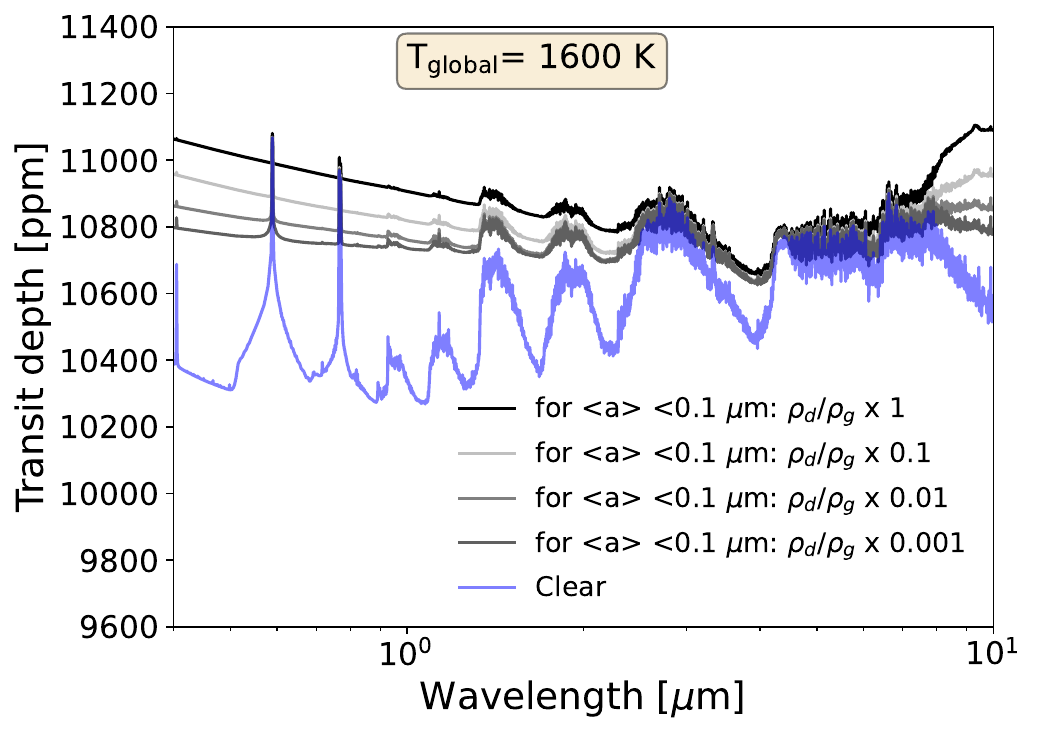}
        \includegraphics[width=0.3\textwidth]{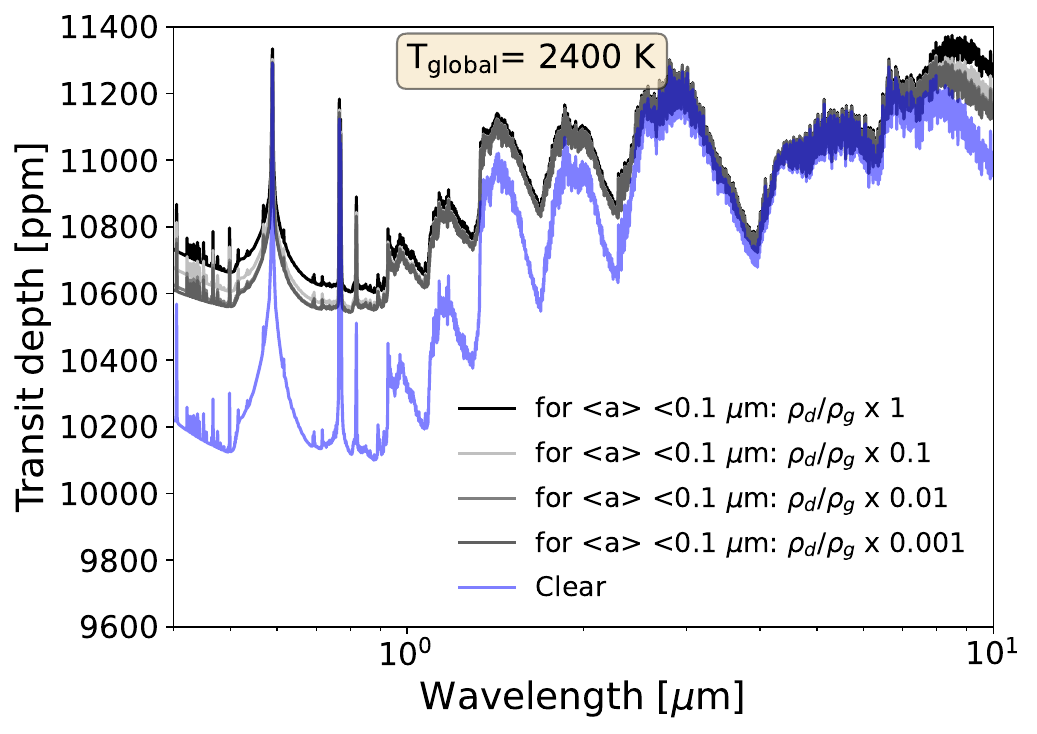}
         \includegraphics[width=0.3\textwidth]{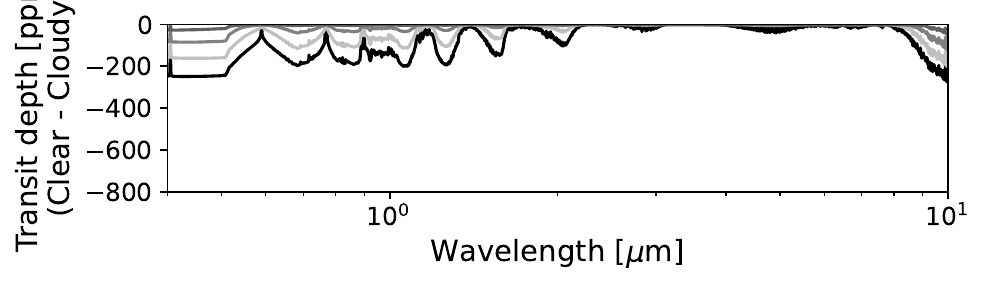}
        \includegraphics[width=0.3\textwidth]{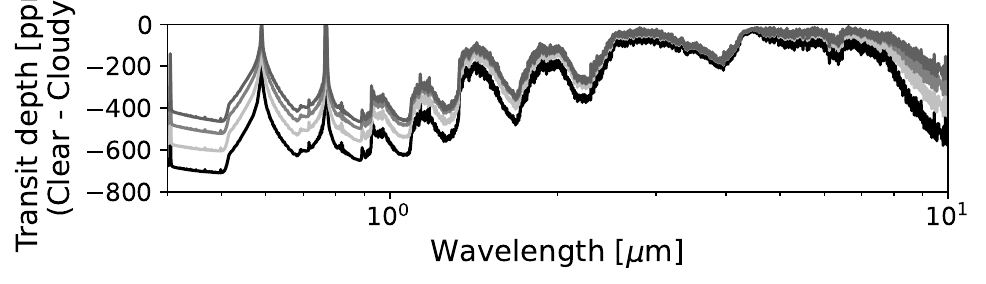}
        \includegraphics[width=0.3\textwidth]{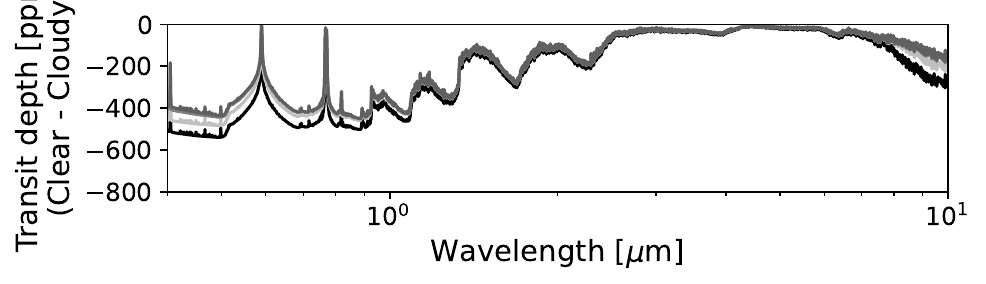}
        
    \caption{Equatorial transit depth differences between cloudy and clear scenarios for three climate regimes. Top: Transit depths averaged over the equatorial morning and evening terminator (latitude $\theta=0^{\circ}$) for tidally locked gas planets with
    T$_{\rm global}$=800~K, 1600~K, 2400~K orbiting a G-type main sequence star. Results for cloud-free calculations (blue) and for different cloud mass loads , $\rho_g/\rho_d(z)$, (gray) in atmospheric layers where $\langle a \rangle <  0.1~\mu$m  are shown (light: $10^{-3}\rho_g/\rho_d(z)$, middle:  $10^{-2}\rho_g/\rho_d(z)$, dark: $0.1\rho_g/\rho_d(z)$).  The black line shows Fe-free results with the full cloud mass load, $\rho_g/\rho_d(z)$. Bottom: Difference between the cloud-free and all cloud opacity cases.}
    \label{fig: TDepth_3Climates}
\end{figure*}

In transmission spectroscopy, it is often assumed that a single atmospheric column is representative for the whole limb region or that one column per limb can be assumed \citep{Goyal2018,Line2016,Parmentier2018,Ahrer2023,Feinstein2023,Baeyens2021}. The single column per limb assumption was applied by \citet{Carone2023} in exploring limb asymmetries in WASP-39b. We test this assumption by first calculating transit depths (Eq.~\ref{eq:tautransit}) and evening-to-morning limb asymmetries for various cloud scenarios only taking into account the equatorial morning and evening terminator. Hence, only regions covering the equator in latitude within $\pm 12.5^{\circ}$ are considered. In longitude, the morning (longitude: $-90^{\circ}$) and evening terminator (longitude: $+90^{\circ}$) encompass cells within $\pm 7.5^{\circ}$ at each limb.  Figure~\ref{fig: TDepth_3Climates} shows the results for the mean transit depth for the three climate regimes, taking into account both limbs for various cloud scenarios, 
testing assumptions for the cloud mass load, $\rho_g/\rho_d(z)$,
in the upper atmosphere. 

\smallskip
\noindent
\paragraph{Clouds vs. no clouds:}
The average transit depths 
generally show large differences of $\geq$ 100~ppm between the cloud-free and cloudy scenarios in the optical wavelength range for all climate regimes (Fig.\ref{fig: TDepth_3Climates}). The differences are largest for the intermediately hot Jupiter (Fig.\ref{fig: TDepth_3Climates}, middle) with 700~ppm, less strong for the ultra-hot Jupiter (\ref{fig: TDepth_3Climates}, right) and smallest for the warm Jupiter (\ref{fig: TDepth_3Climates}, left). 
The variations across different cloud opacity calculations indicating sensitivity to submicron cloud particles are largest for the intermediately hot Jupiter (Fig.\ref{fig: TDepth_3Climates} middle), second largest for the warm Jupiter (Fig.\ref{fig: TDepth_3Climates} left) and the smallest for the ultra-hot Jupiter (Fig.\ref{fig: TDepth_3Climates} right).

The relatively large sensitivity of the warm Jupiter transmission spectra to variations in submicron particle density can be attributed to the dominance of such small cloud particles for most of the atmosphere at such relatively cool temperatures ($T_{\rm global}=800$~K) compared to the hotter planets (Figure~\ref{fig: 3DClimates}, bottom row).

An atmosphere dominated by sub-micron cloud particles is in line with observations that suggest for $T_{\rm global}=600-800$~K highly muted spectral features \citep{Brande2024,Raissa2021}. Such a scenario 
provides an alternative or is at least complementary to the impact of photochemical hazes \citep[e.g.][]{Ohno2024}.

\begin{figure*}
    \centering
    \includegraphics[width=0.3\textwidth]{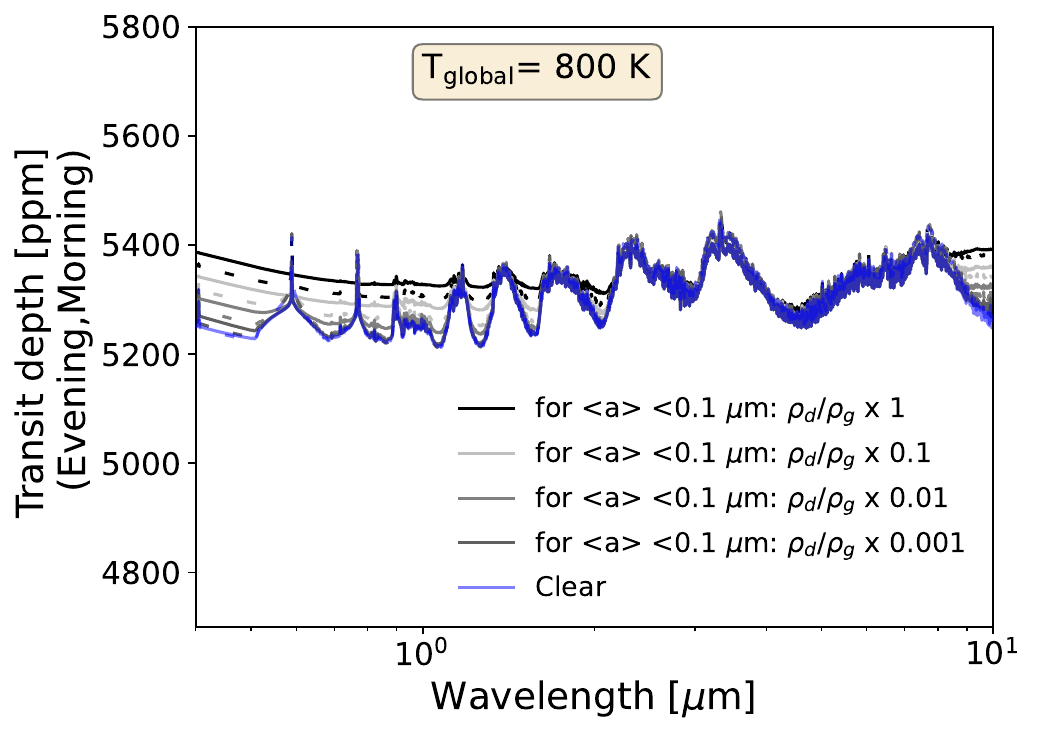}
        \includegraphics[width=0.3\textwidth]{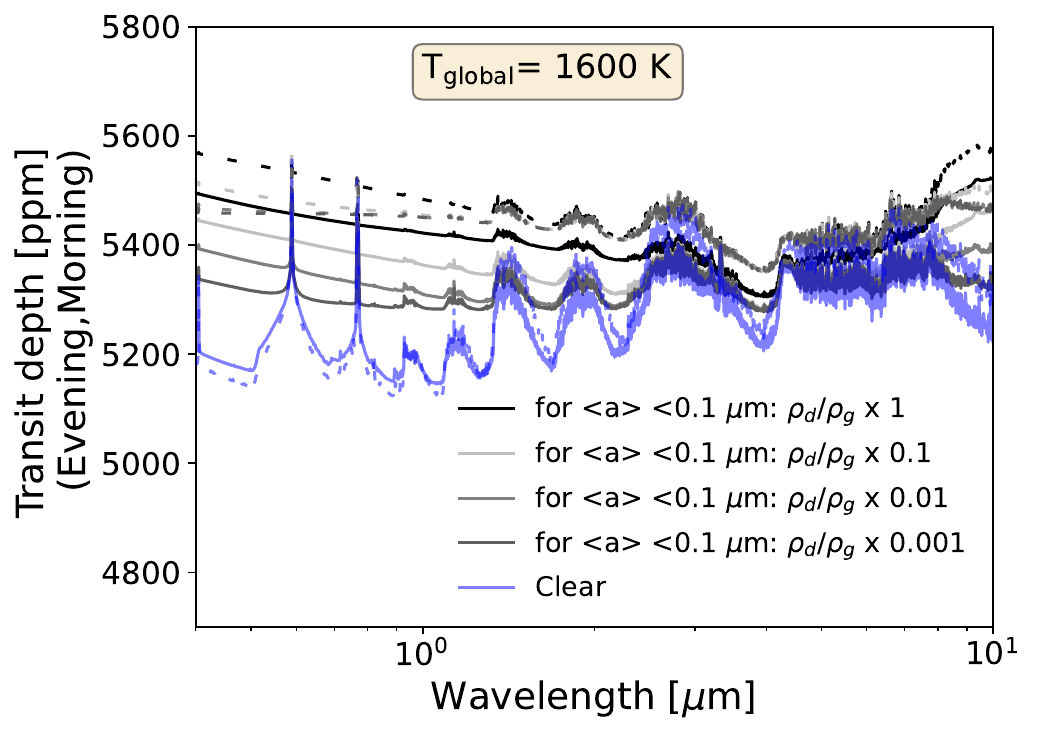}
        \includegraphics[width=0.3\textwidth]{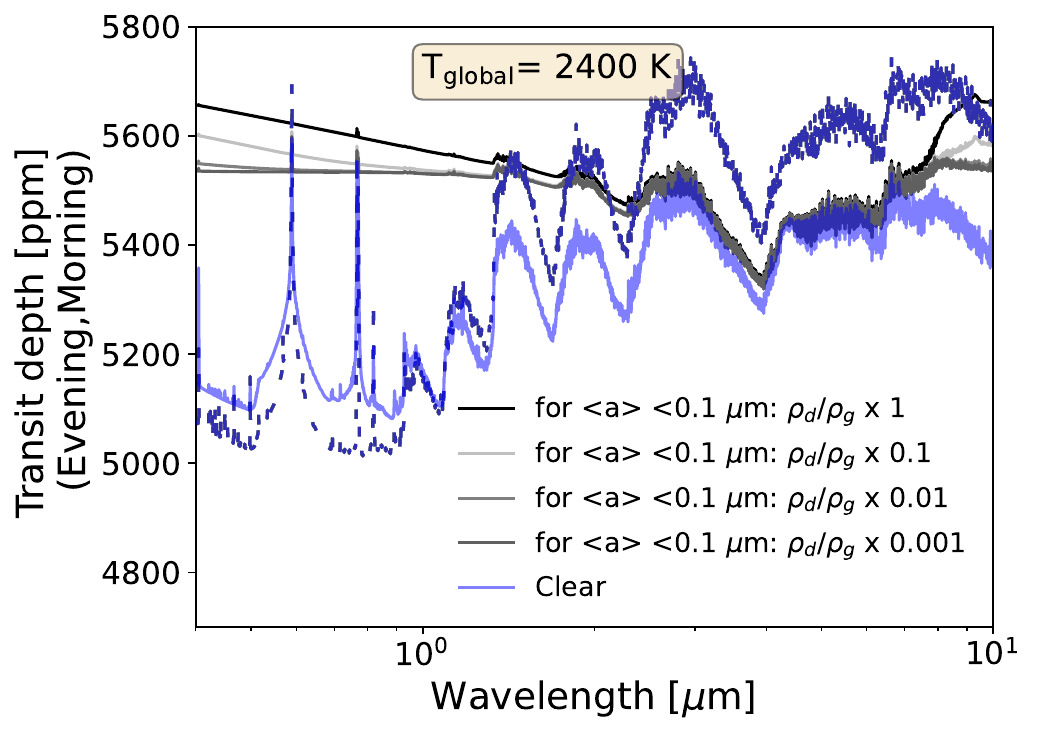}
        \includegraphics[width=0.3\textwidth]{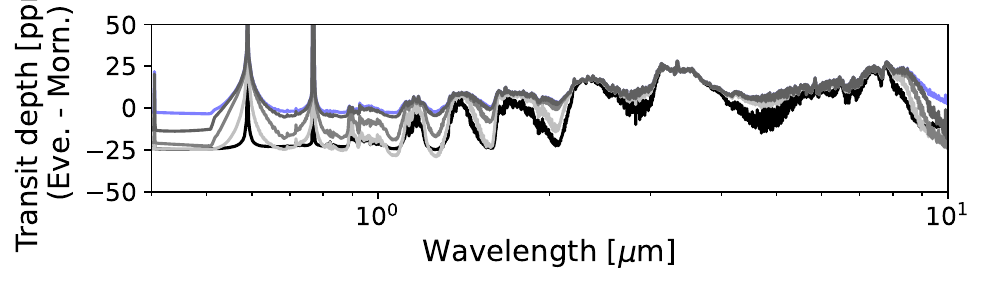}
        \includegraphics[width=0.3\textwidth]{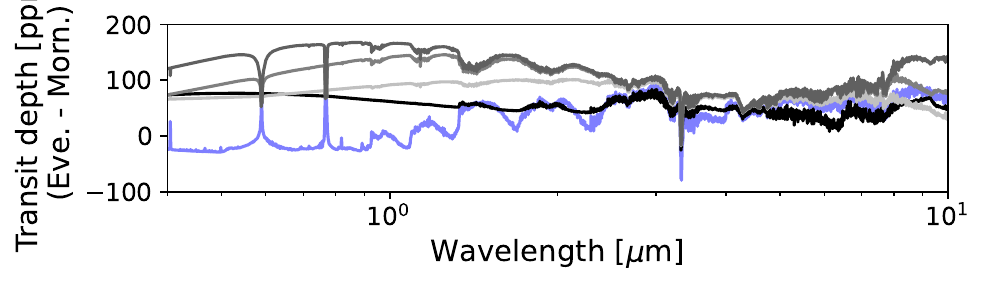}
        \includegraphics[width=0.3\textwidth]{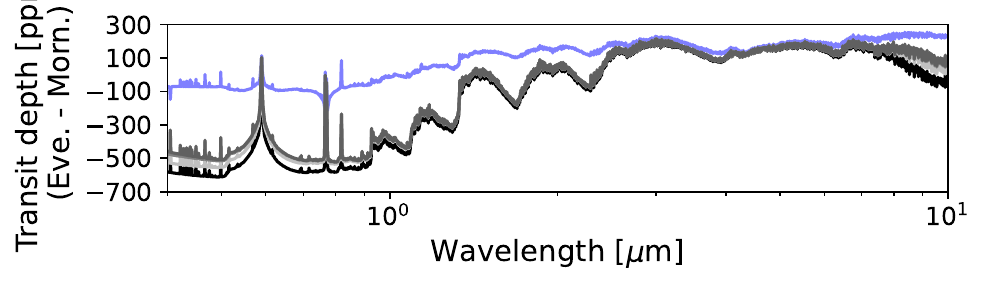}

    \caption{Equatorial transit asymmetries for three climate regimes. Top: Individual evening and morning equatorial (latitude $\theta=0^{\circ}$)  transit depths (dashed and solid) for tidally locked gas planet with T$_{\rm global}$=800~K, 1600~K, 2400~K orbiting a G main sequence star. Results for cloud-free calculations (blue) and for different cloud mass loads, $\rho_g/\rho_d(z)$, (gray) in atmospheric layers where $\langle a \rangle <  0.1~\mu$m  are shown  (light: $10^{-3}\rho_g/\rho_d(z)$, middle:  $10^{-2}\rho_g/\rho_d(z)$, dark: $0.1\rho_g/\rho_d(z)$). The black lines shows Fe-free results with full $\rho_g/\rho_d(z)$. Please note that in the top rightmost panel, dashed lines lie on top of each other.  Bottom: Difference between the evening and morning terminator transit depths.} 
    \label{fig: TDepth_Asym_3Climates}
\end{figure*}

\smallskip
\noindent
\paragraph{The inverse transit depth effect:}
When inspecting the differences between cloud-free and cloud opacity calculations for the individual equatorial morning and evening terminator (Fig.~\ref{fig: TDepth_Asym_3Climates}), it is immediately evident from the results in the optical wavelength range (0.3-1~$\mu$m) that limb differences in temperature and cloud coverage compete with each other. By comparing the evening terminator depths of the ultra-hot Jupiter (Fig.~\ref{fig: TDepth_Asym_3Climates}, right) with the intermediately hot Jupiter (Fig.~\ref{fig: TDepth_Asym_3Climates} middle) and warm Jupiter (Fig.~\ref{fig: TDepth_Asym_3Climates}, left), it is also evident that a higher local gas temperature generally leads to a smaller and not a  larger transit depth in the optical.

The reason for this behavior are the Rayleigh scattering slope and the strong (temperature-dependent) pressure-broadening of the sodium and potassium absorption bands between 0.5-0.8~$\mu$m (See also \citealt[][Fig.9]{Powell2019})\footnote{The importance of the temperature and pressure depending broadening of alkali metals absorption lines that reveal relatively deep pressure layers in hot planets/brown dwarfs in the optical has been pointed out already by \citet{Burrows2000}. However, absolute transit depth differences are difficult to identify due instrumental offsets of several 100~ppm  even with JWST\citep{Carter2024}. Repeated precise and accurate PLATO observations on the same field may, however, be able to yield comparison of average transit depth differences between planets (Sect.~\ref{sec: PLATO},\citealt{Rauer2025})}. Hence, the transit depth of a clear atmosphere is generally smaller for a higher local gas temperature. This inverse transit depth effect with gas temperature is also evident by comparing the cloud-free evening and morning transmission spectra for each planet. As a consequence, evening to morning transit depth differences are for clear scenarios negative in the optical, albeit comparatively small ($\Delta T_{\rm Depth} \approx -25$~ppm, Fig.~\ref{fig: TDepth_Asym_3Climates} bottom). 

On the other hand, the evening transmission spectrum of an ultra-hot Jupiter is always larger compared to the morning transmission spectrum in the infrared $>1 \mu$m, resulting in moderately positive evening-to-morning transit depth differences of 100-200~ppm (Fig.~\ref{fig: TDepth_Asym_3Climates}, right). In addition, the ultra-hot Jupiter evening terminator has a larger transit depth compared to the intermediately hot Jupiter evening spectrum (Fig.~\ref{fig: TDepth_Asym_3Climates}, middle). Thus, in the infrared, the transit depth limb asymmetries are indeed mostly driven by the horizontal temperature contrast.

When cloud are taken into account, transit depth asymmetries between the equatorial evening and morning limb are predominantly driven by the morning cloud for the ultra-hot Jupiter in the optical (Fig.~\ref{fig: TDepth_Asym_3Climates} right): The contrast between a cloudy morning and a clear evening terminator (Fig.~\ref{fig: 3DClimates}, right) results in a large negative evening-to-morning transit asymmetry $\Delta T_{\rm Depth} =-500$~ppm with variations of up to $100$~ppm between different cloud opacity calculations. 

The warm Jupiter evening/morning spectra (Fig.~\ref{fig: TDepth_Asym_3Climates}, left) show a comparable trend in asymmetries as described above for the ultra hot Jupiter albeit with smaller amplitudes, at least in the hierarchical approach. Generally, transit asymmetries between -25 to +25~ppm are predicted. The biggest asymmetries arise for the warm Jupiter between 3 - 4 $\mu$m in the methane band.

\smallskip
\noindent
\paragraph{The impact of upper-atmosphere cloud mass load on the amplification of terminator asymmetry:} For the intermediately hot Jupiter ( T$_{\rm global}=1600$K) where both equatorial limbs are cloud covered (Fig.~\ref{fig: 3DClimates}, middle), a more complex temperature and cloud dependency emerges for the cloudy evening and morning transmission spectra (Fig.~\ref{fig: TDepth_Asym_3Climates} middle):

In contrast to the other climate regimes, the warmer cloudy evening terminator of such an intermediately-hot Jupiter has always a larger transit depth compared to the cooler morning terminator, in particular between $0.3-3 \mu$m. Thus, for cloudy scenarios moderately positive evening to morning transit depth differences, $T_{\rm Depth}=100-150$~ppm, arise for the whole wavelength range (Fig.~\ref{fig: TDepth_Asym_3Climates} bottom row, middle). In addition, the transit depth differences between the evening and morning terminator is larger for the cloud opacity calculations (Fig.~\ref{fig: TDepth_Asym_3Climates} bottom row, middle: black and gray lines) compared to the cloud-free calculation (blue lines). This shows that the mere presence of clouds (even without local gas temperature feedback can amplify the local gas temperature differences between both limbs compared to clear scenarios. }

The cloud amplification tendency holds true for the whole wavelength range, but is particularly large for $\leq 3 \mu$m. In addition, variations of $\lesssim$ 100~ppm across different cloud opacity calculations arise for this wavelength range (Fig.~\ref{fig: TDepth_Asym_3Climates} bottom row, middle). Hence, evening/morning transmission spectra for the intermediately hot Jupiter are particular sensitive to the cloud dust to gas mass ratio $\rho_d/\rho_g(z)$ in atmospheric layers for which the mean cloud particle size $\langle a \rangle< 0.1~\mu$m.

An inspection of the scattering slope for the cloudy evening and morning transmission spectra (Fig.~\ref{fig: TDepth_Asym_3Climates} top row, middle) is informative. The evening terminator is displaying a flat scattering slope in the optical already for $10^{-2} \rho_d/\rho_g(z)$ 
(Fig.~\ref{fig: TDepth_Asym_3Climates} top row, middle: middle gray dashed line), and a reduction to $10^{-3} \rho_d/\rho_g(z)$ leaves still enough sub-micron particles at the cooler morning terminator that a moderately slanted scattering slope in the optical appears
(middle gray and dark gray solid lines). Therefore, the differences between the evening and morning transmission spectra are maximal for a flat evening and slanted morning terminator as a consequence of asymmetric cloud particle size distribution (Fig.~\ref{fig: 3DClimates} bottom row, middle).

Conversely, the cloud particle size distribution is more similar across the morning and evening limb for the warm Jupiter (Fig.~\ref{fig: 3DClimates} bottom left). Thus, only small differences arise between scattering slopes of the evening and morning transmission spectra for different cloud opacity calculations (Fig.~\ref{fig: TDepth_Asym_3Climates} top row, left). The dominance of sub-micron cloud particles in the 800~K Jupiter simulation also explains why clouds tend to diminish the transit depth asymmetries compared to a clear atmosphere model in the infrared (Fig.~\ref{fig: TDepth_Asym_3Climates} bottom row, left). 

In any case, the equatorial cloud model comparison is highly informative and sheds light on the intricate effects that the full cloud complexity can have on transit depth asymmetries at different wavelengths. A full three dimensional planet can display, however, also changes in temperature and cloud properties in latitude - in particular in the ultra-hot Jupiter climate regime.

\subsection{High latitudes affecting transit depth calculations}
\label{sec: High latitudes}

\begin{figure*}
    \centering
    \includegraphics[width=0.3\textwidth]{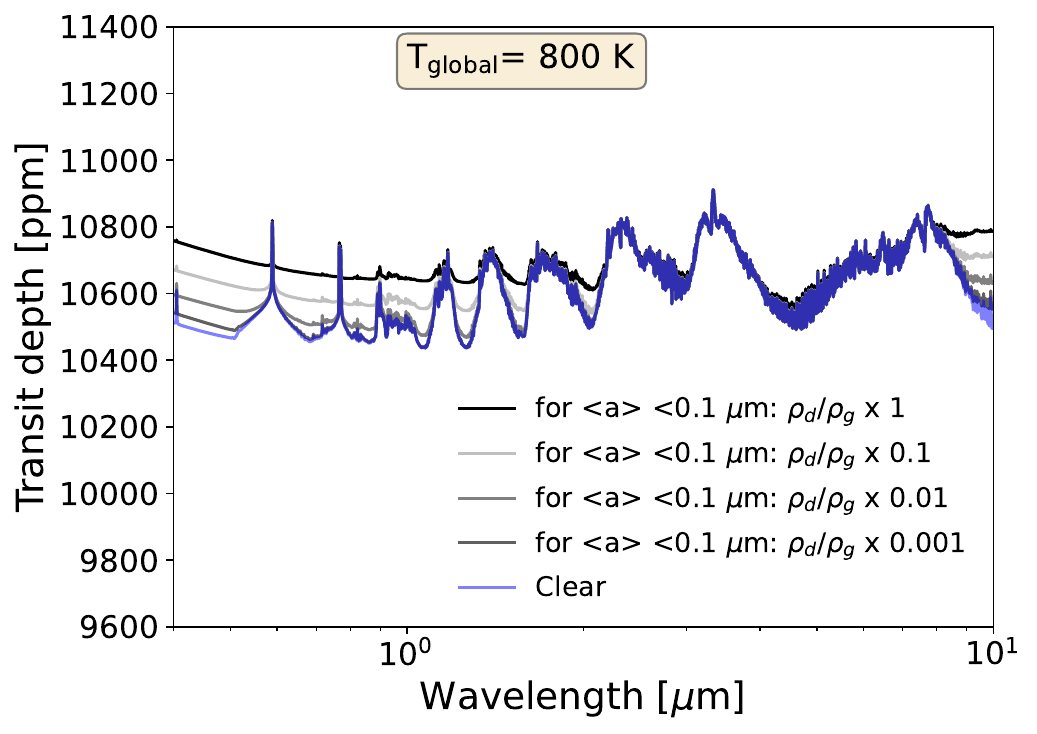}
        \includegraphics[width=0.3\textwidth]{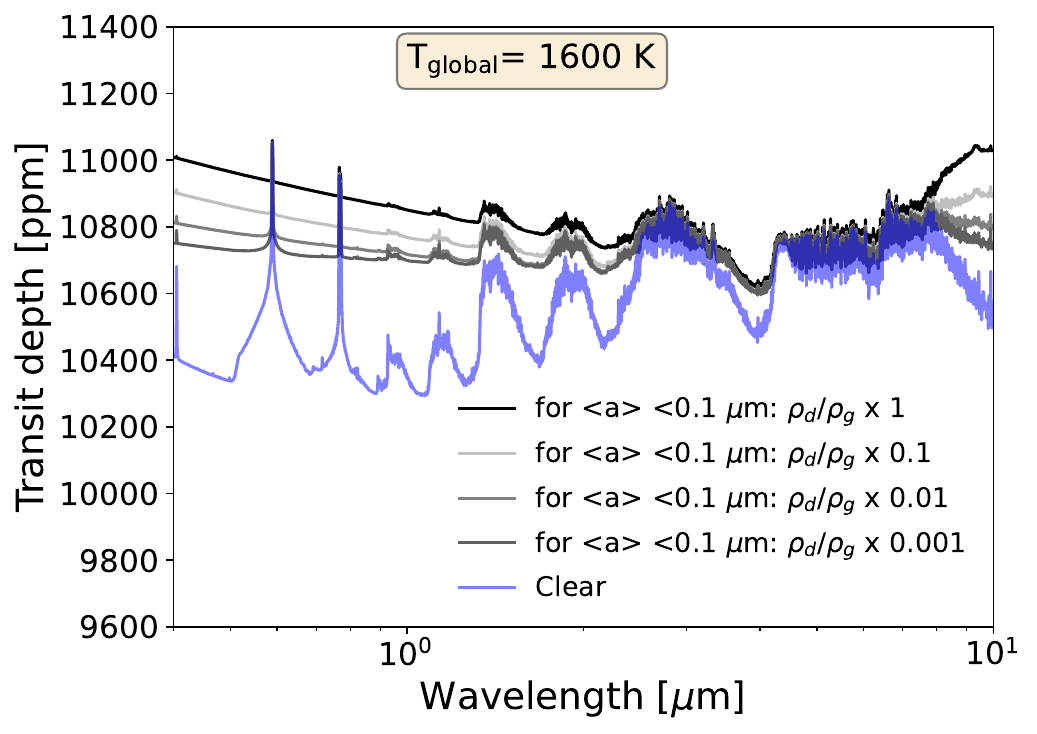}
        \includegraphics[width=0.3\textwidth]{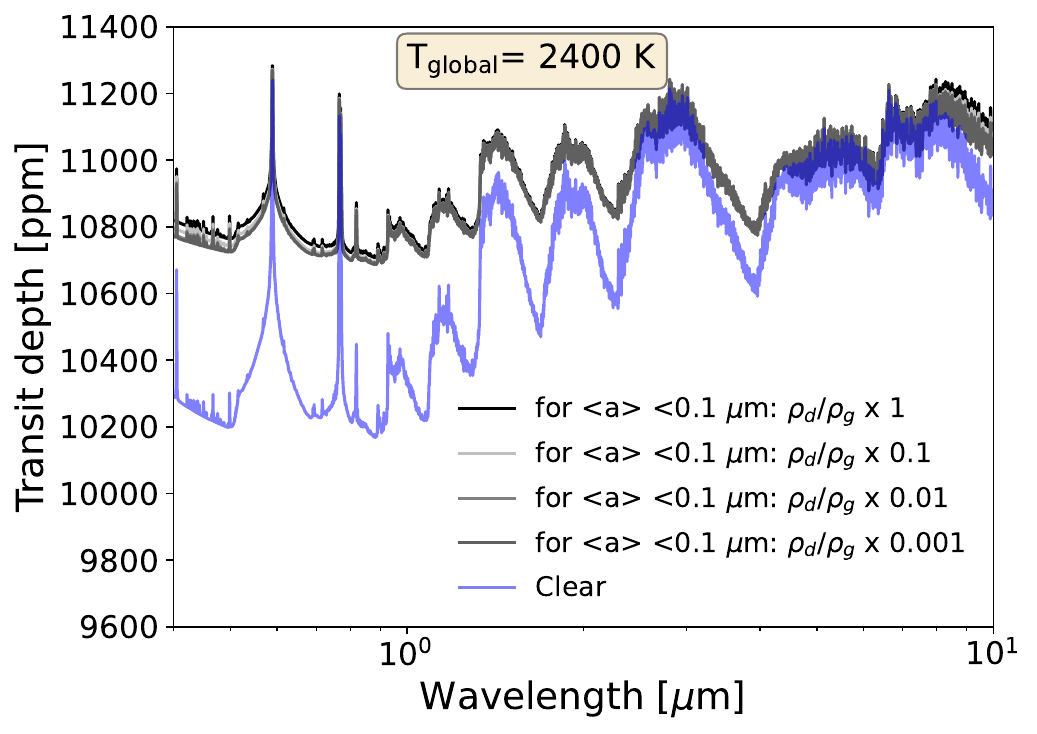}
         \includegraphics[width=0.3\textwidth]{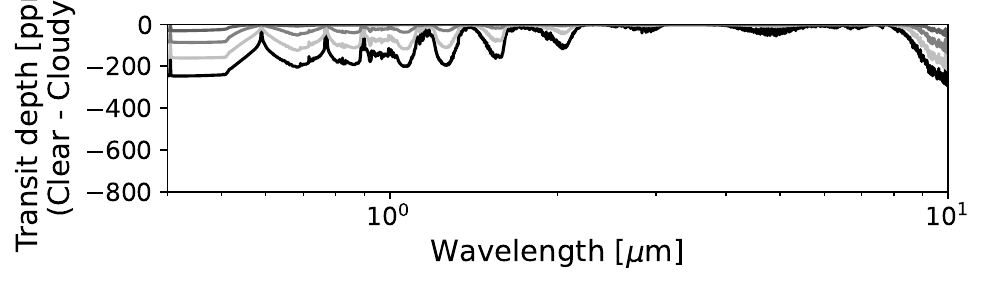}
        \includegraphics[width=0.3\textwidth]{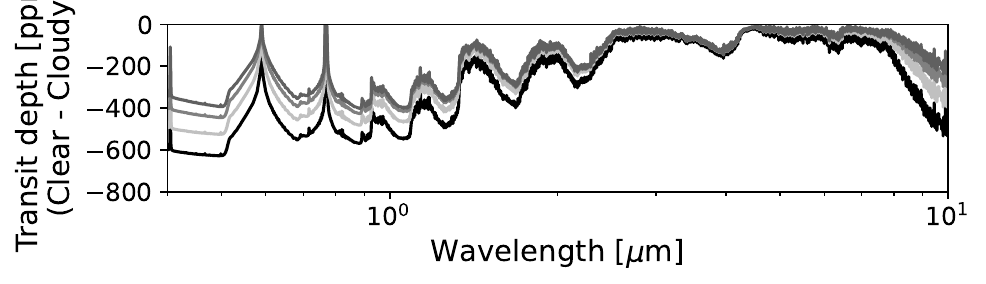}
        \includegraphics[width=0.3\textwidth]{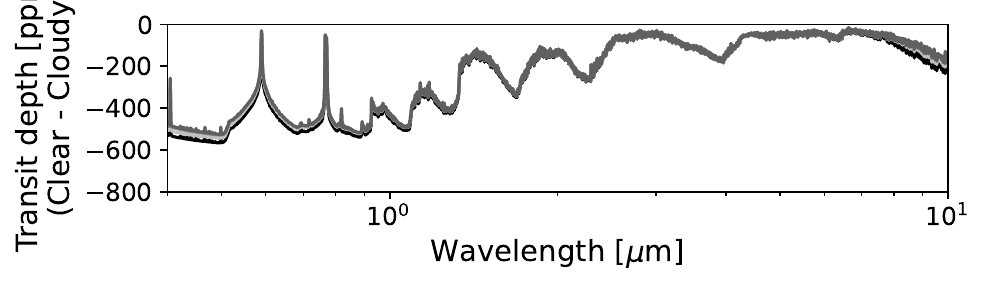}
        
    \caption{Latitudinally averaged transit depth differences between cloudy and clear scenarios for three climate regimes. Top: Transit depths averaged over the evening and morning terminator for tidally locked gas planet with $T_{\rm global}$=800~K, 1600~K,2400~K orbiting a G main sequence star denoted by dashed and solid lines, respectively. The following latitudes were combined for each limb in the calculations: $\theta =0^{\circ},\pm 23^{\circ}, \pm 45^{\circ},  \pm 68^{\circ}, \pm 86^{\circ}$). Results for cloud-free calculations are shown in blue and for different cloud mass loads, $\rho_g/\rho_d(z)$, in atmospheric layers where $\langle a \rangle <  0.1~\mu$m  are shown in gray (light: $10^{-3}\rho_g/\rho_d(z)$, middle:  $10^{-2}\rho_g/\rho_d(z)$, dark: $0.1\rho_g/\rho_d(z)$). The black lines shows Fe-free results with full $\rho_g/\rho_d(z)$. Bottom: Difference between the cloud-free and all cloudy opacity calculations.}
    \label{fig: TDepth_3Climates_all_Lats}
\end{figure*}

\begin{figure*}
    \centering
    \includegraphics[width=0.3\textwidth]{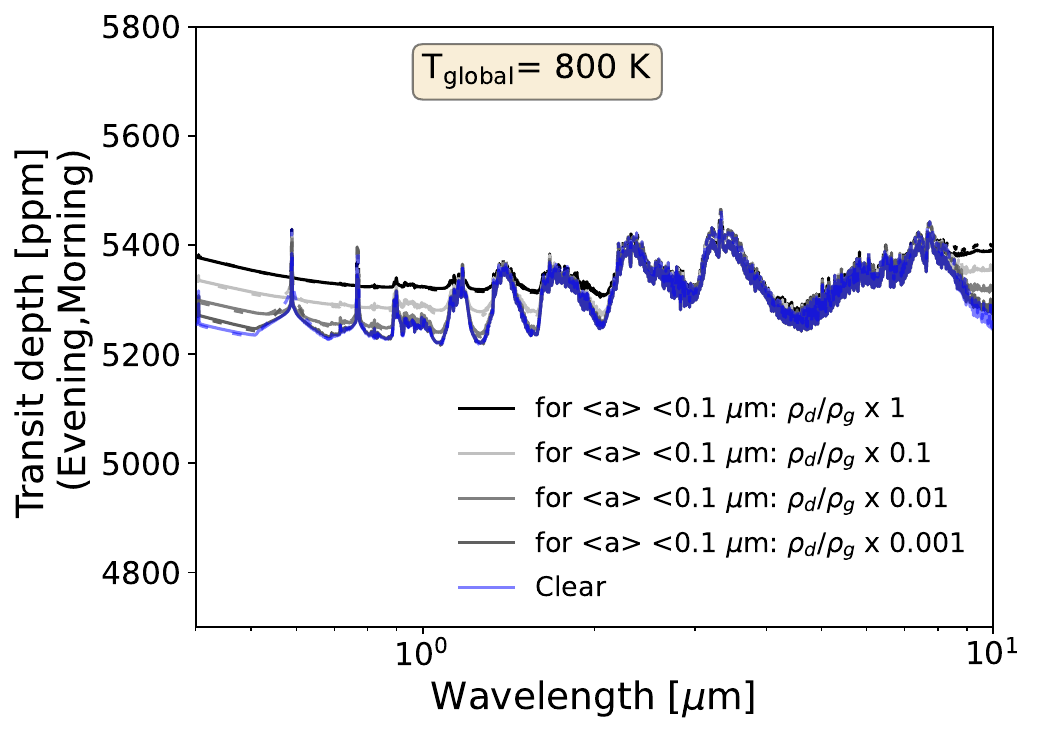}
        \includegraphics[width=0.3\textwidth]{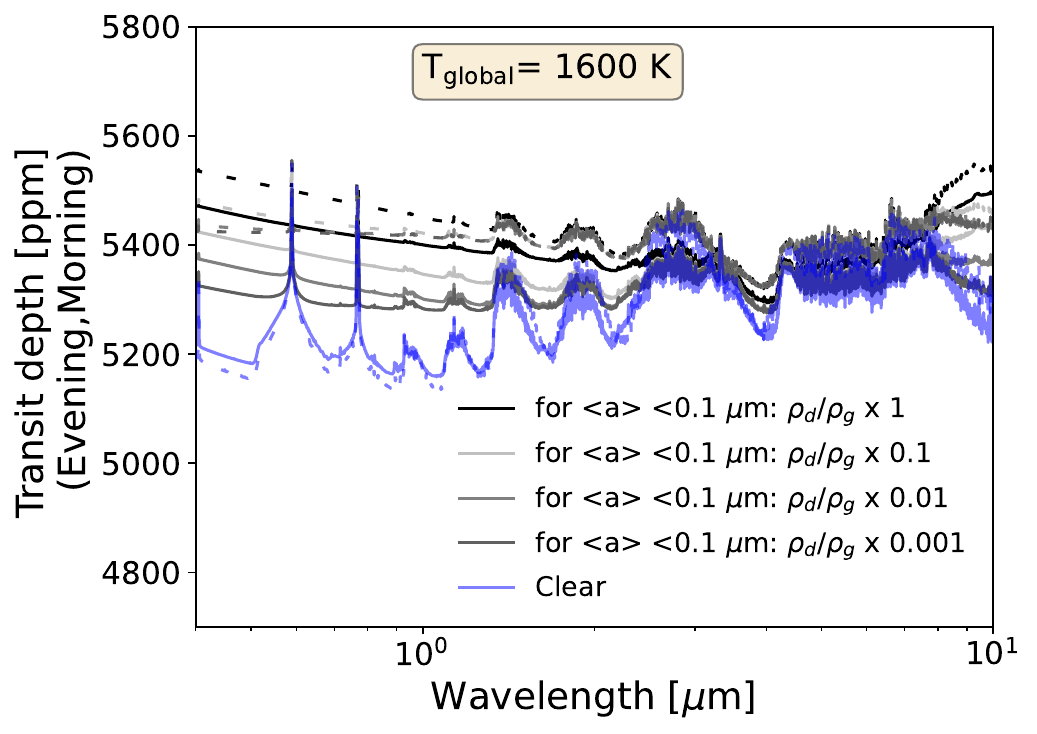}
        \includegraphics[width=0.3\textwidth]{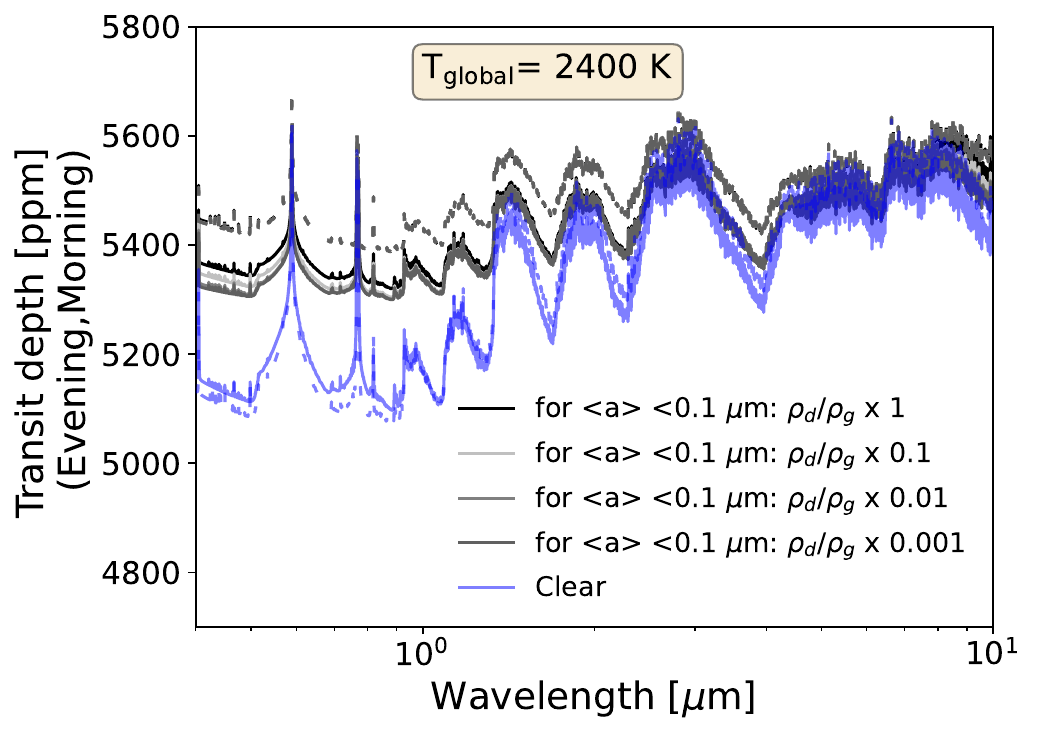}
        \includegraphics[width=0.3\textwidth]{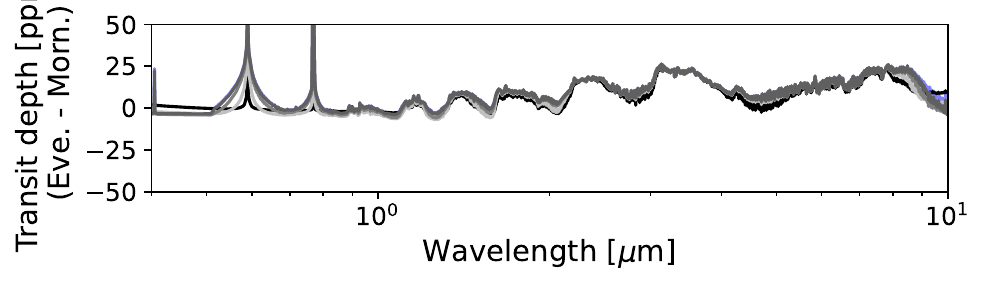}
        \includegraphics[width=0.3\textwidth]{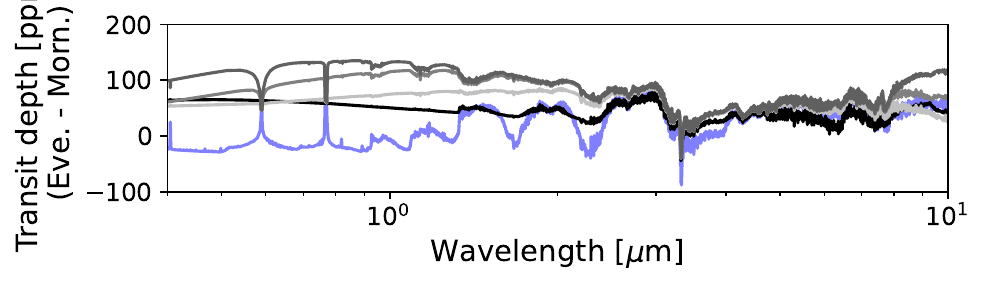}
        \includegraphics[width=0.3\textwidth]{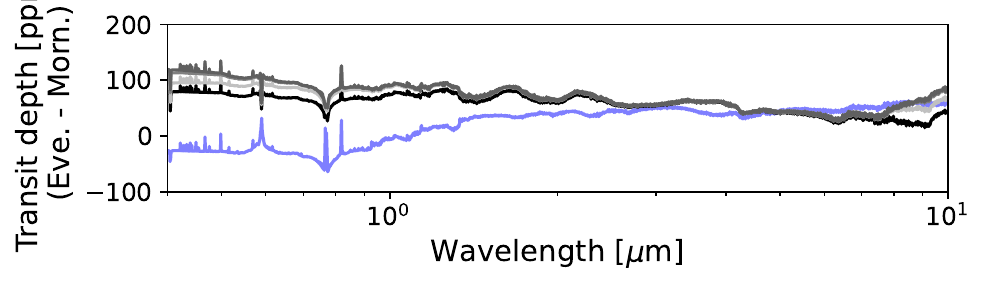}

    \caption{Latitudinally averaged transit asymmetries for three climate regimes. Top: Individual evening and morning transit depths for tidally locked gas planets with $T_{\rm global}=$800~K, 1600~K,2400~K orbiting a G main sequence host star are shown as dashed and solid lines, respectively. The following latitudes were taken into account at each limb for the calculations: $\theta =0^{\circ},\pm 23^{\circ}, \pm 45^{\circ},  \pm 68^{\circ}, \pm 86^{\circ}$).  Results for cloud-free calculations are shown in blue and for different cloud mass loads, $\rho_g/\rho_d(z)$, in atmospheric layers where $\langle a \rangle <  0.1~\mu$m  are shown in gray (light: $10^{-3}\rho_g/\rho_d(z)$, middle:  $10^{-2}\rho_g/\rho_d(z)$, dark: $0.1\rho_g/\rho_d(z)$). The black lines shows Fe-free results with full $\rho_g/\rho_d(z)$. Bottom panel: Differences between evening and morning terminator transit depths.}
    \label{fig: TDepth_Asym_3Climates_all_Lats}
\end{figure*}

Do temperature and cloud differences at higher latitudes matter for transit depth consideration? This is explored by considering $\theta=0 \pm 12.5^{\circ},\pm 23^{\circ}, \pm 45^{\circ},  \pm 68^{\circ}, \pm 86^{\circ}$.  Each transit depth at the morning, $\rm T_{\rm Morning}(\theta)$, and evening terminator, $\rm T_{\rm Evening}(\theta)$, is assumed to be representative of a segment covering $\pm 7.5^{\circ}$ in longitude and $\pm 22.5^{\circ}$ in latitude from the selected mid point in latitude and longitude. The sum of the latitude averaged morning and evening transit depth is equal to the planetary averaged transit depth that accounts for all latitudes.

For different cloud scenarios and three climate regimes the planetary averaged transit depth (Fig.~\ref{fig: TDepth_3Climates_all_Lats}), the individual latitude averaged morning and evening transit depth and asymmetries are studied (Fig.~\ref{fig: TDepth_Asym_3Climates_all_Lats}).

For the warm Jupiter, differences between the clear and cloudy atmosphere scenarios are not significantly changed when taking into account higher latitudes (Fig.~\ref{fig: TDepth_3Climates_all_Lats} left) compared to the equatorial transit depths (Fig.~\ref{fig: TDepth_3Climates} left). Transit asymmetries are, however, affected: The latitudinally averaged transit asymmetries (Fig.~\ref{fig: TDepth_Asym_3Climates_all_Lats} left) are much smaller compared to the equatorial transit asymmetries (Fig.~\ref{fig: TDepth_Asym_3Climates_all_Lats} left). Thus, averaging over all latitudes, or spatially un-resolved observation, would misleadingly suggest a more uniform cloud coverage. 

The small moderately positive gas temperature differences in the infrared, however, including the \ce{CH4} peak between 3-4 $\mu$m persist for both transit asymmetry calculations (Figs.~\ref{fig: TDepth_Asym_3Climates} left and \ref{fig: TDepth_Asym_3Climates_all_Lats} left). The changes in \ce{CH4} abundances reflect local gas temperature differences between the limbs that preserved even when higher latitudes are considered. In the planetary averaged transit, the optical wavelength regime remains the most promising to identify different cloud scenarios (Fig.~\ref{fig: TDepth_3Climates_all_Lats}, left).

The results for the intermediately hot Jupiter do not change significantly when higher latitudes are considered, neither for the planetary average (Figs.~\ref{fig: TDepth_3Climates} middle and \ref{fig: TDepth_3Climates_all_Lats} middle) nor for transit asymmetries (Figs.~\ref{fig: TDepth_Asym_3Climates} middle and \ref{fig: TDepth_Asym_3Climates_all_Lats} middle). For this planet climate regime, the assumption is valid that the temperature and cloud extent at the equatorial morning and evening terminators are representative also for higher latitudes, applied e.g. in \citet{Carone2023}.

For the ultra-hot Jupiterclimate regime, the planetary average transit depths as well as the transit asymmetries changes significantly when higher latitudes are considered. In  this  superrotating climate regime, only the equatorial atmosphere region is cloud-free (Fig.~\ref{fig: 3DClimates}, right). Adding higher latitudes thus strongly changes the transit depth asymmetries in the optical: $\Delta T_{\rm Depth}$ changes from -500~ppm (Fig.~\ref{fig: TDepth_Asym_3Climates} bottom row right) to +100~ppm (Fig.~\ref{fig: TDepth_Asym_3Climates_all_Lats} bottom row right). The change from a negative to a positive evening-to-morning transit contrast $\Delta T_{\rm Depth}$ indicates that adding higher latitudes changes the driving physical mechanism. Instead of the high vertical extent of the cloud at the colder morning limb, the larger scale height of the warmer evening terminator drives the asymmetry. Clouds further amplify the transit asymmetries due to the horizontal temperature contrast as can be seen by comparing the cloudy calculations (black and gray lines) to the clear atmosphere scenario (blue line) across the whole investigated wavelength range (Fig.~\ref{fig: TDepth_Asym_3Climates_all_Lats} bottom row right).

Qualitatively, the latitude averaged transit asymmetries for the ultra-hot Jupiter are similar to those of the intermediately hot Jupiter scenario (Fig.~\ref{fig: TDepth_Asym_3Climates_all_Lats} bottom row middle). Quantitatively, the ultra-hot Jupiter transit asymmetry amplitudes are smaller: $\Delta T_{\rm Depth}=$ 50- 100~ppm.  In addition, both, the planetary average transit depth (Fig.~\ref{fig: TDepth_3Climates_all_Lats} bottom row left) and the transit asymmetries (Fig.~\ref{fig: TDepth_Asym_3Climates_all_Lats} bottom row left) are less sensitive to upper-atmosphere-cloud scenarios. Variations in between cloud scenarios are at most 100~ppm at 0.4~$\mu$m. Apparently, a partly cloud-free evening limb reduces the cloud amplification effect compared to the intermediately hot Jupiter case with a completely cloudy evening limb.

In summary, we find that considering higher latitudes matters the least for the intermediately hot Jupiter. For such planets, latitudinal variations in temperature and cloud properties are negligible. For the warm Jupiter, including higher latitudes suppresses the signs of asymmetric cloud particle distributions that are more prominent at the equatorial region. Considering higher latitudes matters the most for the ultra-hot Jupiter. Here, the partially cloud-free evening terminator will impact the observed latitudinally averaged evening transit depth predominantly in the optical. Thus, it may be possible to identify the latitudinal temperature and cloud coverage at the evening terminator for ultra-hot Jupiters via evening-to-morning transit asymmetries in the optical in comparison with 3D cloud-climate models. 

The intermediately hot Jupiter climate regime consistently emerges as the most sensitive to upper atmosphere cloud scenarios in the optical in both, the planetary average transit depth and transit asymmetries. Thus, planets in this climate regime are the most promising to identify the cloud-temperature feedback effect. 

\section{Observability with PLATO, CHEOPS, TESS and JWST}

\label{sec:cloudimpact}

A coherent study of variations in transit depth and transit asymmetries across the PLATO, CHEOPS, TESS, and JWST wavelengths
($\lambda=0.33\,\ldots\,10\mu$m)  for three example  planets (T$_{\rm global}=800, 1600, 2400$K) orbiting a G-type stars was presented Sect.~\ref{sec: Three cases}. These examples represent three distinct climate regimes: warm, intermediately hot and ultra-hot gas giant planets. Section~\ref{sec:cloudimpact} explores transit depths and limb asymmetries for the whole global temperature range of the 3D AFGKM \texttt{ExoRad} GCM grid (T$_{\rm global}=800\,\ldots\,2600$K) for telescopes that cover the optical wavelengths (PLATO, CHEOPS, TESS; Sec.~\ref{sec: PLATO}), the near infrared range (HST, JWST/NIRSpec; Sec.~\ref{sec: NIR}) and larger wavelengths (JWST/MIRI; Sec.~\ref{sec: MIRI}).

\subsection{Transit depths in PLATO, CHEOPS, and TESS}
\label{sec: PLATO}
Section~\ref{sec: WASP39b} has shown that both, average transit depths as well as evening to morning transit depth asymmetries are particularly sensitive to the amount of cloud particles with mean sizes $\langle a \rangle <  0.1~\mu$m 
in the upper atmosphere in the optical spectral range.
This wavelength interval is covered by the upcoming PLATO space mission, as well as by CHEOPS and TESS:

-- CHEOPS covers the wavelengths $\lambda= 0.33\,\ldots\,1.1\mu$m and has shown its potential to characterize the atmospheres of selected hot to ultra-hot Jupiters (\citealt{Benz2021,Brandeker2022,Deline2025}). 

-- TESS covers the wavelength $\lambda= 0.6\,\ldots\,1.1\mu$m and is  complementary to CHEOPS for atmosphere characterization \citep[e.g.][]{Brandeker2022,Deline2025}.

-- PLATO has the capability to characterize the atmospheres of at least 200 gas giants covering a diverse set of bulk density and global temperatures \citep{Rauer2025}. The normal cameras cover the wavelengths $\lambda= 0.5\,\ldots\,1\mu$m. In addition, the two fast-cams split the flux into a red (0.5~$\mu$m -0.68~$\mu$m) and blue band (0.68~$\mu$m - 1~$\mu$m)\citep{Grenfell2020,Rauer2025}. The blue filter is thus ideally centered on the pressure broadened Na line part of the optical transmission spectrum (see e.g. Fig.~\ref{fig: TDepth_3Climates}).

\begin{figure}
%\centering
\includegraphics[width=1\linewidth]{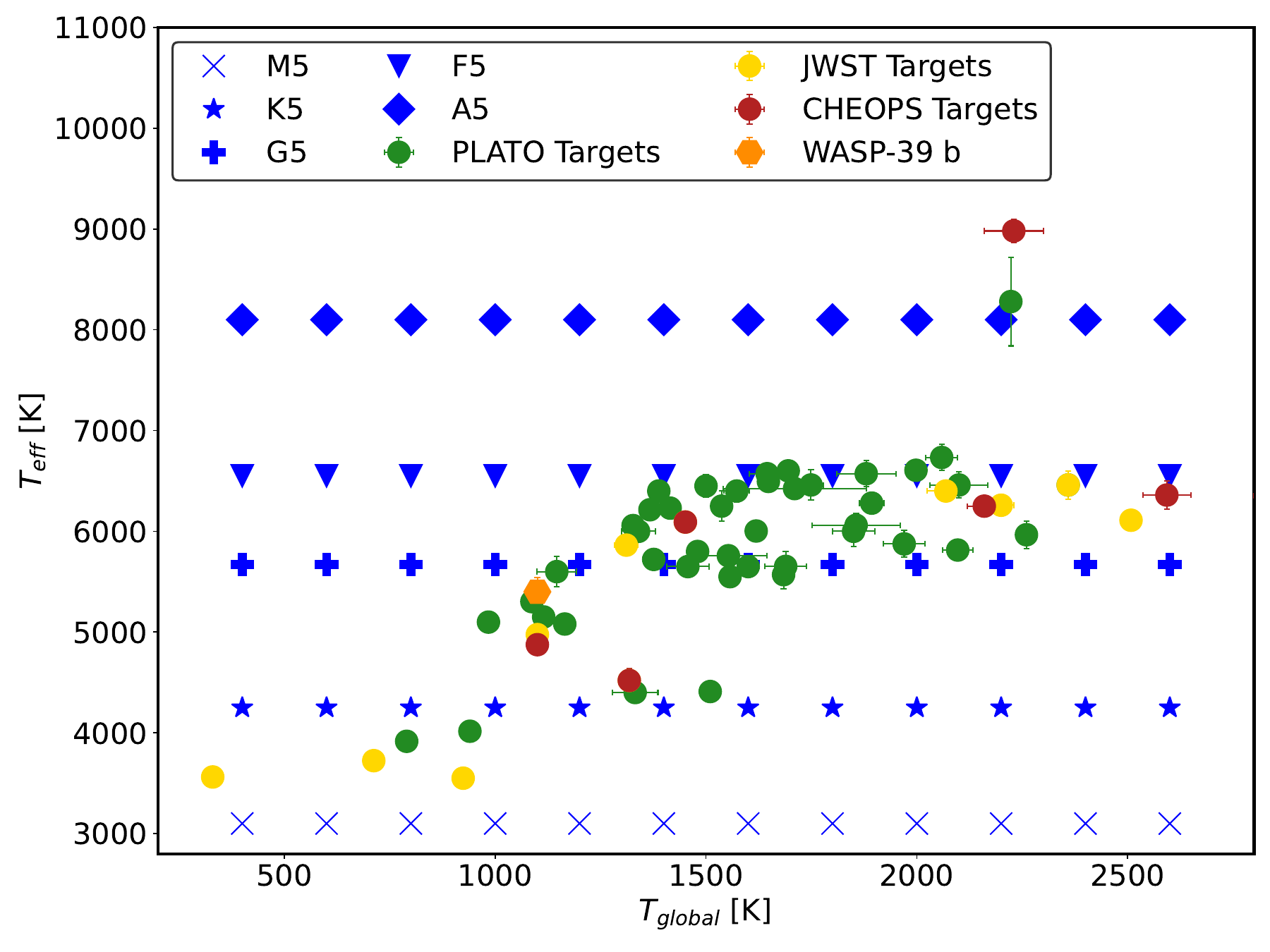}
\caption{
3D \texttt{ExoRad} GCM grid of 60 simulated planets (blue) for different host stars from M to A ($T_{\rm eff}$ [K]) and global planetary temperatures $T_{\rm global}$ [K] (\citealt{Plaschzug2025}).
Observation targets for JWST (yellow), PLATO (green, \citealt{Nascimbeni2025}), and CHEOPS (red) are overlayed. WASP-39b  is highlighted in dark orange.}
\label{fig: PLATO_targets}
\end{figure}

PLATO will provide accurate and precise transit depth measurements in up to three bands (white, blue, red) in the LOPS2 field over 2 years \citep{Rauer2025,Grenfell2020} (Fig.~\ref{fig: PLATO_Band}). Here, PLATO will mostly observe intermediately hot to ultra-hot gas giants around G and F stars (Fig.~\ref{fig: PLATO_targets} green diamonds). This section will therefore explore the potential of the PLATO space mission to study the diversity of 3D clouds in a large ensemble of warm to ultra-hot Jupiters via observations of transit asymmetries. In addition, CHEOPS and TESS may provide complementary information for specific planets.

\begin{figure*}
    \centering
        \includegraphics[width=0.45\textwidth]{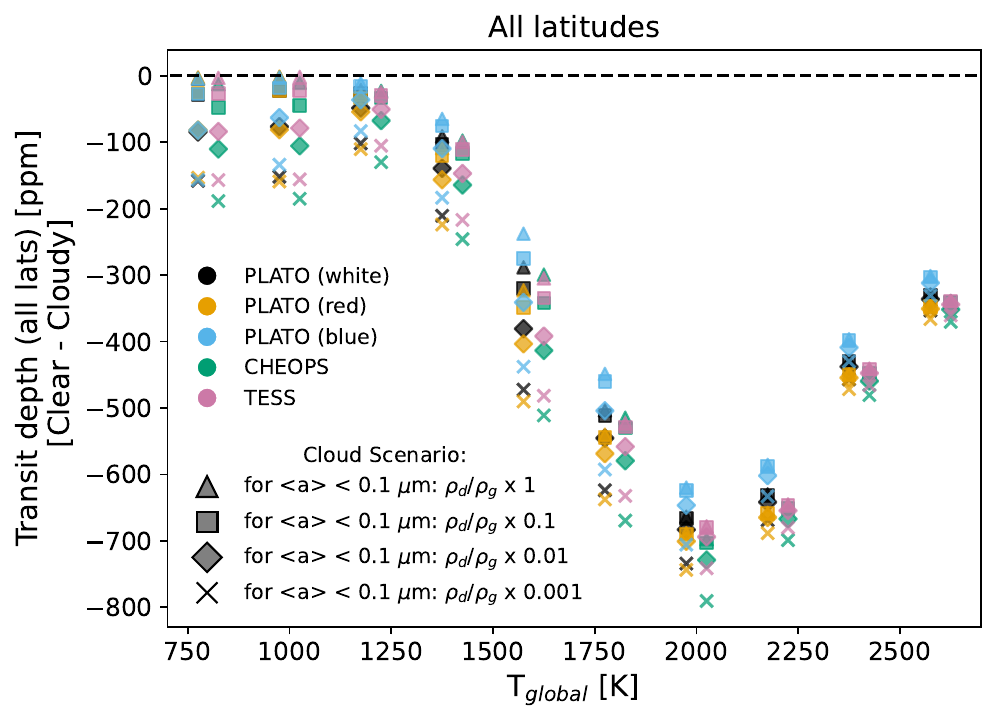}
    \includegraphics[width=0.45\textwidth]{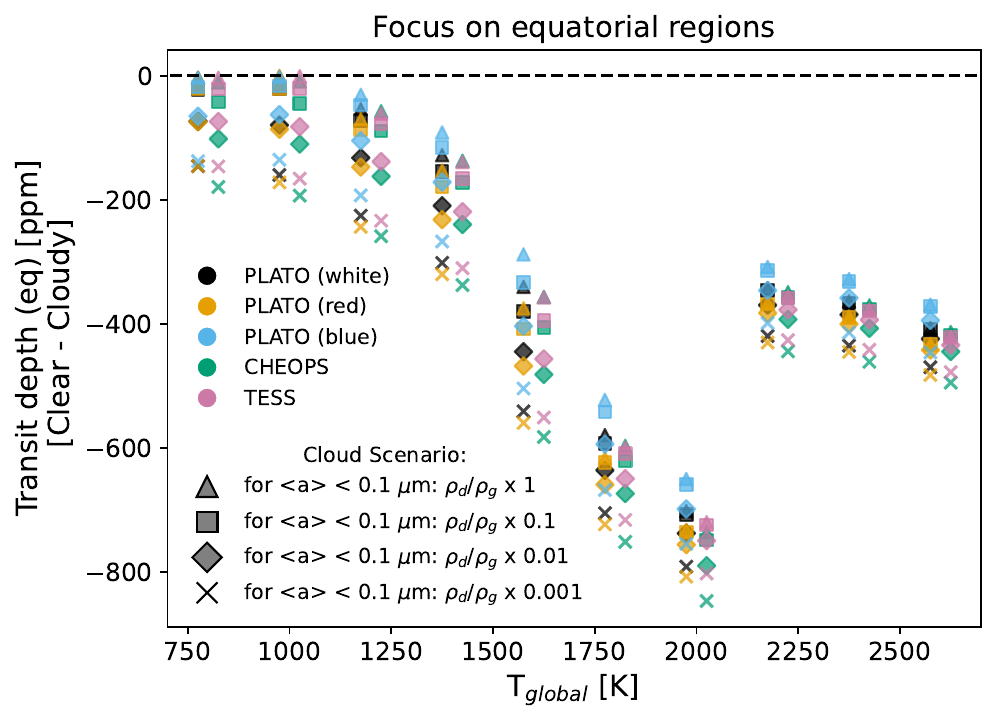}
    
        \includegraphics[width=0.45\textwidth]{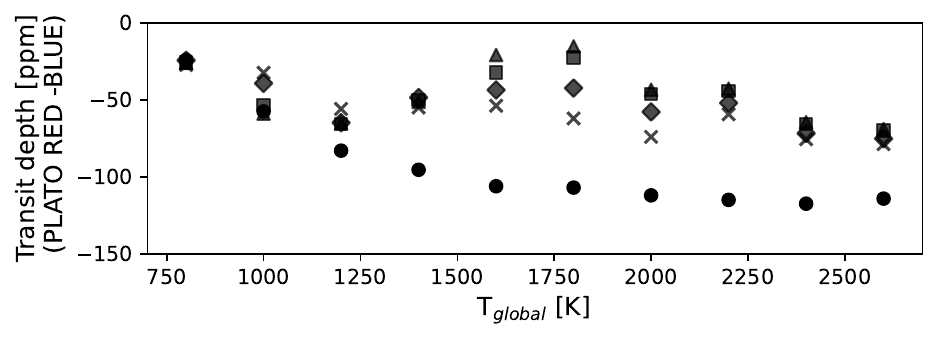}
            \includegraphics[width=0.45\textwidth]{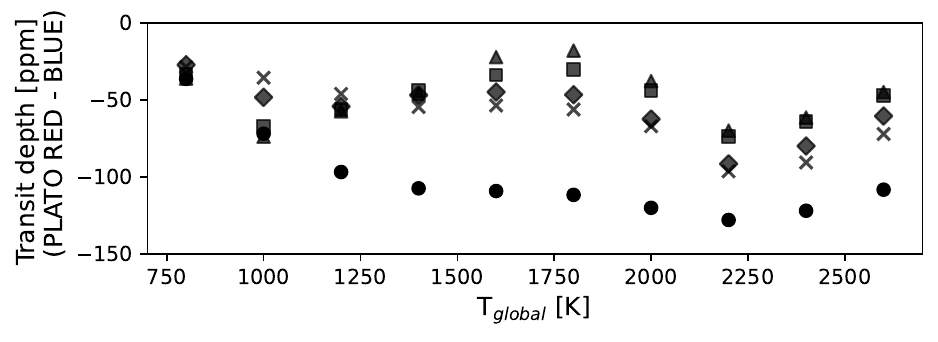}

    \caption{Transit depth differences between cloudy and clear scenarios. Top: Differences between (planetary) average transit depths of clear and cloudy atmosphere scenarios for tidally locked planets with $T_{\rm global}= 800$~K$\ldots 2600$~K orbiting G main sequence stars (left: all latitudes are used, right: only equatorial information is used).  The cloud mass loads, $\rho_g/\rho_d(z)$, in atmospheric layers where $\langle a \rangle <  0.1~\mu$m  are denoted by different markers (crosses: $10^{-3}\rho_g/\rho_d(z)$, diamonds:  $10^{-2}\rho_g/\rho_d(z)$, squares: $0.1\rho_g/\rho_d(z)$). Triangles denote Fe-free results with full $\rho_g/\rho_d(z)$. Colors denote different telescopes with optical band passes (black: PLATO's white band, golden: PLATO's red filter, blue: PLATO's blue filter, green: CHEOPS, pink: TESS. The PLATO data has been offset horizontally by -25~K and the TESS/CHEOPS data by +25~K, respectively, for better visibility.). Bottom: Average transit depths differences between PLATO's red and blue filter for cloudy (crosses, diamonds, squares, triangles) and clear atmosphere calculations (circles).}
    \label{fig: TDepth_G_PLATO_Clear_Cloud}
\end{figure*}

\paragraph{Average transit depth scenarios:} Figure~\ref{fig: TDepth_G_PLATO_Clear_Cloud}
presents the planetary average transit depth results (sum of morning and evening transit depth) using all latitude (left) and using only equatorial information (right) for the whole 3D AFGKM \texttt{ExoRad} GCM grid. The top row in both figures show the difference between the average transit depth differences assuming a clear atmosphere and different scenarios of cloud mass loads in the upper atmosphere for p$_{\rm gas}< p_{\rm gas}(\langle a \rangle <  0.1)~\mu$m. The differences in average transit depths between clear and cloudy atmospheres generally increase with global temperature until the ultra-hot temperature regime is reached at T$_{\rm global}=2200$K for G-type host stars. For the hottest global temperatures in the grid, the evening terminator that contributes 50\% to the planetary average transit depth is seen as partly cloudy, when the cloudy higher latitudes are taken into account (latitude averaged, Fig.~\ref{fig: TDepth_G_PLATO_Clear_Cloud} top left) or even completely cloud-free (only equatorial information, Fig.~\ref{fig: TDepth_G_PLATO_Clear_Cloud} top right). Consequently, the transit depth differences between cloudy and clear atmospheres decrease gradually with higher global temperature when all latitudes are taken into account (Fig.~\ref{fig: TDepth_G_PLATO_Clear_Cloud}, top left). Conversely, the transition between the intermediately hot regime and the ultra-hot Jupiter regime occurring at $T_{\rm global}=$2000~K $\rightarrow$ 2200 K is characterized by a sharper decrease (by 200 ppm, from -800 to -400 ppm) for the equatorial average transit depth (Fig.~\ref{fig: TDepth_G_PLATO_Clear_Cloud}, top left).

CHEOPS (Fig.~\ref{fig: TDepth_G_PLATO_Clear_Cloud} top, green symbols) with its extension towards shorter wavelengths compared to  PLATO (Fig.~\ref{fig: TDepth_G_PLATO_Clear_Cloud} upper row, black symbols) and TESS (Fig.~\ref{fig: TDepth_G_PLATO_Clear_Cloud} top, pink symbols) \footnote{See Figs.~\ref{fig: CHEOPS_Band} and \ref{fig: PLATO_Band} for the bandpass description.} tends to yield the largest differences between clear and cloudy atmosphere scenarios at the transition between intermediately hot to ultra-hot Jupiters (T$_{\rm global}=1800\,\ldots\,2200$K). Multiple, precise transit depth observations of $<100$~ppm accuracy with CHEOPS thus may complement PLATO observations for specific planets at the intermediately hot to ultra-hot Jupiter climate threshold. TESS is comparable to PLATO's red filter on the fast cameras albeit with lower precision which makes synergies less obvious.

The PLATO fast cameras' red and blue filter could be in principle used to obtain a very low resolution (two band) transmission spectrum. This is shown in Fig.~\ref{fig: TDepth_G_PLATO_Clear_Cloud} (bottom row), where the differences between the average transit depth in PLATO's red and blue bandpass, respectively, are calculated for the clear and cloudy atmosphere scenarios.  An atmosphere that is cloud-free at both limbs would be immediately apparent for global temperatures $T_{\rm global}>1200~K$ as it would result in an average transit depth that is deeper by at least 100~ppm in the blue band compared to the red band and corresponds to the required accuracy calculated by \citet{Grenfell2020} for G stars.

Cloudy scenarios flatten the scattering slope observed in PLATO's red and blue band and would thus consequently yield lower signals. 75-90~ppm differences may be reached for some cloud scenarios in the ultra-hot Jupiter regime, in particular if the evening limb is cloud-free (Fig.~\ref{fig: TDepth_G_PLATO_Clear_Cloud} bottom left).

In summary, precise PLATO average transit depths can shed light on the evening cloud limb coverage in PLATO's white band, which can be complemented with observations in PLATO's red and blue filter, provided the accuracy of the latter is smaller than 75~ppm.

\begin{figure*}
    \centering

        \includegraphics[width=0.45\textwidth]{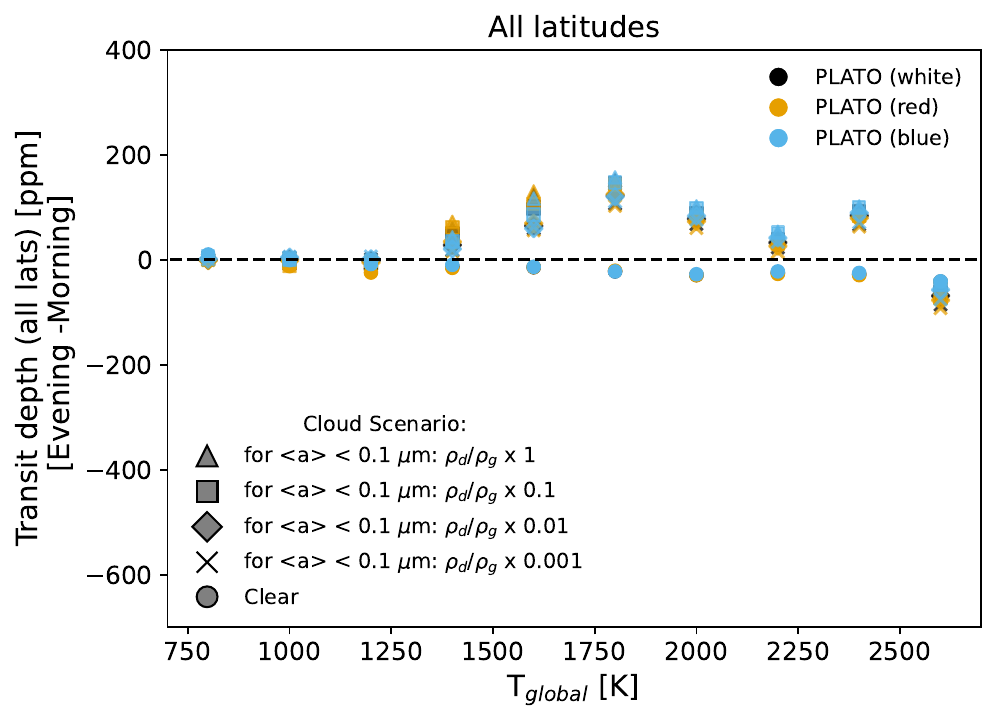}
            \includegraphics[width=0.45\textwidth]{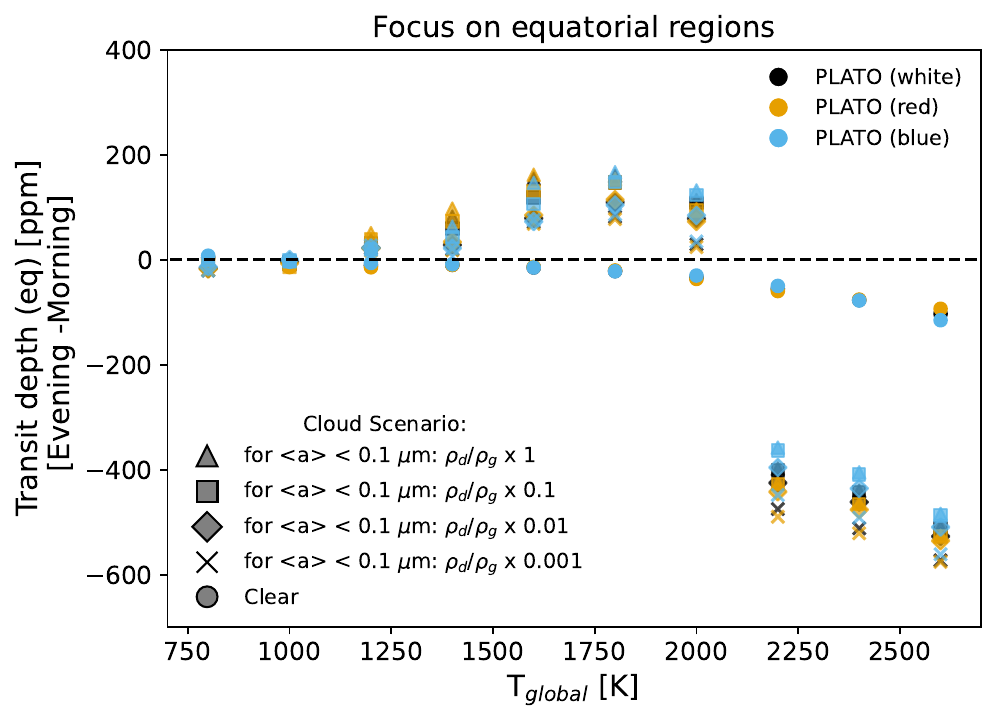}

 \includegraphics[width=0.45\textwidth]{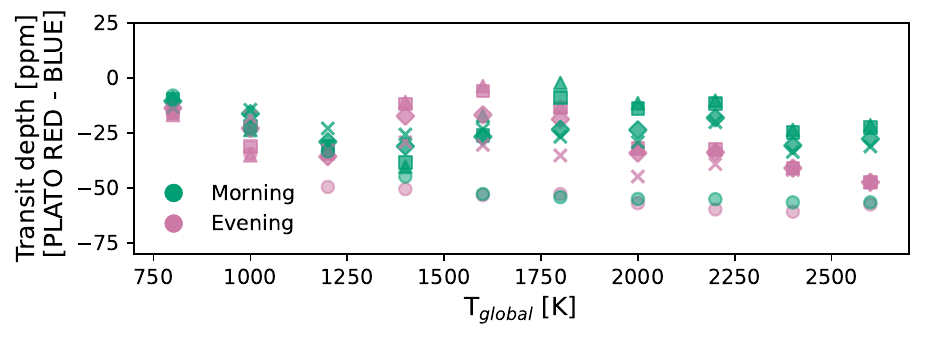}
            \includegraphics[width=0.45\textwidth]{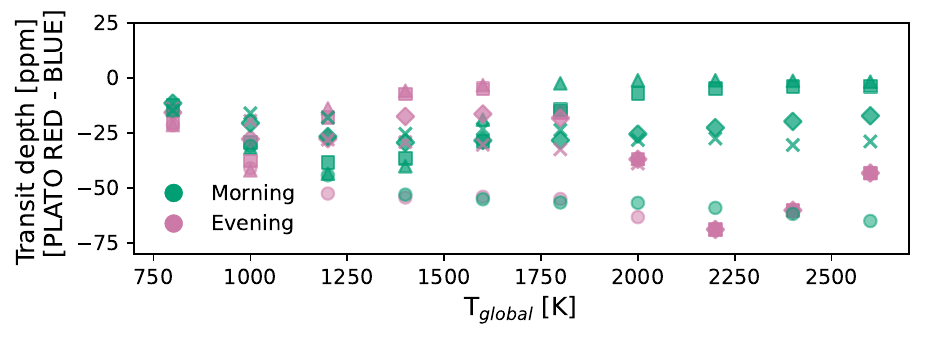}
    \caption{Transit asymmetry calculations from different cloud scenarios for PLATO. Top: Differences between the individual evening and morning transit depths (transit asymmetries) for tidally locked planets with $T_{\rm global}= 800$~K$\ldots 2600$~K orbiting G main sequence stars (left: all latitudes are used, right: only equatorial information is used). Results for clear atmosphere calculations are indicated by circles. Cloudy scenarios with different cloud mass loads, $\rho_g/\rho_d(z)$, in atmospheric layers where $\langle a \rangle <  0.1~\mu$m  are denoted by different markers (crosses: $10^{-3}\rho_g/\rho_d(z)$, diamonds:  $10^{-2}\rho_g/\rho_d(z)$, squares: $0.1\rho_g/\rho_d(z)$). Triangles denote Fe-free results with full $\rho_g/\rho_d(z)$. Colors denote different PLATO observation  modes (black: PLATO's white band, golden: PLATO's red filter, blue: PLATO's blue filter). Bottom: Evening/Morning transit depth differences between PLATO fast cameras' red and blue filter (red symbols/green symbols)).}
    \label{fig: TDepth_G_PLATO_Diff}
\end{figure*}

\paragraph{Evening-to-morning transit depth differences:} 
Transit depth asymmetries calculations for the different PLATO observation modes (Fig.~\ref{fig: TDepth_G_PLATO_Diff}, top) show that clear atmosphere models have very little evening-to-morning transit depth contrast between $0.5 \ldots 1 \mu$m for all global temperatures. Conversely, for all cloudy scenarios, the transit asymmetry (i.e., the evening - morning difference) tends to increase to about 150~ppm with global temperature in the intermediately hot climate regime. For T$_{\rm global}\geq 2200$~K, the transit asymmetry depends on the evening limb cloud coverage.

If the evening limb is completely cloud-free in the ultra-hot Jupiter regime, the morning terminator may appear much larger than the evening terminator, as is evidenced most clearly when only equatorial information is used, resulting in evening-to-morning transit depth differences of up to -500~ppm (Fig.~\ref{fig: TDepth_G_PLATO_Diff}, top right). In the presented work, however, the cloudy higher latitudes at the evening limb have a strong impact on the observed transit asymmetry. Higher latitudes only gradually clear up with increasing temperature (Fig.~\ref{fig: TDepth_G_PLATO_Diff}, top left), leading instead to a gradual decrease in positive transit depth asymmetries, until the cloud morning top effect, i.e. the vertical extent of the morning cloud, starts to dominate for $T_{\rm global}=2600$~K, resulting in a small negative transit asymmetry (-50~ppm). 

The transit limb asymmetries in all PLATO observation modes (Fig.~\ref{fig: TDepth_G_PLATO_Diff} top; white band: black, red filter: golden, blue filter: blue ) follow the same trend. Thus, no qualitatively difference is expected from comparing evening-to-morning transit depth differences in the red and blue filter (see Appendix~\ref{sec: PLATO red and blue asym} for complete set of figures.)

The scattering slope between PLATO's red and blue filter can, however, differ for a cloudy morning terminator (flatter slope) and a cloud-free evening terminator (steeper slope). Thus, a two-band transmission spectrum applied separately for the evening and morning terminator may provide additional evidence for strongly asymmetric cloud coverage in the ultra-hot Jupiter climate regime (Fig.~\ref{fig: TDepth_G_PLATO_Diff}, bottom row).

 For $T_{\rm global}>2000$K, the scattering slope is in the equatorial transit depth calculations indeed flatter for the cloudy morning terminator scenarios ($<40$~ppm, Fig.~\ref{fig: TDepth_G_PLATO_Diff}, bottom left: green symbols except circles) compared to the evening terminator with  differences between ($40-70$~ppm, Fig.~\ref{fig: TDepth_G_PLATO_Diff}, bottom left: pink symbols except circles).
 
 In clear atmosphere scenarios, both limbs exhibit a similarly steep slope for all $T_{\rm global}>1200$K. That is also true, when all latitudes are included (50-70~ppm, Fig.~\ref{fig: TDepth_G_PLATO_Diff} bottom right: circles).
 
 When both limbs are at least partly cloud covered, the flattening of the scattering slope results in small transit depth differences between the red and blue filter for both, morning (green symbols) and evening terminator (red symbols). That is true for the equatorial calculations for $T_{\rm global}\leq 2000$~K ($<40$~ppm, Fig.~\ref{fig: TDepth_G_PLATO_Diff} bottom right: all symbols except circles) and for the transit depth calculations that include higher latitudes for all temperatures (Fig.~\ref{fig: TDepth_G_PLATO_Diff}, bottom left: all symbols except circles). The latter are representative of the predicted transit asymmetries arising from the 3D GCM grid.

Measurable planetary average transit depth differences in PLATO's red and blue filter of 50\,\ldots\,100 ppm
were considered realistic by \citet{Grenfell2020} for F and G stars. For their analyses, the flat bottom of a full transit was used that typically covers $\geq 1$~hour of the transit lightcurve. Measuring morning and evening transit depths separately to extract the scattering slope via two band transmission spectroscopy with PLATO' fast cameras requires analyzing the shorter transit ingress and egress part of the lightcurve. Ingress/egress duration is about 30~mins for WASP-39b \citep{Espinoza2024}. In addition, stellar limb darkening needs to be taken into account.

Here, we highlight the large potential of PLATO observations based on a  conservative hierarchical cloud model that underestimates the transit asymmetry for WASP-39b by a factor of two. In addition, the present results assume  relatively small 1 R$_{\rm Jup}$-sized planets. In the ultra-hot Jupiter regime, the planetary radii are $>$1 R$_{\rm Jup}$
causing an  increased  red and blue transit depth signal to noise ratio.  Based on the results presented here, it is therefore reasonable to invest in more detailed transit depth observations in particular for ultra-hot Jupiters with PLATO's normal as well as with the PLATO fast cameras.

\begin{figure}[ht]
    \centering
        \includegraphics[width=0.45\textwidth]{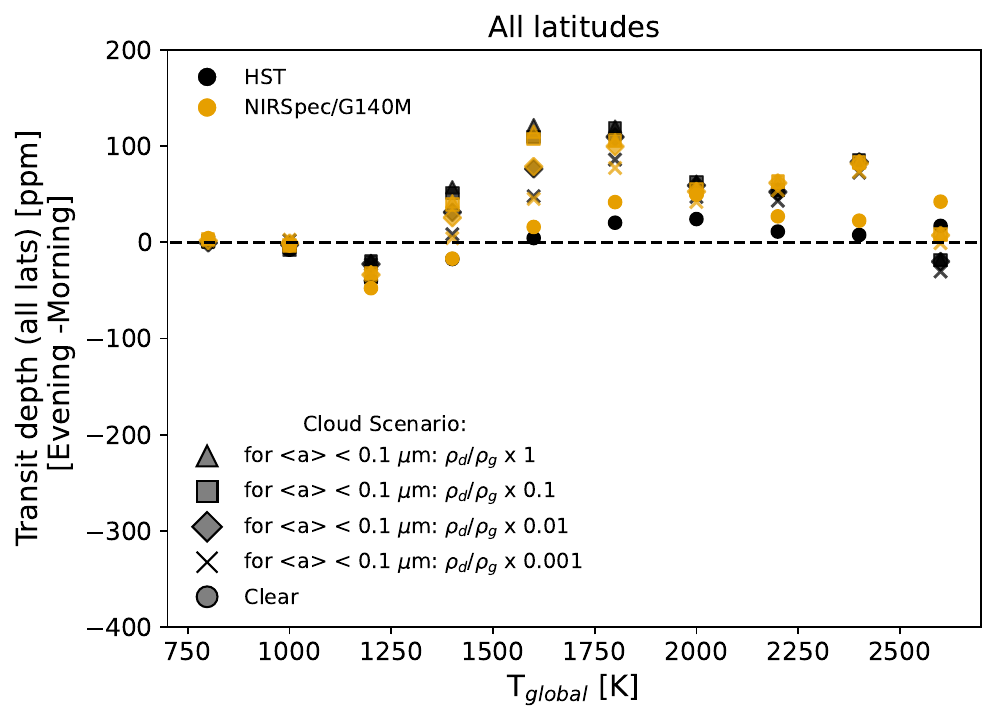}
        
    \caption{Transit asymmetry calculations from different cloud scenarios in the near infrared. Differences between the individual evening and morning transit depths (transit asymmetries) for tidally locked planets with $T_{\rm global}= 800$~K$\ldots 2600$~K orbiting G main sequence stars (all latitudes are used). Results for clear atmosphere calculations are indicated by circles. Cloudy scenarios with different cloud mass loads, $\rho_g/\rho_d(z)$, in atmospheric layers where $\langle a \rangle <  0.1~\mu$m  are denoted by different markers (crosses: $10^{-3}\rho_g/\rho_d(z)$, diamonds:  $10^{-2}\rho_g/\rho_d(z)$, squares: $0.1\rho_g/\rho_d(z)$). Triangles denote Fe-free results with full $\rho_g/\rho_d(z)$. Colors denote values integrated in the HST/WFC3IR (black) and NIRSpec/G140M (golden) wavelengths.}
    \label{fig: TDepth_G_HST_Diff}
\end{figure}

\subsection{The ideal PLATO - HST - JWST wavelength 
window for transit limb asymmetry observations}
\label{sec: NIR}

The James Webb space telescope (JWST) probes in the infra-red mainly the chemistry of an exoplanet gas phase and provides indication for cloud material contributions \citep{Dyrek2024,Grant2023}

As shown in Sect.~\ref{sec: WASP39b}, observations of WASP~39b  for $\lambda=3\,\ldots\,5\mu$m with NIRSpec and/or Prism 
suggest an iron conundrum since cloud particles must be iron free to reproduce the observed spectrum with our forward modelling framework. This wavelength range is, however, less sensitive to the upper atmosphere cloud particles. Conversely, observations in the optical, $\lambda=0.5\,\ldots\,1\mu$m, are very sensitive to the upper atmosphere sub-micron %condensate
cloud particles.

From our climate overview studies (Figs.~\ref{fig: TDepth_3Climates} -  \ref{fig: TDepth_Asym_3Climates_all_Lats}), $\lambda=0.6\,\ldots\,2.8\mu$m emerges as the wavelength interval that is well suited to tie together transit depth asymmetry observations with PLATO and JWST across the hot and ultra-hot Jupiter population. This is %also
the wavelength interval covered by JWST with NIRISS and NIRSpec/Prism and also partly by NIRSpec/G140M and HST/WFC3/IR.

To demonstrate the PLATO-JWST-HST synergy, we focus here solely on transit depth asymmetries, i.e. evening to morning transit depth differences, $\Delta T_{Depth}$, for which all latitudes are considered. The calculations with only equatorial information are shown in Sect.~\ref{sec: Equatorial}. The individual transit morning/evening transmission spectra are each integrated over two wavelength bands: $\lambda_{\rm HST}=0.8\,\ldots\,1.7\mu$m and $\lambda_{\rm NIRSpec}=1.0\,\ldots\,1.87~\mu$m. The NIRSpec/G140M wavelength range is used to highlight the importance of the near infrared region to connect complex cloud properties as observed in the optical and the infrared. For actual observations, NIRSpec PRISM or NIRISS/SOSS observations are more appropriate.

Figure~\ref{fig: TDepth_G_HST_Diff} shows that transit depth asymmetries should follow a very similar trend in the HST and NIRSpec band compared to the PLATO band for cloudy gas giant atmospheres: The evening to morning transit depth difference is positive and increases with global temperature in the intermediately hot climate regime. The difference is decreasing in the ultra-hot Jupiter regime when all laltitudes (Fig.~\ref{fig: TDepth_G_HST_Diff}) are considered.  The transit asymmetries values of $\lesssim$150~ppm in the intermediately hot Jupiter regime are also similar between HST, NIRSpec/G140M and the PLATO band. The trend in negative transit asymmetry as seen when only equatorial information is used is also preserved between the PLATO, and HST, NIRSpec/G140M wavelength regime (Fig.~\ref{fig: TDepth_G_HST_Diff_eq}).

At T$_{\rm global}=1600$K, a particularly large differences in cloud scenarios occur with a maximum difference of 100~ppm between scenarios. In this respect, HST and NIRspec/G140M asymmetry observations would be complementary to the PLATO red and blue filter band that shows the strongest contrast  between cloud scenarios at T$_{\rm global}=1200$K. Therefore, with a sufficient accuracy ($<50$~ppm), cloud scenarios and the difference between cloud-free and partially cloudy evening terminator scenarios could be identified.

The qualitative similarities of transit asymmetries between HST, NIRSpec and PLATO may be used 
for population studies across the intermediately hot and ultra-hot Jupiter regime. A combination of these observations would be ideal to also calibrate against stellar variability impact that is expected to decrease between 1 and 2~$\mu$m \citep{Kostogryz2025}.

\subsection{A complementary wavelength window into detailed cloud properties with JWST/MIRI}
\label{sec: MIRI}

Figures~\ref{fig: TDepth_Asym_3Climates} and~\ref{fig: TDepth_Asym_3Climates_all_Lats} show that transit asymmetries  of $\lesssim$200~ppm can occur at 10~$\mu$m. That is in the MIRI wavelength range.  It has been shown, however, that the uncertainty of JWST/MIRI transmission spectra can be larger than 100 ppm for $> 8$~$\mu$m \citep[e.g.][]{Grant2023}. We thus explore in how far the wavelength regions $5\,\ldots\,8$~$\mu$m  and $8\,\ldots\,10$~$\mu$, are helpful to elucidate cloud properties to complement observations in the optical.

\begin{figure}[ht]
    \centering
        \includegraphics[width=0.45\textwidth]{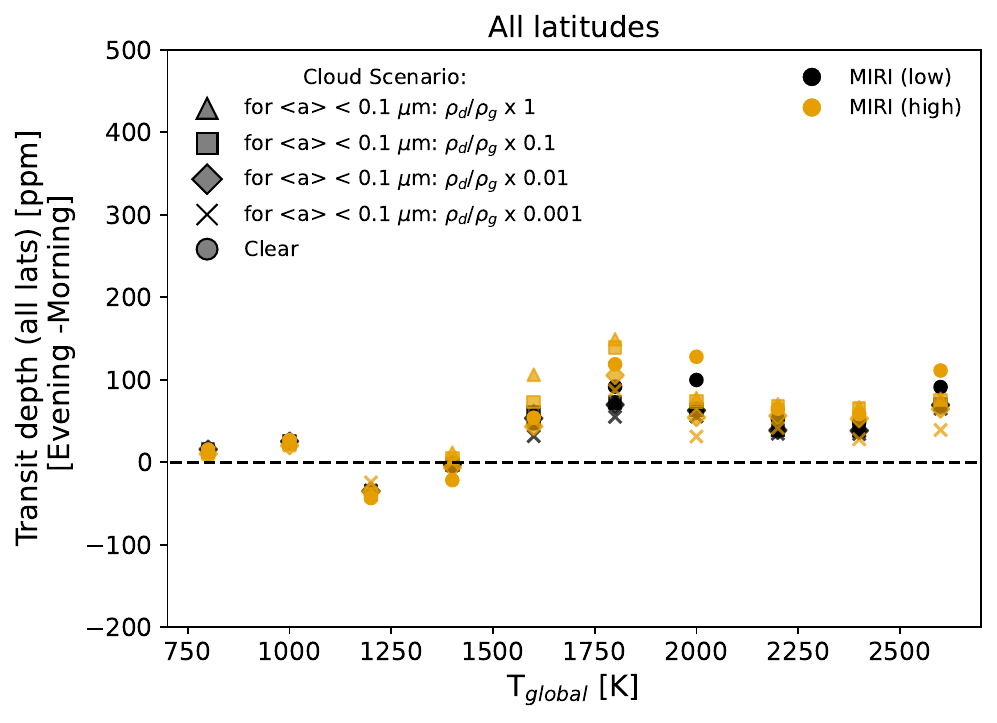}
        
    \caption{Transit asymmetry calculations from different cloud scenarios for MIRI. Differences between the individual evening and morning transit depths (transit asymmetries) for tidally locked planets with $T_{\rm global}= 800$~K$\ldots 2600$~K orbiting G main sequence stars (all latitudes are used). Results for clear atmosphere calculations are indicated by circles. Cloudy scenarios with different cloud mass loads, $\rho_g/\rho_d(z)$, in atmospheric layers where $\langle a \rangle <  0.1~\mu$m  are denoted by different markers (crosses: $10^{-3}\rho_g/\rho_d(z)$, diamonds:  $10^{-2}\rho_g/\rho_d(z)$, squares: $0.1\rho_g/\rho_d(z)$). Triangles denote Fe-free results with full $\rho_g/\rho_d(z)$. Colors denote values integrated between 5-8~$\mu$m (black) and 8-10~$\mu$m (golden).}
    \label{fig: TDepth_G_MIRI_Diff}
\end{figure}

When all latitudes are considered (Fig.~\ref{fig: TDepth_G_MIRI_Diff}), the evening terminator will be at least partly cloud covered even for the hottest $T_{\rm global}$. The largest transit asymmetries (100~ppm) occur in between cloud scenarios for T$_{\rm global}=1200\,\ldots\,1800$ K. A 50~ppm precision in transit asymmetry between evening to morning terminator would thus ideally complement similar observations in HST or NIRSpec or the PLATO band.

The shorter MIRI wavelengths ($5\,\ldots\,8\mu$m) are less sensitive to detailed cloud scenarios, but would still be informative in population studies to constrain the cloud coverage scenario of the evening terminator in the ultra-hot Jupiter regime. With a partially cloudy evening terminator (Fig.~\ref{fig: TDepth_G_MIRI_Diff}), the evening to morning transit asymmetry ranges between $50\,\ldots\,100$~ppm for T$_{\rm global}=1800\,\ldots\,2600$~K.

\section{Dependence on host star spectral type (AFGKM)}
\label{sec: all}
Previous sections have shown that average transit depths as well as transit depth differences between evening and morning limb may be used to observationally constrain climate regimes of gaseous exoplanets for G type host stars. A coherent combination of PLATO and potentially also CHEOPS observations with JWST and HST provides ideal synergies in validating evening limb cloud coverage for ultra-hot Jupiters.
The present section focuses mainly on transit depth asymmetries between the evening and morning limb and presents the results for the whole ensemble of planets that orbit A, F, G, K and M-type host stars  
to provide an overview of the impact of clouds on transit depth and transit variations for different host stars at various critical wavelength ranges identified so far.

\begin{figure}
    \centering
        \includegraphics[width=0.45\textwidth]{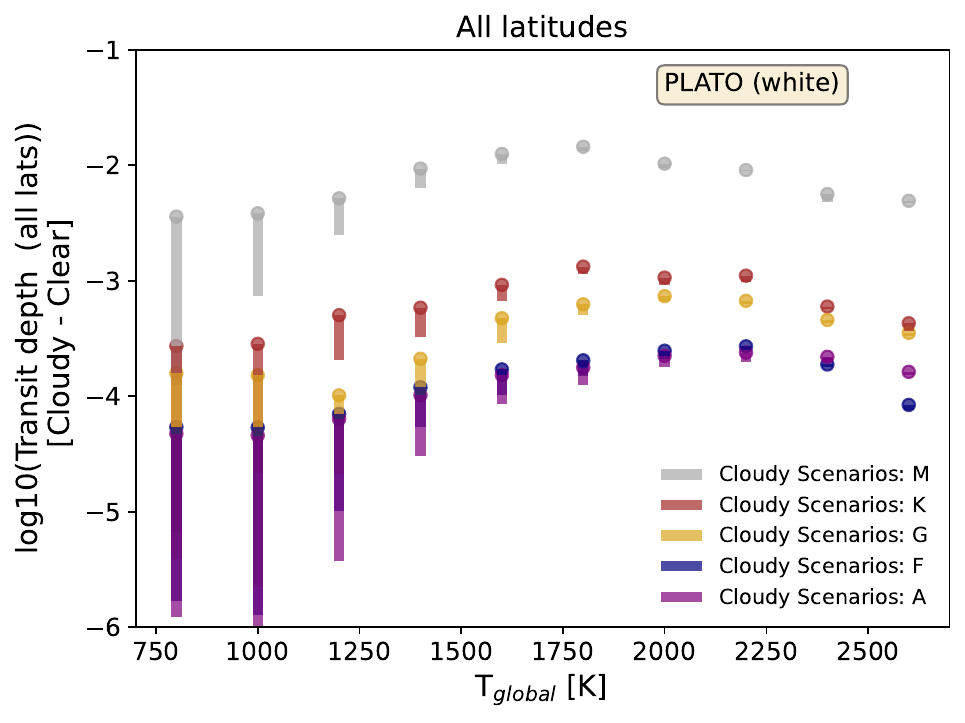}
  
            \includegraphics[width=0.45\textwidth]{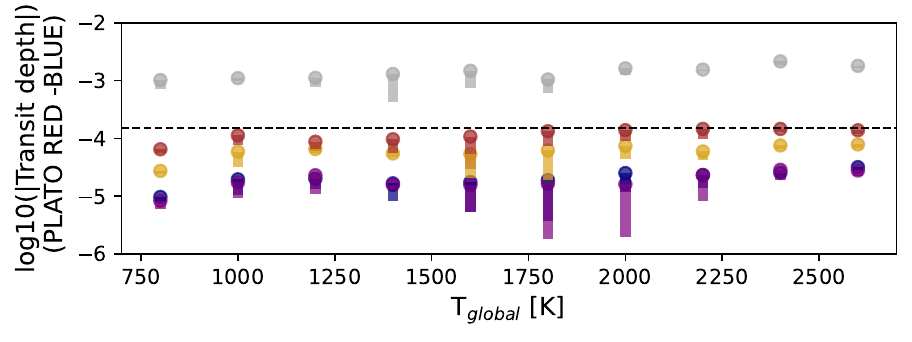}

    \caption{Difference between clear and cloudy transit depth calculations around diverse host stars for PLATO. Top: Differences between latitudinally averaged sums of morning and evening transit depths of cloud scenarios for observations in PLATO's white band for tidally locked planets with $T_{\rm global}= 800$~K$\ldots 2600$~K. Colored stripes indicate the range for cloud scenarios (with cloud mass load, $\rho_g/\rho_d(z)$ for atmospheric layers, where $\langle a \rangle <  0.1~\mu$m is scaled by values between 1 and $10^{-3}$) around different main sequence host stars (purple: A, dark blue: F, golden: G, orange: K, gray: M). Circles denote the maximum values for clarity. Bottom: Range of absolute differences between average transit depths in PLATO's red and blue for diverse cloudy atmosphere scenarios.}
    \label{fig: TDepth_All_PLATO_Clear_Cloud}
\end{figure}

\begin{figure*}
    \centering
       \includegraphics[width=0.45\textwidth]{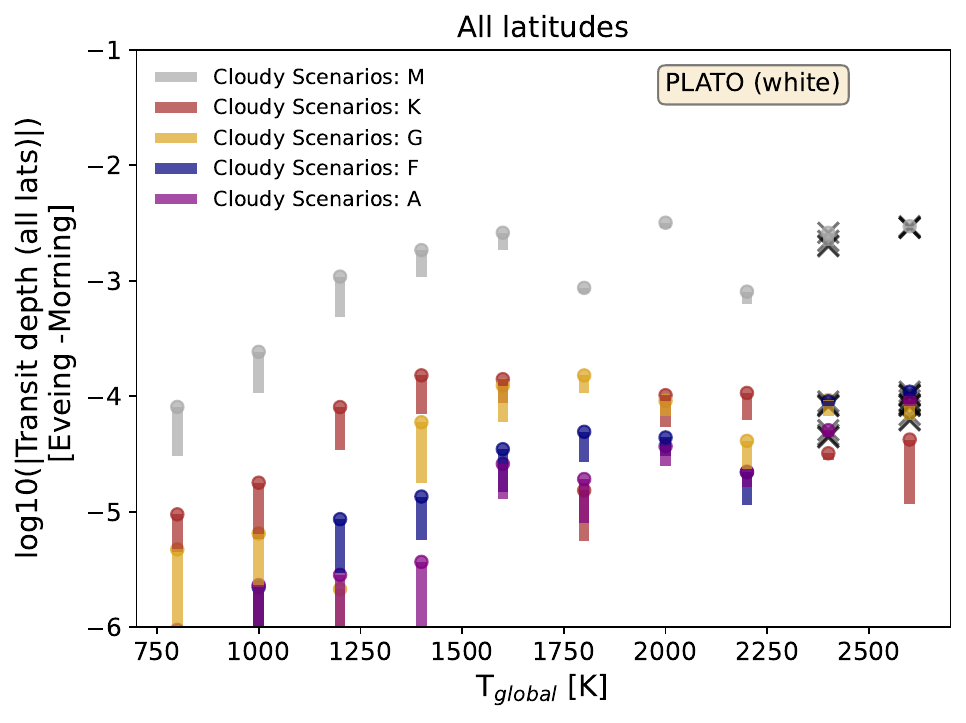}
    \includegraphics[width=0.45\textwidth]{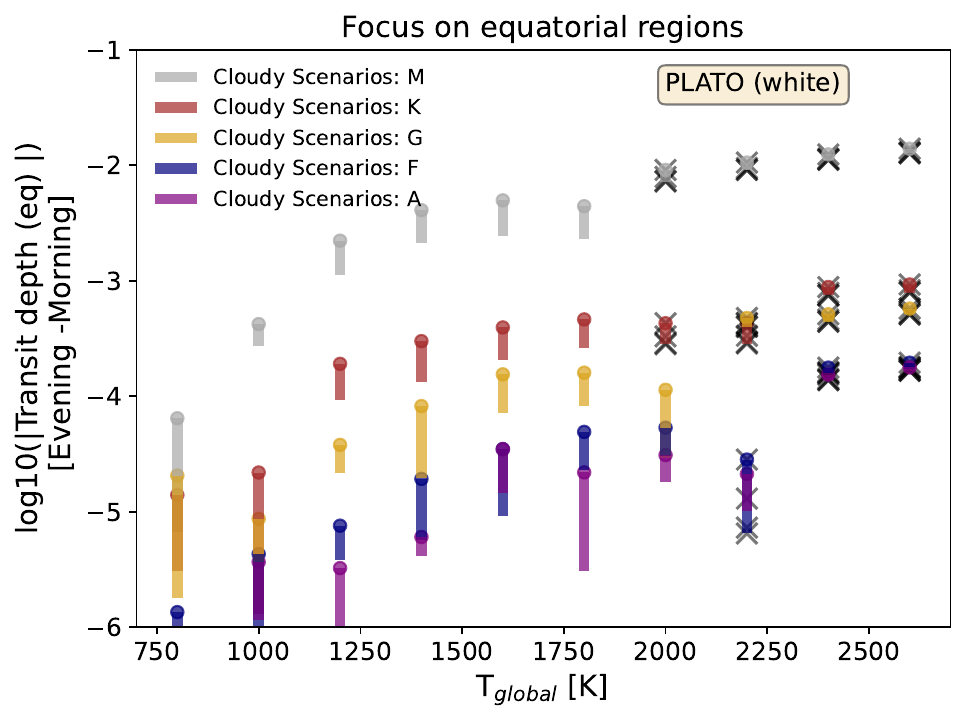}
    
         \includegraphics[width=0.45\textwidth]{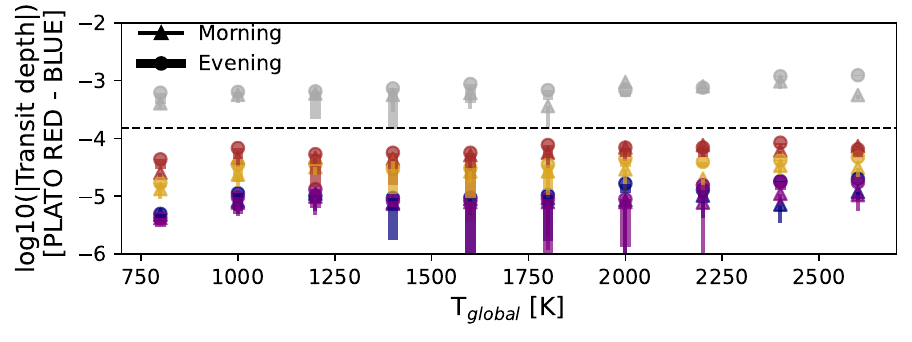}
            \includegraphics[width=0.45\textwidth]{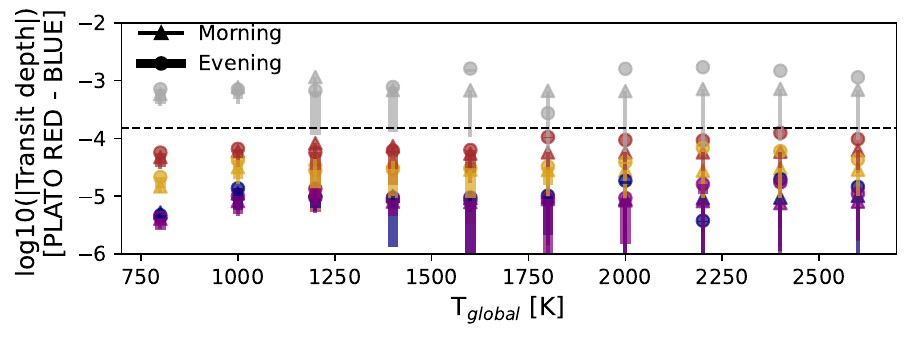}

    \caption{Transit asymmetry calculations for planets around diverse host stars for PLATO. Top: Differences between the individual evening and morning transit depths (transit asymmetries) of cloudy atmosphere scenarios for observation in PLATO's white band for tidally locked planets with $T_{\rm global}= 800$~K$\ldots 2600$~K (left: all latitudes are used, right: only equatorial information is used). Colored stripes indicate the range for cloud scenarios (with cloud mass load, $\rho_g/\rho_d(z)$ for atmospheric layers, where $\langle a \rangle <  0.1~\mu$m is scaled by values between 1 and $10^{-3}$) around different main sequence host stars (purple: A, dark blue: F, golden: G, orange: K, gray: M). Circles denote the maximum values and crosses indicate negative evening-morning transit asymmetries for clarity. Bottom: Differences for individual  evening (thick lines with circle) and morning (thin lines with triangle) transit depths between PLATO's red and blue filter. The dashed line indicates a transit depth signal of 150~ppm.}
    \label{fig: TDepth_All_PLATO_Diff}
\end{figure*}

For PLATO, the mission goal to provide accurate and precise transit depths is taken into account. Thus, the average transit depth differences between cloudy and clear atmosphere scenarios (Fig.~\ref{fig: TDepth_All_PLATO_Clear_Cloud}, top) for the whole 3D AFGKM \texttt{ExoRad} GCM grid, as well as the differences in average transit depth in PLATO's red and blue filter for each scenario (Fig.~\ref{fig: TDepth_All_PLATO_Diff}, bottom) are presented. For the latter, it is beneficial that individual transits can be observed simultaneously in the fast cameras in both bands, so that a very precise two-band transmission spectrum can be extracted. Also here synthetic transit asymmetry calculations that take into account both, all latitudes and only equatorial regions are shown.

\begin{figure}[ht]
    \centering
        \includegraphics[width=0.45\textwidth]{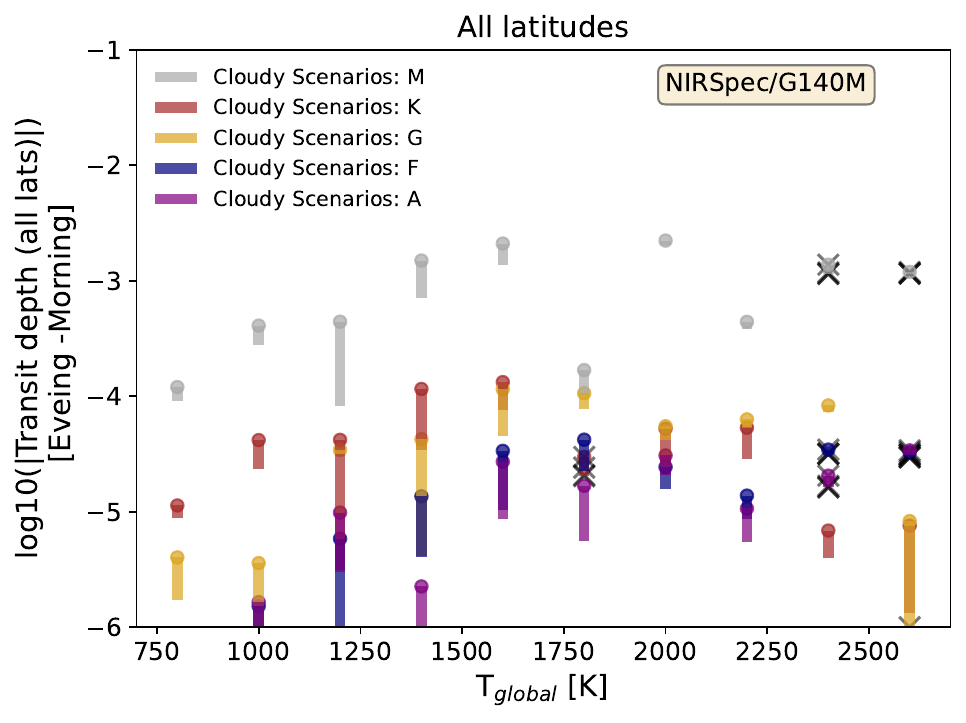}
        
    \caption{Transit asymmetry calculations for planets around diverse host stars in the near infrared. Differences between the individual evening and morning transit depths (transit asymmetries) of cloudy atmosphere scenarios for observation in the integrated NIRSpec/G140M wavelength range for tidally locked planets with $T_{\rm global}= 800$~K$\ldots 2600$~K (left: only equatorial information is used, right: all latitudes are used). Colored stripes indicate the range for cloud scenarios (with cloud mass load, $\rho_g/\rho_d(z)$ for atmospheric layers, where $\langle a \rangle <  0.1~\mu$m is scaled by values between 1 and $10^{-3}$) around different main sequence host stars (purple: A, dark blue: F, golden: G, orange: K, gray: M). Circles denote the maximum values and crosses indicate negative evening-morning transit asymmetries for clarity.}
    \label{fig: TDepth_All_NIR_Diff}
\end{figure}

For other instruments, only transit depth asymmetries between the evening and morning limb for the whole ensemble of planets that orbit A, F, G, K and M-type host stars  are studied. In addition, only the latitudinally averaged transit calculations are shown that best represent transit observations. The equatorial transit calculations that are still informative about the latitudinal temperature and cloud variations are shown in Sect.~\ref{sec: Equatorial}.
This section aims to provide an overview of the impact of clouds on transit depth and transit variations for different host stars at various critical wavelength ranges identified so far. Scenarios with negative evening-to-transit-depth asymmetries in the optical and near infrared range are highlighted in Figures~\ref{fig: TDepth_All_PLATO_Diff}-\ref{fig: TDepth_All_NIR_Diff} (black crosses).

\begin{figure*}[ht]
    \centering
        \includegraphics[width=0.45\textwidth]{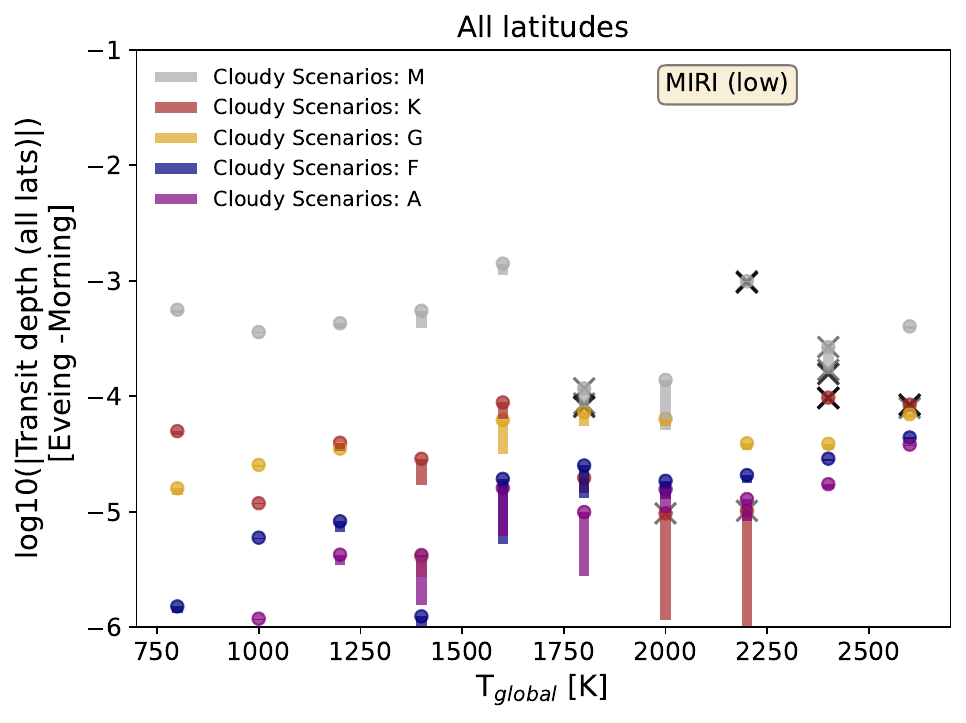}
 \includegraphics[width=0.45\textwidth]{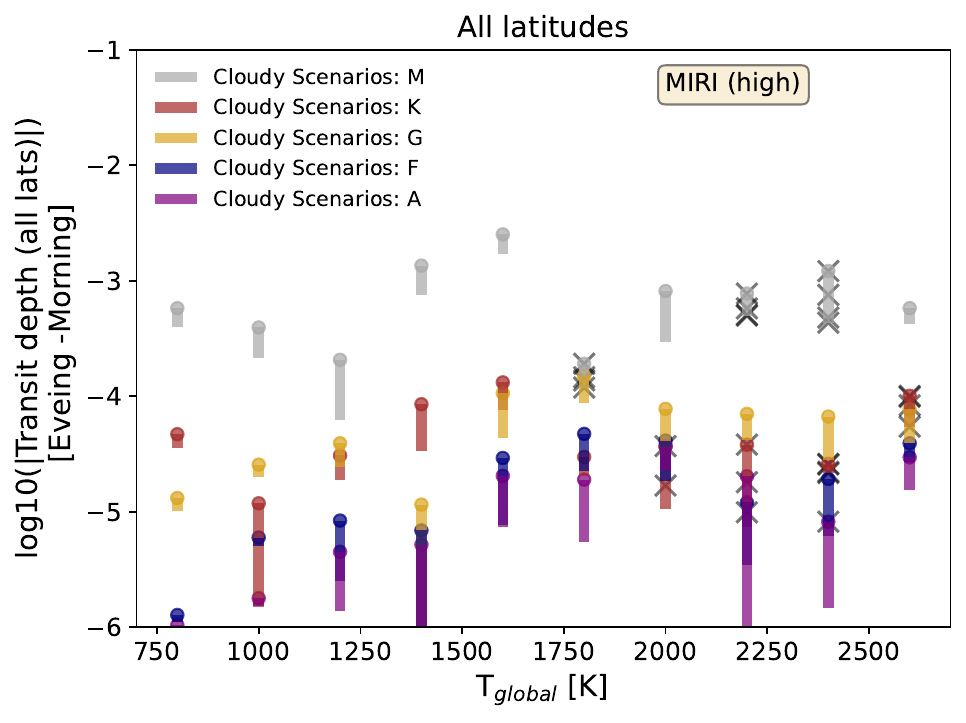}        
    \caption{Transit asymmetry calculations for planets around diverse host stars for MIRI. Differences between the individual evening and morning transit depths (transit asymmetries) of cloudy atmosphere scenarios for observation in the integrated wavelength range between 5 -8~$\mu$m for tidally locked planets with $T_{\rm global}= 800$~K$\ldots 2600$~K (left: only equatorial information is used, right: all latitudes are used). Colored stripes indicate the range for cloud scenarios (with cloud mass load, $\rho_g/\rho_d(z)$ for atmospheric layers, where $\langle a \rangle <  0.1~\mu$m is scaled by values between 1 and $10^{-3}$) around different main sequence host stars (purple: A, dark blue: F, golden: G, orange: K, gray: M). Circles denote the maximum values  for clarity.}
    \label{fig: TDepth_All_MIRI_Diff}
\end{figure*}

\subsection{Trends in the optical range for different host stars}
\label{ss:optical}
For all host stars, large  negative transit asymmetries ($T_{\rm Depth}<-10^{-4}$, Fig.~\ref{fig: TDepth_All_PLATO_Diff}, right) can occur in the PLATO white band consistently for sufficiently large global temperatures in the ultra-hot Jupiter regime if the evening terminator is cloud-free: For G-type host stars for T$_{\rm global}\geq2000$~K, 
for F and A host stars T$_{\rm global}\geq2200$~K, for M and K stars this transitions occur at cooler temperatures, T$_{\rm global}=1600\,\ldots\,1800$~K. 

\paragraph{M stars:} Jupiter-sized gas planets around M dwarf stars (e.g. TOI-6894~b, \citealt{Bryant2025}) consistently show the largest contrast in transit depths across the whole  wavelength ranges ($0.33\,\ldots\,10~\mu$m) 
studied, thus also in the optical (Figs.~\ref{fig: TDepth_All_PLATO_Clear_Cloud} and ~\ref{fig: TDepth_All_PLATO_Diff} gray lines). In the optical, however, PLATO observations of low luminosity stars  (e.g. $m_v=16$ for TOI-6894~b) are not favorable \citep{Grenfell2020} as PLATO is optimized to observe luminous stars with $m_V < 11$ \citep{Rauer2025}. 

\paragraph{K stars:} Gas giant planets around K stars represent the ideal 'sweet spot' between relative low optical luminosity allowing to identify transit depth variations with only 50~\% lower accuracy compared to F and G type stars in the PLATO bands (\citealt{Grenfell2020}).
K-type host stars are 30~\% smaller than the G host star which is sufficient to consistently increase the model differences (between clear and cloudy scenarios) in the planetary average transit depths and transit depth asymmetries to reach values of 150~ppm in PLATO's red and blue filter (see Figs.~\ref{fig: TDepth_All_PLATO_Clear_Cloud} and \ref{fig: TDepth_All_PLATO_Diff} orange bars for ultra-hot climates with $T_{\rm global}\geq 2200$~K). It is evident from the population overview from the PLATO LOPS2 field (Fig.~\ref{fig: PLATO_targets}), however, that while intermediately hot gas giants are observed for these host stars, their stellar effective temperature, T$_{\rm eff}$ [K], is too low to irradiate exoplanets to global temperatures T$_{\rm global}> 1600$K. 
The
planetary radii of the models utilized here are set to $R_{\rm Pl}=1 R_{\rm Jup}$ which is a rather conservative assumption: Planets like HIP 65 A b with its large radius $R_{\rm Pl}=2 R_{\rm Jup}$ around a $m_V=11.1$ K-type star would thus be ideal targets to yield maximum transit limb differences. This particular planet is, however, not part of the PLATO LOPS2 field \citep{Nielson2020} but was observed with TESS \citep{Wang2024}. While there are no other strongly inflated planets around bright K stars in the PLATO LOPS2 field, PLATO will allow with its initial 2 years continuous observation strategy to collect multiple transit observations of suitable host stars. With a precision of 50~ppm, cloud scenarios may be constrained with PLATO even for relatively cool (T$_{\rm global}\approx 800$~K) planets. HIP 65 A b may, however, be a worthwhile target for combined CHEOPS and TESS observations.

The full transition to the ultra-hot climate regime, for which the evening limb becomes locally too hot 
for cloud formation, would occur for K star planets with T$_{\rm global}=1800\rightarrow2000$~K (Fig.~\ref{fig: TDepth_All_PLATO_Diff}, right: Black crosses indicate a negative evening-morning difference and thus a cloud-free evening terminator $T_{\rm Depth}$ in orange bars). These planets do not exist, however, in the PLATO LOPS2 field (Fig.~\ref{fig: PLATO_targets}).
Observing the hottest detected K planets (T$_{\rm global}=1400\,\ldots\,1600$~K ) that lie just before the threshold of the climate transition in PLATO's white band is still informative.

Due to the relatively short orbital periods of K star planets and thus narrower wind jets compared to G star planets, already for T$_{\rm global}=1000\,\ldots\,1600$~K latitudinal evening temperature gradients occur.
Hence, noticeable differences arise between transit asymmetry calculations for which only equatorial regions are considered (Fig.~\ref{fig: TDepth_All_PLATO_Diff}, right: orange bars) compared to a scenario, for which higher latitudes  are taken into account (Fig.~\ref{fig: TDepth_All_PLATO_Diff} left: orange bars). For equatorial calculations, transit asymmetries steadily increase with increasing global temperatures, that is, when no latitudinal changes in cloud and temperature are considered. The latitude averaged asymmetries 
for T$_{\rm global}=1400\,\ldots\,1600$~K  are not increasing with higher temperatures indicating that temperature/cloud asymmetries between the equatorial at higher latitudes counteract the cloud temperature amplification effect. The relatively strong impact of higher latitudes for K star planets in this temperature range results from their comparatively short orbital periods and thus a narrower equatorial wind jet compared to G and F stars, leading to a less efficient heating of the evening terminator at higher latitudes \cite[][Fig.5]{Plaschzug2025}.

\paragraph{G stars:} For these sufficiently bright host stars, a large number of planets with inflated radii that cover the all three climate regimes, warm, intermediately hot and ultra-hot, can be expected.
PLATO transit depth asymmetry observations for T$_{\rm global}=800\,\ldots\,1200$~K and and ultra hot Jupiters should, hence,  be examined for this population. Here, in particular the transition between hot to ultra-hot Jupiters (T$_{\rm global}=1800\rightarrow 2000$~K) is of interest: The jump from positive evening to morning transit depth difference to negative difference for a clear evening limb scenario (Fig.~ \ref{fig: TDepth_All_PLATO_Diff} top right: orange bar) is an ideal {\it smoking gun} to constrain not only cloud scenarios but even the latitudinal extent of evening limb clouds. The expected large transit asymmetry signal of  -1000~ppm is large already for a conservative, $1 R_{\rm Jup}$-sized  planets.

\paragraph{F stars:}
A F-type star is $\approx$50\% larger than a G-type star. This reduces the transit depth. 
For the evening-to-morning difference still a signal of -150 to -200~ppm is expected for the cloud-free evening scenario with T$_{\rm global}= 2400\,\ldots\,2600$~K (Fig.~\ref{fig: TDepth_All_PLATO_Diff} top row: blue bar). Largely inflated ultra-hot Jupiters like WASP-121b in PLATO's LOPS2 field with R$_{\rm Planet}=1.865~R_{\rm Jup}$  and T$_{\rm global}\approx 2360$~K would thus be ideal for the study of evening limb cloud coverage for comparison with G star planets. 

Planets around F stars are for similar global temperatures located at larger distances and, hence have larger rotation periods compared to planets that orbit G stars. The differences in orbital period impacts the efficiency of the superrotating jet as a major factor shaping asymmetric cloud coverage \citep[e.g.][]{Helling2023}. One consequence of this climate difference is that the transition
from intermediately hot to ultra-hot Jupiters is shifted towards hotter global temperatures for F stars compared to G stars,  to T$_{\rm global}=2000\,\ldots\,2200$~K (\citealt{Plaschzug2025}).

\paragraph{A stars:}
A large stellar radius of 1.79 $R_{\rm Sun}$ is assumed for A-type stars. Thus for T$_{\rm global} <2200$~K, the expected transit asymmetry signal, even in the PLATO white band is small: $\lesssim 75$~ppm (Fig.~\ref{fig: TDepth_All_PLATO_Diff} top: purple bar). Only for a clear evening terminator and T$_{\rm global} > 2000$~K, larger asymmetries of $\lesssim-150$~ppm are expected (Fig.~\ref{fig: TDepth_All_PLATO_Diff} top right: purple bar).
Kelt-9b  represents this category of host stars as a potential PLATO target. The unusual brightness of its A-type host star ($m_V=7.6$) and the large planet radius (1.94~$R_{\rm Jup}$) would still make it worthwhile to observe transit depth asymmetries; in particular, since Kelt-9b lies with T$_{\rm global}\approx 2200$~K at the intermediately hot to ultra-hot climate regime threshold. Here, the relatively large differences of the evening terminator transit depth in PLATO's red and blue filter compared to the morning terminator would also be sufficient 
to identify if the evening terminator is still partially cloudy or cloud-free compared to the morning terminator (Fig.~\ref{fig: TDepth_All_PLATO_Diff}, bottom).

\subsection{Trends in the near infrared for different host stars}
\label{sec: NIR_all}
The results for the transit asymmetries in the JWST/NIRSpec range $\lambda_{\rm NIRSpec}=1.0\,\ldots\,1.87~\mu$m for all host stars and $T_{\rm global}=800$~K\,$\ldots 2600$~K are briefly summarized here (Fig.~\ref{fig: TDepth_All_NIR_Diff}). This wavelength range complements observations in the optical (Fig.~\ref{fig: TDepth_All_PLATO_Diff}) and  with MIRI (Fig.~\ref{fig: TDepth_All_MIRI_Diff}).

\paragraph{M stars:} The transit asymmetry signal is predicted to be $>100$~ppm for T$_{\rm global}=800\,\ldots\, 1000$~K, at least for some cloud scenarios (Fig.~\ref{fig: TDepth_All_NIR_Diff} gray bars). Thus, these wavelength ranges can ideally complement the MIRI observations of M planets for  $\lambda_{\rm MIRI}=8\,\ldots\,10\mu$m (see next Section~\ref{sec: MIRI_all}). Planets with T$_{\rm global} > 1000$~K around small M stars have not been found (Fig.~\ref{fig: PLATO_targets}), likely because they would be engulfed by tidal friction shortly after formation, because particularly short semi major axes ($a < 0.007$~AU) would be required around an 0.16 solar mass star to yield T$_{\rm global} > 1000$~K \citep{Plaschzug2025}.

\paragraph{K stars:} For planets with $T_{\rm global}=1400$~K,$\ldots, 1600$~K comparatively large transit depth asymmetries of several 100~ppms are expected and would complement both, MIRI (next Sect.~\ref{sec: MIRI_all}) and optical observations (previous subsection Sec.~\ref{ss:optical}) to investigate the impact of higher latitudes (Fig.~\ref{fig: TDepth_All_NIR_Diff}: orange bars) as the evening terminator is gradually warmed by the equatorial wind jet with higher global temperatures, but before evening cloud clearing can occur that would require $T_{\rm global}> 1800$~K. 

\paragraph{G stars:} For these host stars (Fig.~\ref{fig: TDepth_All_NIR_Diff}, golden bars), the impact of cloudy higher latitudes on the evening terminator and thus the transit asymmetry matter as outlined in Sect.~\ref{ss:optical}. Their signature is in the NIR smaller for T$_{\rm global} \geq 2200$~K. With sufficient accuracy, NIRSpec observations could still complement observations in the optical, in particular, when negative transit asymmetries occur that would indicate a completely cloud-free evening terminator. These scenarios occur in our calculations when only equatorial information is used (Fig.~\ref{fig: TDepth_All_NIR_Diff_eq}, golden bars).

\paragraph{F and A stars:} NIRSpec observations would be more challenging compared to G stars. Here, observations in the ultra-hot climate regime T$_{\rm global}\geq 2400$~K would require to resolve transit depth asymmetry of -20 to -50~ppm (Fig.~\ref{fig: TDepth_All_NIR_Diff}: blue (F) and purple (A) bars) indicating a partly cloud-free evening terminator, in particular, in combination with PLATO observations. Since the transition to the ultra-hot Jupiter regime occurs 
at higher global temperatures
compared to G stars, such observations may facilitate a broader climate population study across diverse host stars.

\subsection{Trends in the MIRI spectral range for different host stars}
\label{sec: MIRI_all}
Results for two JWST/MIRI  wavelength ranges are presented: $\lambda_{\rm Low}=5\,\ldots\,8~\mu$m (Fig.~\ref{fig: TDepth_All_MIRI_Diff}: left) and $\lambda_{\rm High}=8\,\ldots\,10~\mu$m (Fig.~\ref{fig: TDepth_All_MIRI_Diff}: right).

\noindent
\paragraph{M stars:} At $\lambda_{\rm High} =8\,\ldots\,10~\mu$m, M-type stars %would
typically exhibit a higher luminosity ($m_K<12$) then in the optical wavelength  (Sect.~\ref{ss:optical}). %Thus, 
MIRI observations would be very favorable for planets around M-type host stars, for which transit depths asymmetries 
$T_{\rm Depth}>1000$~ppm are predicted (Fig.~\ref{fig: TDepth_All_MIRI_Diff}:  gray bars).

In the shorter JWST/MIRI
wavelength ranges, transit depth asymmetry gives access to the cloud particle distributions down to planetary temperature of T$_{\rm global}\approx800$~K and possibly lower. In the lower MIRI wavelengths ($\lambda_{\rm Low}= 5 \ldots 8 \mu$m, Fig.~\ref{fig: TDepth_All_MIRI_Diff} left,  gray bars) cloud scenarios yield for planets  T$_{\rm global}<1250$~K less variations. Thus, for M-type host stars, larger wavelengths are more favorable, provided the accuracy is $\lesssim 200$~ppm.

\paragraph{K stars:} For these colder host stars, planets with 
T$_{\rm global}\lesssim1600$~K are observed (Fig.~\ref{fig: PLATO_targets}).  Observations  (Fig.~\ref{fig: TDepth_All_MIRI_Diff} right) at $\lambda_{\rm High} =8\,\ldots\,10~\mu$m with accuracies $\lesssim 200$~ppm may allow to constrain for such planets the cloud mass load. The hottest known K star planets lie in temperature just before the ultra-hot Jupiter climate regime threshold. With the climate transition the cloud distribution changes from a completely cloud covered evening terminator to a scenario, for which the evening limb becomes locally too hot for cloud formation. 

This transition occurs for planets orbiting K-type stars for T$_{\rm global}=1800\rightarrow2000$~K, that is at colder temperatures compared to G, F and A stars due the relative short orbital periods of hot K dwarf planets (Fig.~\ref{fig: TDepth_All_PLATO_Diff}). 
The shorter orbital periods compared to planets around G stars with similar global temperatures also reduce the width of the equatorial wind jet,  resulting in less efficient heating of the higher latitudes at the evening terminator (see previous section~\ref{sec: NIR_all}).
Observations for $\lambda_{\rm Low}= 5 \ldots 8 \mu$m are less informative for cloud particle distributions and would require accuracies of $\ll 100$~ppm to constrain cloud scenarios for T$_{\rm global}=1400\,\ldots\,1600$~K (Fig.~\ref{fig: TDepth_All_MIRI_Diff} left).

\paragraph{G stars:} While the transit asymmetry signal is reduced for planets that orbit G-type host stars compared to the K and M dwarfs, the expected signal is still predicted to be between 75 - 300 ppm for T$_{\rm global}>1600$~K for both, $\lambda_{\rm Low}$ (Figs.~\ref{fig: TDepth_All_MIRI_Diff} left, golden bars) and $\lambda_{\rm High}$  (Figs.~\ref{fig: TDepth_All_MIRI_Diff} right, golden bars). While lower MIRI wavelengths may thus be useful to constrain the cloud property transitions from intermediately hot to ultra-hot climate regime for $5\,\ldots\,8~\mu$m with accuracies $< 100$~ppm, a higher noise level is expected for $8\,\ldots\,10~\mu$m, making these large wavelengths less usable for G type host stars. 

\paragraph{F and A stars:}
Even for global temperatures of 2600~K, the expected transit depth asymmetry is expected to be $\lesssim 100$~ppm in the whole MIRI wavelength range (Fig.~\ref{fig: TDepth_All_MIRI_Diff} blue and purple bars for F and A stars). Thus, no large variation in cloud scenarios are expected to be detectable.

\section{Discussion}
\label{sec: discus}

\subsection{Lessons learned from kinetic cloud modelling}

JWST has so far provided a wealth of data that has the potential to broaden our understanding of chemistry, dynamics and cloud formation in warm to ultrahot Jupiters. For example, \citet{Samra2023}  and \citet{Carone2023} suggest that vertical mixing may need to be less efficient than derived in simulations informed by the vertical velocities derived from \texttt{ExoRad} \citep{Helling2023,Helling2021Book,Baeyens2021}. The reduction in mixing by a factor of 100 yields better agreement with WASP-96b \citep{Samra2023} and WASP-39b, where the location of the optically thick cloud top ($\tau_{\rm cloud}(\lambda)=1$) yielded $10^{-2}\,\ldots\,5\cdot10^{-3}$~bar for \citep{Carone2023} in agreement with the cloud top values based on 1D retrieval models \citep{Ahrer2023,Feinstein2023,Alderson2023}.
Section~\ref{sec: WASP39b} confirms that reduced mixing provides a better fit for observational data for the JWST ERS target WASP-39b when synthetic transmission spectra are compared to observation. In addition, a reduced Fe-content of the cloud material mix and of the cloud mass load in the upper atmosphere
 are necessary to improve agreement with the full set of observational data for WASP-39b for $\lambda=0.3\,\ldots\,5.25\mu$m based on one 3D kinetic cloud model.

\citet{Komacek2022} utilized cloud tracer studies in GCMs to show that vertical particle mixing in irradiated tidally locked gas giants can be locally suppressed. Cloud particles were implemented in a simplified set-up (formation when supersaturation factor S is equal to 1)  as either co-moving with the gas or to permanently settle out of the atmosphere. Previous GCM studies on vertical mixing for disequilibrium chemistry show that the high irradiation in hot Jupiters suppresses convection in contrast to brown dwarfs \citep{Komacek2019,Baeyens2021}. 
\citet{Carone2020} diagnosed that vertical transport of momentum is driven by coupled planetary waves than overturning, another indication of the strong difference between the dynamical regime of externally irradiated hot Jupiters and self-luminous brown dwarfs as also outlined in \citet{Showman2020}.

In addition,
the present work finds that 
the silicate cloud materials have to be free of iron atoms in the atmosphere regions accessible to transit observations, in agreement with other JWST observations \citep{Grant2023,Dyrek2024}. This is a counterintuitive result since iron atoms make up about 6\% of the total element content in a solar metallicity atmosphere \citep{Asplund2009} and have also been found in the upper atmospheres of ultra-hot Jupiters \citep[e.g.][]{Sing2019} and as \ce{FeSiO3}[s] in the clouds of brown dwarfs \citep{Suarez2023}. The lack of even the smallest amount of iron-species in the silicate cloud deck
upper cloud regions suggests either a stronger decoupling between an upper silicate and a lower iron dominated cloud region that emerges from both, in the \texttt{IWF Graz} and \texttt{ExoLyn} kinetic cloud models, or the absence of a lower iron dominated cloud layer e.g. by an efficient locking of the iron in either the innermost atmosphere or even in the planets interior.

Moreover, a reduction of cloud mass load is required at atmospheric heights where the mean particle sizes  $<0.1~\mu$m. That is, where p$_{\rm gas}<10^{-3}$~bar for intermediately hot Jupiter atmospheres like WASP-39b (Fig.~\ref{fig: 3DClimates}, middle column). This is the atmospheric region where the formation of cloud condensation nuclei is the dominating cloud formation process that ultimately determines how many condensate cloud particles form. Hence, reducing this population of small particles mimics either a decreased nucleation efficiency (leading to less but bigger cloud particles that gravitationally settle faster below the observable optical depth $\mathcal{T}(\lambda)=1$-level (Eq.\ref{eq:tautransit})) or completely absent nucleation (or cloud formation) in these atmospheric regions, indicating a very inefficient mixing into the upper atmospheres. Option one leads to bigger cloud particles that settle deeper into the lower atmosphere. The more efficient cloud settling is consistent with the need for Fe-reduced cloud material compositions where iron gets displaces deep into the inner atmosphere or a planetary interior from which it can not be brought up again for further participating in condensation processes.

This work, hence highlights the potential to 
understand the complex interplay between atmospheric dynamics and cloud processes with detailed observations ranging from the optical to the infrared. The WASP-39b test case shows that insights can be gained over the whole physically relevant modeling domain: From the top, where cloud nucleation through complex cluster formation may dominate, to the bottom where material may be locked into the interior.

\subsection{Cloud impact on transit depth asymmetries}

\begin{figure}
    \centering
        \includegraphics[width=0.97\linewidth]{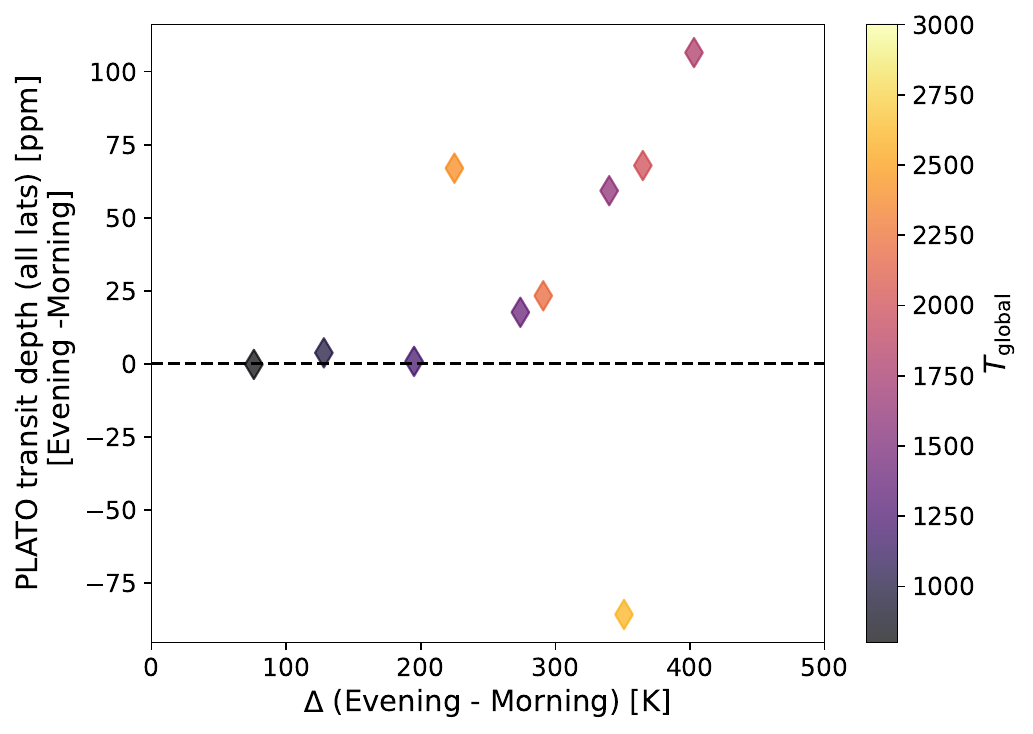}
    \caption{Transit asymmetry trend with global temperature. PLATO white band transit depth asymmetries versus evening to morning temperature differences at $p_{\rm gas}=10^{-3}$~bar for tidally locked planets with $T_{\rm global}= 800$~K$\ldots 2600$~K orbiting a G host star for a cloud scenario with cloud mass load, $\rho_g/\rho_d(z)$ scaled by 0.01 for atmospheric layers, where $\langle a \rangle <  0.1~\mu$m.  Both properties are latitudinally averaged.}
    \label{fig:Plato_G_basic_climate}
\end{figure}

The hierarchical cloud modelling framework applied here to planets with the same radius and metallicity allows to compare transit depth asymmetries for cloud-free and cloudy scenarios based on
the same underlying 3D gas temperature field. The \texttt{IWF Graz} modelling framework thus allows to
unambiguously identify if transit asymmetries stem from differences in scale height, and thus local gas temperature, or asymmetric cloud coverage.

For planets in the intermediately hot temperature regime (T$_{\rm global}=1200\,\ldots\,1800$~K orbiting G-type stars), both limbs are cloud covered and moderate gas temperature difference of a few 100~K are predicted between morning and evening limb (Fig.~\ref{fig:Plato_G_basic_climate}). For such planets, the scale height differences (hence, local gas temperatures) are the major driver for the transit depth asymmetries as indicated by the positive sign of the evening-to-morning transit depth differences: the warm evening limb is larger compared to the cool morning limb. Cloud coverage, however, always amplifies the positive transit asymmetries, which is particular significant (by about 100~ppm for G; e.g. Fig.~\ref{fig: TDepth_All_PLATO_Diff} golden bars) and even larger for K stars (Fig.~\ref{fig: TDepth_All_PLATO_Diff} orange bars) in the optical wavelength range. Limb asymmetries for $\lambda=2\,\ldots\,5~\mu$m, where NIRSpec PRISM probed transit asymmetries for WASP-39b (Sec.~\ref{sec: WASP39b}), are expected to be relatively small compared to both, shorter and longer wavelengths (Figs.~\ref{fig: TDepth_Asym_3Climates} middle \& \ref{fig: TDepth_Asym_3Climates_all_Lats} middle). Moreover, transit depth asymmetries in PLATO's white band can be directly related to evening-to-morning transit asymmetries when both terminators are cloudy (Fig.~\ref{fig:Plato_G_basic_climate}, the two hottest planets have partly cloud-free evening terminators). Our modeling framework, hence, allows to combine the observational properties with characteristics that describe the climate state of a planet. 

The cloud amplification of scale height differences vanishes for planets with $T_{\rm global}=800$~K probably due to the dominance of small cloud condensate particles in the whole atmospheric domain that is probed by transmission spectroscopy (Fig.~\ref{fig: 3DClimates}) as well as due to negligible temperature asymmetries between the terminators (Fig.~\ref{fig:Plato_G_basic_climate}). Only micron-size cloud particles thus allow to amplify scale height differences. Evening-to-morning transit depth differences consistently become negative (albeit with a very small transit depth signal $<$50~ppm) for all climate regimes in the optical for a cloud-free atmosphere.
In this case, the Rayleigh scattering slope and Na and K absorption features dominate the transmission spectra and the warmer evening terminator results in deeper transit depths (Fig.~\ref{fig: TDepth_3Climates}, middle top panel).

The predicted transit depth asymmetries are relatively small for intermediately hot and warm climates. They are on average
$\lesssim~100$~ppm for G stars, which is a conservative lower limit (Fig.~\ref{fig:Plato_G_basic_climate}). These predictions should be considered in comparison to the WASP-39b test case, where trends are well recovered qualitatively in the measured NISS transit depth limb asymmetries, but  quantitatively underestimate by a factor of two (Sec.~\ref{ss:oooo}).
Another reason for underestimating the transit signal is the choice of 1 Jupiter sized planets across the whole 3D GCM  grid. The transit depth signal scales with $R_P^2/R_*^2$ (Sec.~\ref{sec:Transmission}) and inserting the relevant parameters for a G planet from the 3D AFGKM \texttt{ExoRad} grid and WASP-39b results in:
\begin{equation}
\underbrace{ \left(\frac{1.27 R_{\rm Jup}}{0.9 R_{\rm Sun}} \right)^2 }_{\rm WASP-39 b}  \underbrace{\left(\frac{0.98 R_{\rm Sun}}{1 R_{\rm Jup}} \right)^2}_{\rm 1/Grid\, Planet} = 1.9 
\end{equation}
Thus, even a small inflation of the planetary radius and a slightly smaller host star can boost the signal by almost a factor of two.  Indeed, we derived for WASP-39b a transit asymmetry difference of 200~ppm for 2-5~$\mu$m (Figs.~\ref{fig:W39b_asym} and \ref{fig:W39b_asym_all_Lats}) which is larger by a factor of two compared to the 1600~K grid planet around a G type star (Figs.~\ref{fig: TDepth_Asym_3Climates} and \ref{fig: TDepth_Asym_3Climates_all_Lats}).

For ultra-hot Jupiters, the impact of horizontally asymmetric cloud coverage comes to the fore. That is apparent
in the optical wavelength range when comparing synthetic transmission spectra that only account for the equatorial region (Fig.~\ref{fig: TDepth_Asym_3Climates}, right panels) compared to transmission spectra that include all latitudes (Fig.~\ref{fig: TDepth_Asym_3Climates_all_Lats}, right panels). The equatorial evening limb is cloud-free, whereas the morning terminator is cloudy for all  latitudes. Apparently, the morning cloud is optically thick at a higher altitude than the cloud-free evening limb, resulting in very large, negative evening-morning transit depth differences in the optical.

When all latitudes are considered, than the impact of clouds at higher latitudes at the evening limb remove the negative transit depth difference in the optical wavelength range. The transit depth asymmetries using all latitudes is, however, generally smaller for the ultra-hot Jupiters compared to the intermediately hot Jupiters despite larger horizontal temperature differences for the ultra-hot Jupiters. Thus, here clouds act both, to amplify and to diminish transit depth asymmetries. Clouds still amplify transit depth differences due to horizontal differences in local gas temperatures, which is apparent by comparing the synthetic transmission spectra (Fig.~\ref{fig: TDepth_Asym_3Climates_all_Lats}, right lower panel). Clear transmission spectra (blue lines) tend to have smaller transit depth asymmetries despite sharing the same 3D temperature field with the cloudy results (black and gray lines). But the amplification is reduced compared to the intermediately hot climate regime (middle lower panel) due to the equatorial 'cloud-hole' at the warmer evening limb, that is, horizontal cloud coverage differences.

\subsection{Combining PLATO-JWST-HST to characterize ultra-hot Jupiter climates}

The large difference in optical transit depth asymmetries for ultra-hot Jupiters are depending on if the evening limb is assumed to be entirely cloud-free or partly cloud covered. This result opens new possibilities for characterizing off-equatorial atmosphere regions with accurate and precise PLATO measurements, where already the normal cameras (white band) would be informative (Sec.~\ref{sec: PLATO}: Figs.~\ref{fig: TDepth_G_PLATO_Diff} (top),~\ref{fig: TDepth_All_PLATO_Diff} and also Figs.~\ref{fig: TDepth_All_PLATO_Clear_Cloud}, \ref{fig: TDepth_All_PLATO_Diff}). 

It may be possible to diagnose the cloud-coverage of the evening limb more directly by probing the scattering slope of the evening terminator of an ultra-hot Jupiter with PLATO's fast cameras, namely by  comparing the transit depth of the evening terminator in the red and blue filter (Figs.~\ref{fig: TDepth_G_PLATO_Diff}, \ref{fig: TDepth_All_PLATO_Diff}, bottom). Identifying a cloud-free evening terminator would require to identify differences of 75~ppm for G stars between both channels.

K star planets with their shorter orbital periods compared to G star planets for the same $T_{\rm global}$ result in climates with narrower wind jets and thus less efficient heating of higher latitudes at the evening terminator \citep[][Fig.5]{Plaschzug2025}. Thus, the impact of higher latitudes may already be measurable for $T_{\rm global}=1400$~K\, $\ldots 1600$~K (Sects. \ref{ss:optical} and ~\ref{sec: NIR_all} for K stars.) in the optical and NIR.

Physically  interpreting observed transit depth asymmetries in the optical  may be complicated by stellar surface inhomogeneities 
\citep{Kostogryz2024,Rackham2018,Rackham2023}. Combining observations that cover the wavelength range between 0.5 -2$\mu$m is here particularly promising (Sec.~\ref{sec: NIR}). Because in the case of a clear evening limb, the evening-to-morning transit depth experiences a 'sign change' between 1-1.5~$\mu$m that can not be confused with stellar surface inhomogeneities e.g. faculae at the limbs that either induce a positive or negative transit asymmetry across the whole wavelength range and not a mixture \citep{Kostogryz2025}. Moreover, the climate signatures for a clear evening limb is expected to be a persistent feature, whereas star spots and faculae will evolve over the course of days, weeks and months. Calculations for the NIRSpec/G140M regime were so far used as a theoretically justified wavelength regime and also for comparison with HST. For real observations, NIRSpec PRISM or NIRISS/SOSS observations provide a wider wavelength range with the added benefit to also constrain the gas phase chemistry.

PLATO with  its accurate, precise transit depth variations will be instrumental to reveal the intricate cloud - climate connection in the ultra-hot Jupiter regime. Already the average transit depth comparison with cloud-free scenarios may be informative (Figs~\ref{fig: TDepth_G_PLATO_Clear_Cloud} and \ref{fig: TDepth_All_PLATO_Clear_Cloud}). Transit depth asymmetries between the morning / evening terminator in PLATO's normal cameras with the added benefit of color information of the scattering slopes from PLATO's fast cameras is also very promising, especially in combination with stellar atmosphere models \citep{Kostogryz2025}.

\subsection{Transmission spectra applied to 3D planets}
Synthetic transmission spectra from 3D climate models come with inherent assumptions that may need to be revisited when confronted with detailed observations. This is different to deriving the scattering and emission properties from, e.g. Monet Carlo simulations (e.g., \citealt{2017A&A...601A..22L}).  
\citet{Caldas2019}, for example,  showed that in many cases the horizontal chemistry gradient is ignored when interpreting either morning or evening limb from transmission observations. Assuming horizontally uniform chemistry may thus lead to underestimating the chemistry contribution of the cold night side to the overall probed morning terminator chemistry. The authors also acknowledge 
the inherent difficulty to translate between complex 3D climate models and transmission spectra. Here, we follow the approach pioneered by \citet{Fortney2010} and used in  e.g. \citet{Parmentier2018,Baeyens2021}:
The gas temperature and equilibrium chemistry  are averaged across a relative broad longitudinal region over each limb (Sect.~\ref{sec:Transmission}). 
This approach partly includes night side chemistry and local gas temperatures, we may still misjudge by how much. This simplification may also contribute to the
underestimation of the transit depth asymmetries for WASP-39b in particular at the morning terminator (Sec~\ref{sec: WASP39b}). Further, \citet{Kenneth2025} pointed out that it is not straight forward to translate the 'ingress/egress' observations and the 'morning/ evening' spectra arising from radiative transfer calculations from the 3D planet as shown here. The calculations shown here are comparable to spectra calculated with the method of \citet{Jones2020,Espinoza2021} in \citet{Espinoza2024}. Differences may arise, however, when comparing to spectra using the method of \citet{Tada2025}.

The present work addressed another source of uncertainty when interpreting transmission spectra: The latitudinal differences in gas temperature and thus cloud coverage.  The inclusion of higher latitudes tends to reduce transit depth asymmetries, albeit to a degree
that can be neglected to first order for intermediately hot climates, that is, for $1000$K$\leq T_{\rm global}\leq 1800$~K for G,F and A host stars. For K star planets, the narrower equatorial wind jets bring latitudinal differences to the fore albeit to a lesser degree than in the ultra-hot Jupiter regime. For warm climates, including higher latitudes strongly reduces transit asymmetries in the hierarchical modelling approach.

For ultra-hot Jupiters, however,  large differences are predicted in particular for the evening terminator. These differences arise because the equatorial regions become cloud-free already for relatively low $\leq T_{\rm global}\sim 1800\,\ldots 2000$~K. The off-equatorial latitudes gradually clear up with even higher global temperatures. The impact of the cloud covered off-equatorial regions leads for optical synthetic transmission spectra to large qualitative changes compared to equator-only spectra because they represent scenarios with cloudy morning and completely cloud-free evening terminators (Figs.~\ref{fig: TDepth_G_PLATO_Clear_Cloud} right, \ref{fig: TDepth_G_PLATO_Diff} right, \ref{fig: TDepth_All_PLATO_Diff} right). Consequently, equatorial transmission spectra yield negative transit evening to morning depth differences with large amplitudes (at least -350~ppm). Transmission spectra that include the cloud covered atmospheric parts at the evening limb at higher latitudes yield positive transit evening to morning depth differences with moderate amplitudes (+100~ppm) (Figs.~\ref{fig: TDepth_G_PLATO_Diff} left, \ref{fig: TDepth_All_PLATO_Diff} left). The impact of off-equatorial regions is also seen to lesser degrees in the near infrared and even partly in the MIRI wavelength ranges (Sect.~\ref{sec: Appendix_all_Transit_diff}).

These results are both, a warning and an opportunity. They serve as a warning against relying on a single atmospheric column per limb when interpreting the atmospheric physics of ultra-hot Jupiters. At the same time, the results indicate that it may be possible to constrain the latitudinal extent of cloud coverage at the evening limb of these planets with PLATO’s precise and accurate optical transmission spectra. PLATO observations could thus provide an unprecedented chance to probe the cloud-climate impact at locations of the planets that are otherwise very difficult to access with low resolution spectroscopy.

\section{Conclusion}
\label{sec: Conclusion}

This work explores how PLATO may be able to characterise different climate regimes for tidally locked gas giants. Transit depth and transit asymmetry changes were studied for a coherent set of 3D GCM models, in particular in the optical wavelength range but also in the IR and near-IR in order to explore synergies between PLATO, CHEOPS, TESS as well as JWST and HST.

Clouds and temperature asymmetries can play a strong role in shaping transit asymmetries. Clouds in the optical wavelength range amplify both, variations in average transit depths that are associated with different climate regimes, as well as evening-morning transit asymmetries. The cloud amplification is so strong that even latitudinal variations at the evening terminator in the ultra-hot climate regime may be diagnosed. Thus, PLATO observations, including PLATO's fast cameras' red and blue filter, would be ideal to bring the climate state of atmospheric regions to the fore that are otherwise very difficult to observe. JWST observations covering the optical to NIR wavelengths (1-2~$\mu$m) would be ideal to complement PLATO observations for K, G,F,A stars. JWST/MIRI (8-10~$\mu$m) provides the most insights about cloud properties for warm M-star planets.

\noindent The following conclusions emerge for different climate regimes:
\vspace{-0.8cm}
\paragraph{Warm climate regime ($T_{\rm global} \leq 1000$~K):}
Transit depth variations and transit asymmetries are minimal for the warm climate regime compared to the other regimes due to the relatively small horizontal temperature and thus cloud variations. We do find, however, a prevalence of submicron cloud particles in the observable atmosphere regions in contrast to the other, hotter climate regimes. JWST transit asymmetry observations around small K and M dwarf stars would be most beneficial. 
 In particular for planets around M dwarfs, the longest JWST/MIRI wavelength range (8-10~$\mu$m) is highly informative for cloud mass load in the upper atmosphere where cloud particles are small. Not only is the transit depth asymmetry large (1000~ppm) in these wavelengths, also cloud scenario variations by several 100~ppm are predicted that are not as prominent for shorter IR wavelengths. For K stars, the shorter MIRI ranges (5-8~$\mu$m) and NIRSpec ranges (1-2~$\mu$m) provide the best chance to constrain at least to first order the 100~ppm transit asymmetries and thus locate the bulk of the cloud mass
For G type stars, in the optical PLATO band in principle the transit depth asymmetry can be used to investigate different cloud scenarios, the required accuracy is, however, 50-75~ppm \citep{Grenfell2020,Rauer2025}. 
We note, however, that we probably underestimate temperature differences and thus transit depth asymmetries in our hierarchical cloud modelling approach.
Cloud temperature feedback may increase such temperature differences as demonstrated for the warm planets HATS-6b \citep{Kiefer2024HATS6b} and observed for WASP-107b \citep{Murphy2024}. 

\paragraph{Intermediately hot climate regime ($<1000$~K~$T_{\rm global}<1800$~K):} 
Latitudinal variations across the morning and evening terminator are minor and in any case much smaller than the differences between each limb for G,F and A host stars. The same assumption was used by \citet{Carone2023} for WASP-39b that lies with a global temperature of $T_{\rm global}=1100$~K in the intermediately hot climate regime. It was further confirmed in Sect.~\ref{sec: WASP39b} that the difference between equatorial and latitudinally average transit asymmetry calculations as used by \citet{Espinoza2024} are negligible for this planet.

Further, the largest evening-to-morning asymmetry variations are expected for this climate regime. A comparison between clear and cloudy transit depth calculations for the same 3D temperature field indicate that clouds at both limbs amplify the scale height differences between the warm extended evening and morning terminator. As a consequence, transit depth asymmetries can reach 100~ppm (G-type stars) between 3-4~$\mu$m. The largest asymmetry signal is predicted for a $1 R_{\rm Jup}$ planet for the optical range and NIR range between 0.6-2~$\mu$m (150-300~ppm, G and K stars). In these wavelengths, 50~ppm differences between scenarios with different cloud to gas mass density for atmosphere layers with small average condensate particle sizes arise. In the infrared, differences between cloud scenarios only arise between 8 - 10~$\mu$m and are less distinctive compared to the optical wavelength range. Simulations for the hottest detected K star planets, $T_{\rm global}=1400-1600$~K show due to their relatively narrow wind jet larger latitudinal differences compared to planets around hotter host stars which may be detectable in transit asymmetries in the optical and NIR.

\paragraph{Ultra-hot climate regime $T_{\rm global}\geq 1800$~K:} 
Strong latitudinal variations in the temperature and cloud coverage occur in particular at the evening terminator. The highly extended evening terminator is cloud-free at the equator but is covered by clouds at higher latitudes.
To probe latitudinal variations at the evening limb, observations of transit evening and morning terminator in the optical are imperative. The high cloud top at the morning terminator dominates in the optical such that the evening-to-morning transit asymmetry differences become negative because the morning terminator appears highly extended compared to the (partly) cloud-free evening terminator. The predicted signal strength is up to -500~ppm G stars and -200 ppm for F and A stars for a cloud-free evening limb in the PLATO band. The contrast between the high morning cloud top and a cloud-free evening limb persists for wavelengths up to 2~$\mu$m. For longer wavelengths ($>2 \mu$m), the evening terminator appears more extended compared to the morning terminator, leading to a smaller positive evening-morning transit depth difference (+300 ppm for G,F,A stars). Here, the morning limb cloud top apparently leads to a reduction of the transit depth asymmetry signal compared to the intermediately hot climate regime, for which both limbs are cloudy.

Thus, PLATO observations can be ideally complemented with observations of ultra-hot Jupiters with JWST/NIRISS and JWST/NIRSpec/Prism to probe the latitudinal extent of the cloud coverage at the evening terminator, that is otherwise not accessible.

Such a combination between optical and near-infrared would also be important to distinguish between asymmetries arising due to the planetary atmosphere or due to faculae at the stellar limbs. While the former can give rise to  sign-'flip' in transit asymmetry, this would not be possible due to magnetic effects at the stellar limb \citep{Kostogryz2024}. 

\paragraph{The use of PLATO to characterize 
climate regime:} Not only PLATO's normal cameras covering 0.5-1~$\mu$m are important to reveal detailed cloud properties. The blue and red filters on PLATO's fast cameras  may reveal the steepness of the scattering slope of the evening terminator (Figs.~\ref{fig: TDepth_G_PLATO_Diff} bottom \& \ref{fig: TDepth_All_PLATO_Diff} bottom, 75 ppm difference at a cloud-free evening terminator of an ultra-hot Jupiter around a G host star).  Thus, observing transit ingress and egress for ultra-hot Jupiter with the fast cameras may provide additional benefit compared to observations with PLATO's normal cameras. It is beneficial that individual transits can be observed simultaneously in the fast cameras in both bands, so that a very precise two-band transmission spectrum can be extracted.

\begin{acknowledgements}
Sven Kiefer is thanked for helpful  discussions. This paper was improved by the insightful and constructive suggestions from the referee.
The simulations (project id 72245) were performed on the Austrian Scientific Computing (ASC) infrastructure, in particular the Vienna Science Cluster (VSC). This work is further part of our science support efforts within the PLATO WPs 116700 and 116800.
This work presents results from the European Space Agency (ESA) space mission PLATO. The PLATO payload, the PLATO Ground Segment and PLATO data processing are joint developments of ESA and the PLATO Mission Consortium (PMC). Funding for the PMC is provided at national levels, in particular by countries participating in the PLATO Multilateral Agreement (Austria, Belgium, Czech Republic, Denmark, France, Germany, Italy, Netherlands, Portugal, Spain, Sweden, Switzerland, Norway, and United Kingdom) and institutions from Brazil. Members of the PLATO Consortium can be found at https://platomission.com/. The ESA PLATO mission website is https://www.cosmos.esa.int/plato. We thank the teams working for PLATO for all their work.
This work presents results from the CHEOPS space mission that is an European Space Agency (ESA) mission in partnership with Switzerland with important contributions to the payload and the ground segment from Austria,
Belgium, France, Germany, Hungary, Italy, Portugal, Spain, Sweden, and
the United Kingdom. Members of the CHEOPS Consortium can be found at the ESA CHEOPS mission webpage  https://www.cosmos.esa.int/cheops. We thank the teams working for CHEOPS for all their work.
\end{acknowledgements}


\bibliographystyle{aa} % style aa.bst
\bibliography{TM}

\begin{appendix}

\section{Stellar parameters}

Here, we list the relevant stellar values of the 3D GCM grid planets. For a more detailed description of the grid planet parameters, we refer the reader to \citet{Plaschzug2025}.

\begin{table}[ht]
\centering
\begin{tabular}{l|c|c|c}
\hline
\bf{Spectral type} & \bf{$\mathbf{T_{\mathrm eff}}$ [K]} & \bf{$\mathbf{R_*}$ [$\mathbf{R_{\mathrm Sun}}$]} & \bf{$\mathbf{M_*}$ [$\mathbf{M_{\mathrm Sun}}$]}\\
\hline
A5V & 8100 & 1.79 & 1.88 \\
F5V & 6550 & 1.47 & 1.33 \\
G5V & 5660 & 0.98 & 0.98 \\
K5V & 4400 & 0.70 & 0.70 \\
M5V & 3060 & 0.20 & 0.16 \\
\hline
\end{tabular}
\caption{Stellar parameters for the 3D \texttt{ExoRad} GCM  grid.}
\label{tab: stellar_params}
\end{table}

\section{PLATO, TESS, CHEOPS band passes}
\label{sec: Pass}

\begin{figure}[ht]
    \centering
    \includegraphics[width=0.5\textwidth]{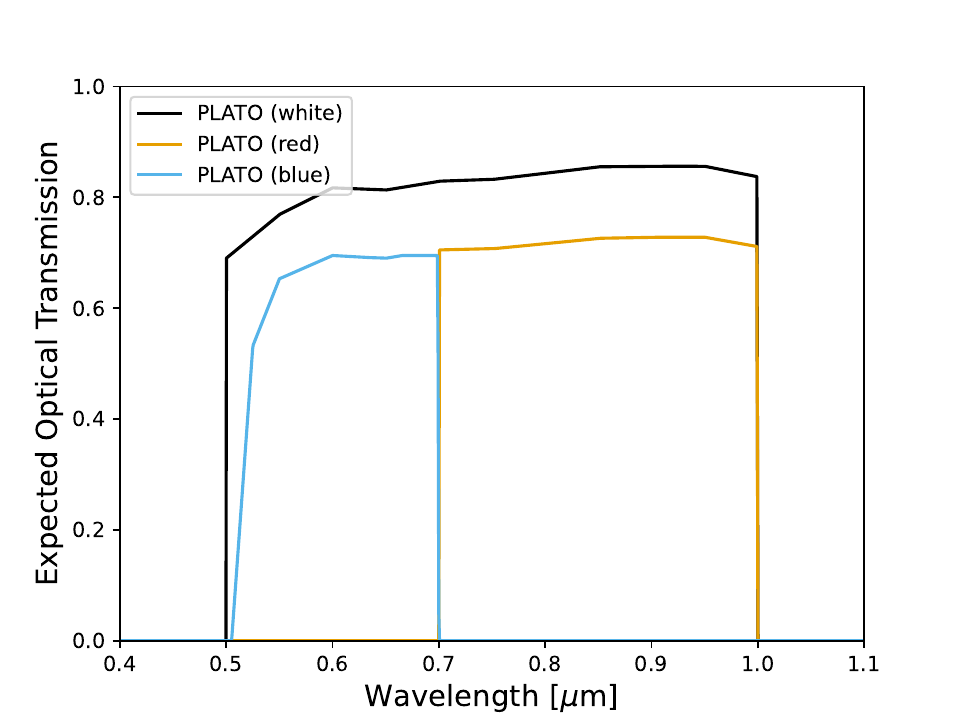}
    \caption{PLATO bands used in this work for the white filter of PLATO's normal cameras (black) and the red (golden) and blue (blue) filter on PLATO's fast cameras \citep{Rauer2025,Grenfell2020}.}
    \label{fig: PLATO_Band}
\end{figure}

\begin{figure}[ht]
    \centering
    \includegraphics[width=0.5\textwidth]{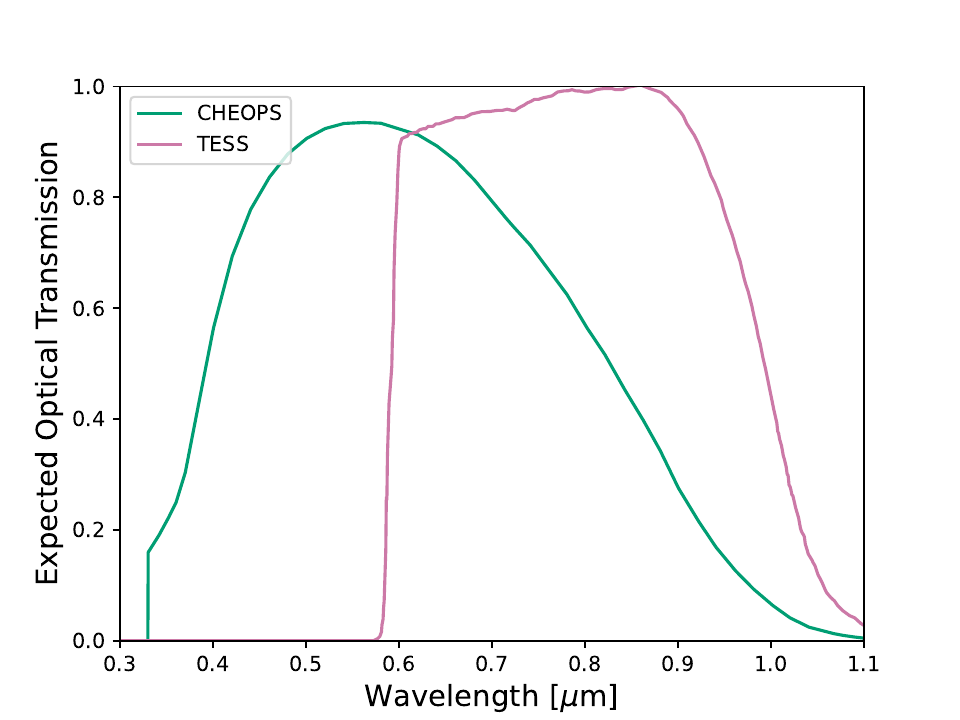}
    \caption{CHEOPS \& TESS bands. Optical transmission functions for CHEOPS (green) and TESS (pink) based on the most recent technical specifications.}
    \label{fig: CHEOPS_Band}
\end{figure}

Here, we show the expected telescope transmission function of the PLATO telescope and the transmission functions of the CHEOPS and TESS space telescope in the optical. PLATO posses apart from the normal cameras that covered the whole optical range between 0.5-1~$\mu$m also two fast cameras, each with a blue and red filter \citep{Rauer2025,Grenfell2020}. The assumption used in this work for PLATO are shown in Fig.~\ref{fig: PLATO_Band}. CHEOPS and TESS passbands are shown in Fig.~\ref{fig: CHEOPS_Band}.  While the PLATO band pass calculations in this work encompass all the relevant wavelength ranges for the normal and fast cameras, the transmission function and quantum efficiencies need to be included in future work, following the work of \citet[][Fig.2]{Jannsen2024}.

These are used to determine the sensitivity of current and upcoming space telescopes to detect transit depth asymmetries in the optical.

\section{PLATO red and blue transit depth limb asymmetries}
\label{sec: PLATO red and blue asym}
\begin{figure}[H]
    \centering
        \includegraphics[width=0.42\textwidth]{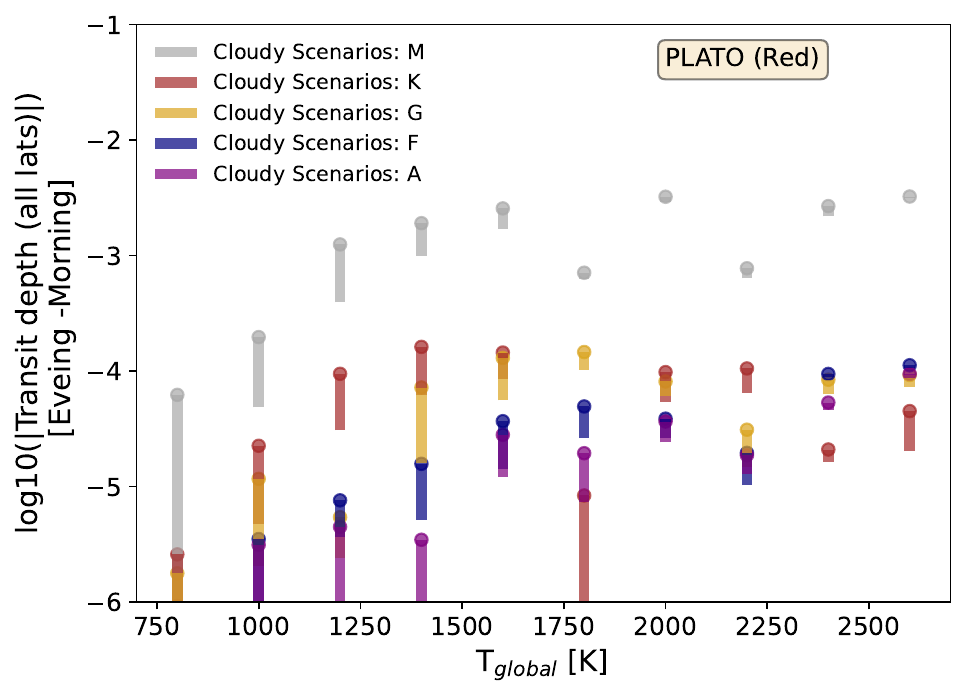}
        \includegraphics[width=0.42\textwidth]{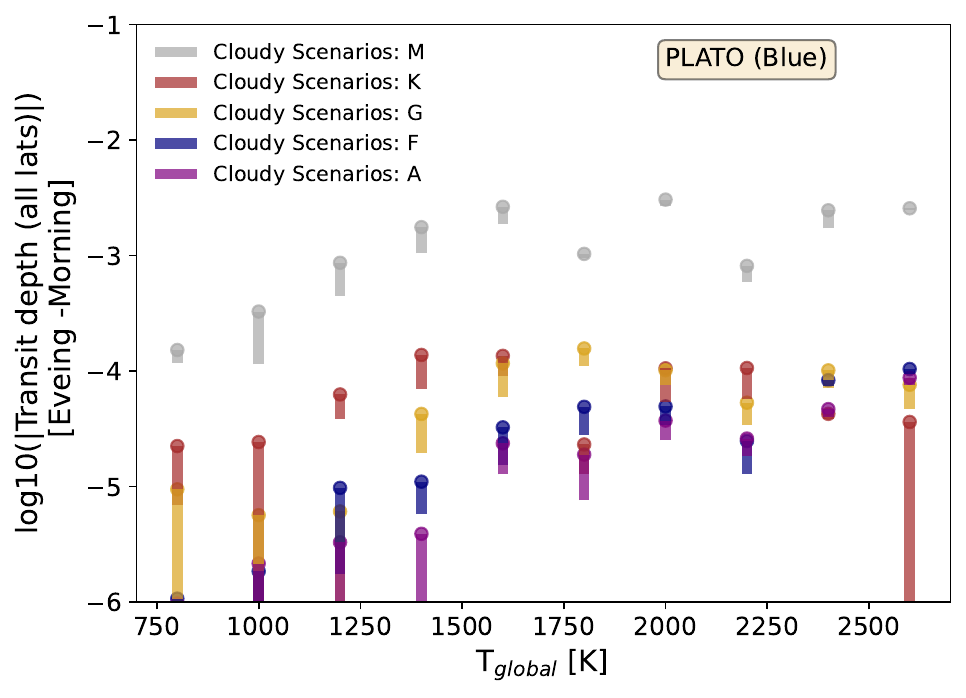}
        
    \caption{Latitudinally averaged transit asymmetries for planets around diverse host stars from different cloud scenarios in PLATO's fast cameras. Results are shown for tidally locked planets with global temperatures between  800 K - 2600 K and G host stars, as seen in the PLATO red filter for different cloud coverage scenarios. Bottom: The same but for PLATO's blue filter}
    \label{fig: TDepth_PLATO_RED_Diff_BLUE_Diff}
\end{figure}

\begin{figure}[ht]
    \centering
    \includegraphics[width=0.42\textwidth]{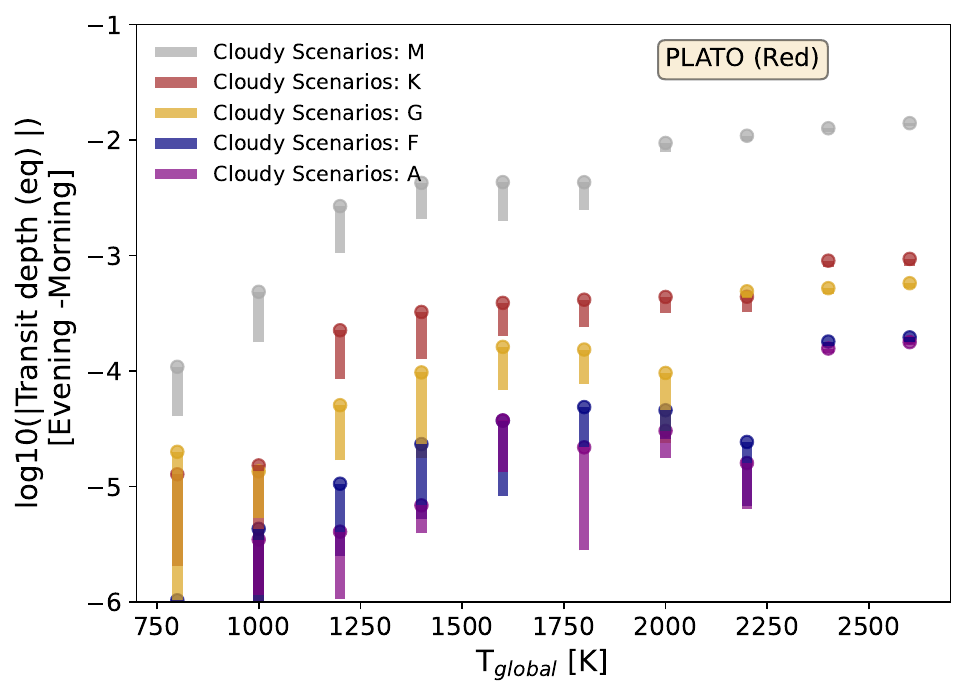}
            \includegraphics[width=0.42\textwidth]{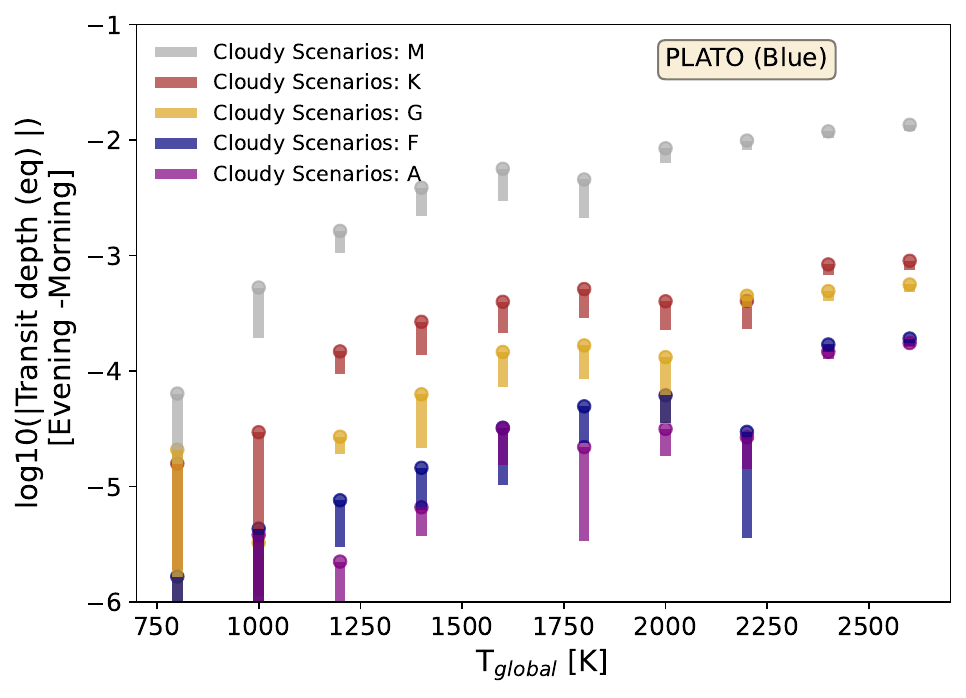}
    \caption{Equatorial transit asymmetries for planets around diverse host stars from different cloud scenarios in PLATO's fast cameras. Top: Results are shown for tidally locked planets with global temperatures between  800 K - 2600 K and G host stars, as seen in the PLATO red filter for different cloud coverage scenarios. Bottom: The same but for PLATO's blue filter}
    \label{fig: TDepth_PLATO_RED_Diff_BLUE_Diff_eq}
\end{figure}

The transit depth asymmetries between both limbs as seen in the red and blue filter of PLATO's fast cameras don't change significantly compared to PLATO's white broad band filter as seen on the normal cameras.

For completeness, we provide here still the evening to morning transit depth differences for the red and blue filter (Figs~\ref{fig: TDepth_PLATO_RED_Diff_BLUE_Diff}\& \ref{fig: TDepth_PLATO_RED_Diff_BLUE_Diff_eq}, top panels). The comparison between the evening-to-morning transit depth differences in the PLATO white band (Fig.~\ref{fig: TDepth_All_PLATO_Diff}, top panel)  shows that the transit depth asymmetries in the red and blue filter are very similar with respect to transit depth signal for diverse host stars and temperatures. Conversely, using the red and blue filters to resolve the individual scattering slope via an two band (R=2) resolution transmission spectrum of the evening and morning limb, respectively, shows a complementary view (Fig.~\ref{fig: TDepth_All_PLATO_Diff}, bottom panel) to the one presented here. The separate scattering slope analysis can reveal directly if one limb is cloud-free or cloudy, because the slope is larger in the first case compared to the latter.

\section{Transit Depth Asymmetries confined to the equatorial regions}
\label{sec: Equatorial}
For the estimate of transit asymmetries, the physically most consistent assumption is to include all latitudes - from the equator to the pole. Still, exploring transit asymmetries due to changes in temperature and cloud properties that emerge from the equatorial regions can still be illuminating. 
In Section~\ref{sec: High latitudes} the difference between the equatorial and full latitude view was shown to be particularly important  for transit asymmetries in the optical in the ultra-hot Jupiter climate regime for G type host stars. For other wavelengths, the differences are less important. For completeness, these other scenarios are collected here.

\begin{figure}[H]
    \centering
    \includegraphics[width=0.45\textwidth]{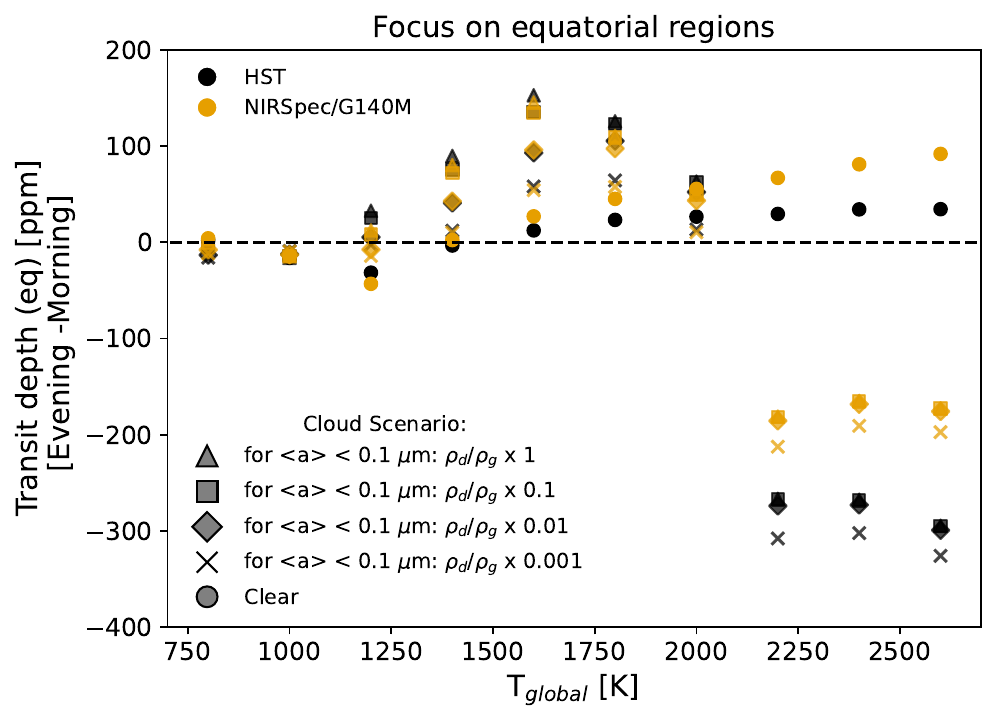}
    \caption{Equatorial transit asymmetries from different cloud scenarios in the near infrared. Results are shown for tidally locked planets with $T_{\rm global}= 800$~K$\ldots 2600$~K orbiting G main sequence stars. Results for clear atmosphere calculations are indicated by circles. Cloudy scenarios with different cloud mass loads, $\rho_g/\rho_d(z)$, in atmospheric layers where $\langle a \rangle <  0.1~\mu$m  are denoted by different markers (crosses: $10^{-3}\rho_g/\rho_d(z)$, diamonds:  $10^{-2}\rho_g/\rho_d(z)$, squares: $0.1\rho_g/\rho_d(z)$). Triangles denote Fe-free results with full $\rho_g/\rho_d(z)$. Colors denote values integrated in the HST/WFC3IR (black) and NIRSpec/G140M (golden) wavelengths.}
    \label{fig: TDepth_G_HST_Diff_eq}
\end{figure}

In the following, equatorial calculations for G planets are shown. Figs.~\ref{fig: TDepth_G_HST_Diff_eq}, \ref{fig: TDepth_G_MIRI_Diff_eq} cover the near infrared (NIR, combining HST and NIRSpec/G140M coverage: $\lambda_{\rm NIR}=0.8\,\ldots\,1.87\mu$m) and MIRI wavelength range ($\lambda_{\rm MIRI}= 5\,\ldots\,10\mu$m).

In Section~\ref{sec: Appendix_all_Transit_diff}, from the whole 3D AFGKM \texttt{ExoRad} GCM grid equatorial calculations are shown for planetary averaged transit differences between cloudy and clear atmospheres in the PLATO band (Fig.~\ref{fig: TDepth_All_PLATO_Clear_Cloud_eq}) and transit asymmetry results for the near infrared and MIRI wavelength ranges (Figs.~\ref{fig: TDepth_All_NIR_Diff_eq}, \ref{fig: TDepth_All_MIRI_Diff_eq}). 

\subsection{G stars}

In the PLATO band there is a strong difference between equatorial and latitudinally averaged transit asymmetries for ultra-hot Jupiters (Sect.~\ref{ss:optical}). In the near infrared the strong negative asymmetry as the signature of a vertically extended morning cloud compared to a clear equatorial evening terminator in the ultra-hot Jupiter regime is preserved. The signal is, however, weaker and results in smaller (-150 to -200 ppm for $T_{\rm global} > 2000$~K, Fig.~\ref{fig: TDepth_G_HST_Diff_eq}) transit depth asymmetry signals compared to the PLATO band ( $\lesssim-500$ ppm, Fig.~\ref{fig: TDepth_G_PLATO_Diff} right).

 A negative evening-to-morning negative transit contrast compared to a positive contrast in longer wavelength would add additional benefit to disentangle atmospheric and stellar surface inhomogeneities as the origin of the transit asymmetry. \citet{Kostogryz2025} show that the impact of the star's magnetic field on transit asymmetries does not change sign from the optical to the infrared.

\begin{figure}[ht]
    \centering

    \includegraphics[width=0.45\textwidth]{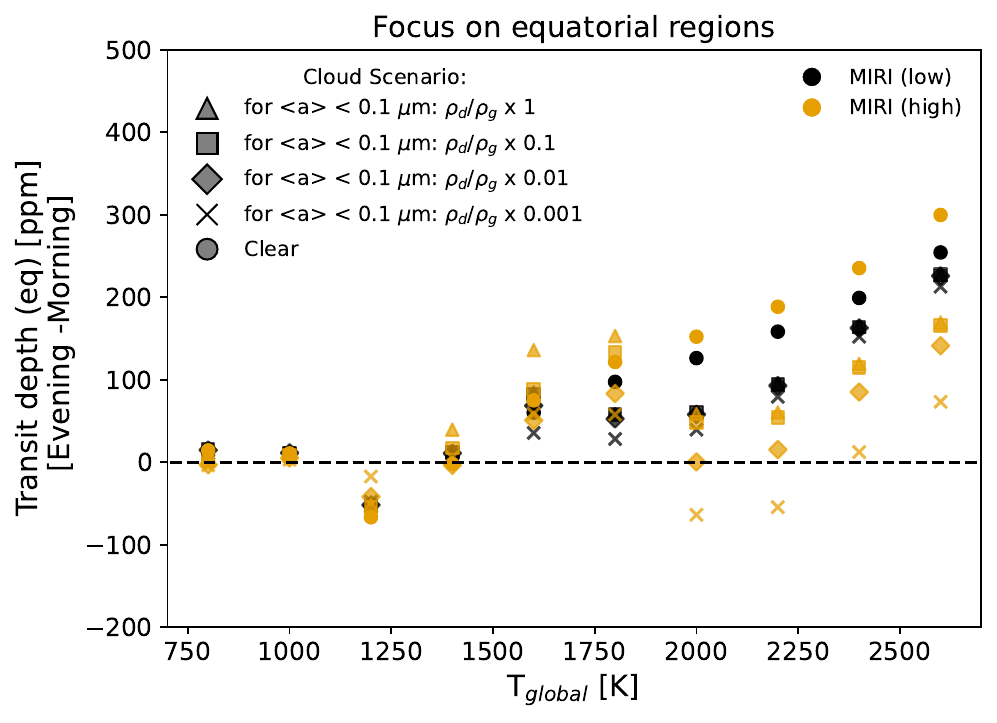}
        
    \caption{Equatorial transit asymmetries from different cloud scenarios for MIRI. Results are shown for tidally locked planets with $T_{\rm global}= 800$~K$\ldots 2600$~K orbiting G main sequence stars (left: only equatorial information is used, right: all latitudes are used). Results for clear atmosphere calculations are indicated by circles. Cloudy scenarios with different cloud mass loads, $\rho_g/\rho_d(z)$, in atmospheric layers where $\langle a \rangle <  0.1~\mu$m  are denoted by different markers (crosses: $10^{-3}\rho_g/\rho_d(z)$, diamonds:  $10^{-2}\rho_g/\rho_d(z)$, squares: $0.1\rho_g/\rho_d(z)$). Triangles denote Fe-free results with full $\rho_g/\rho_d(z)$. Colors denote values integrated between 5-8~$\mu$m (black) and 8-10~$\mu$m (golden).}
    \label{fig: TDepth_G_MIRI_Diff_eq}
\end{figure}

Figure~\ref{fig: TDepth_G_MIRI_Diff_eq} confirms also for equatorial calculations that observations in MIRI's longer wavelength range yields particularly large evening to morning transit asymmetries. A signal strength of $\lesssim 300$~ppm for $T_{\rm global}>2000$~K is derived, if only equatorial regions are taken into account. Complementarily, for the shorter MIRI wavelengths,  a strong increase in evening-to-morning transit asymmetry from $50\,\ldots\,200$~ppm with increasing global temperature is predicted with synthetic equatorial transit asymmetry calculations. Thus even for MIRI, an atmosphere with a particularly large inhomogeneous cloud coverage, that is, with a clear evening and cloudy morning terminator, leave an imprint in transit asymmetries.

\subsection{The whole AFGK-grid}
\label{sec: Appendix_all_Transit_diff}
 
In the following, the insights about equatorial transit asymmetries that reveal particularly large horizontal cloud variations are generalized for diverse host stars.

\begin{figure}[H]
    \centering
            \includegraphics[width=0.45\textwidth]{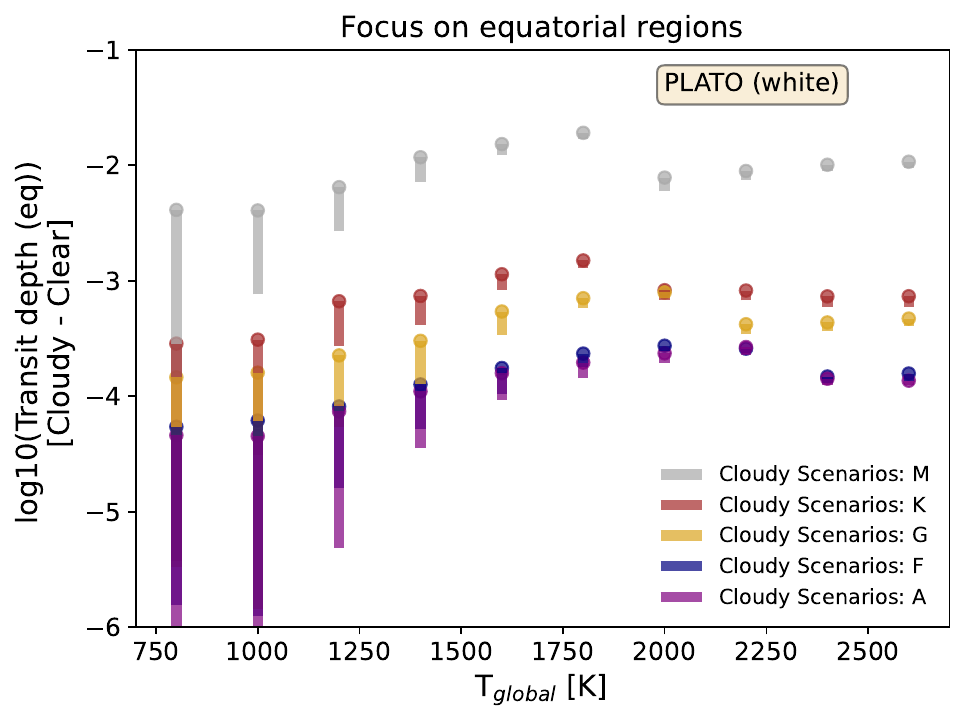} 
        
            \includegraphics[width=0.45\textwidth]{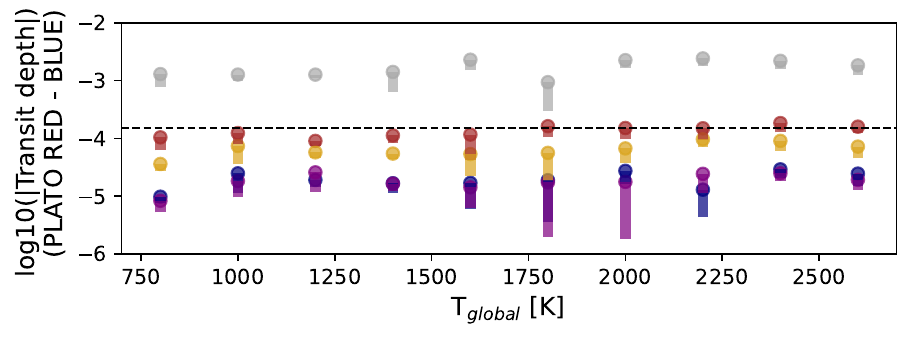}
    \caption{Difference between clear and cloudy equatorial transit depth calculations around diverse host stars for PLATO. Top: Differences for observation in PLATO's white band for tidally locked planets with $T_{\rm global}= 800$~K$\ldots 2600$~K (only equatorial information is used). Colored stripes indicate the range for cloud scenarios (with cloud mass load, $\rho_g/\rho_d(z)$ for atmospheric layers, where $\langle a \rangle <  0.1~\mu$m is scaled by values between 1 and $10^{-3}$) around different main sequence host stars.
    Bottom: Range of absolute differences between average transit depths in PLATO's red and blue.}
    \label{fig: TDepth_All_PLATO_Clear_Cloud_eq}
\end{figure}

\begin{figure}[ht]
    \centering

    \includegraphics[width=0.45\textwidth]{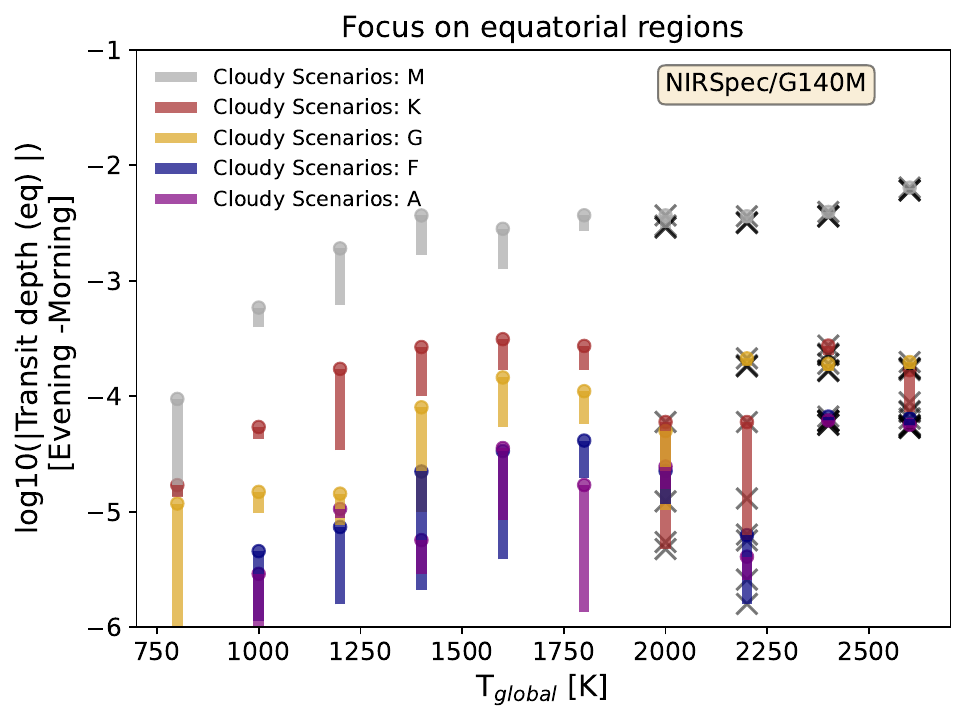}
    \caption{Equatorial transit asymmetry calculations for planets around diverse host stars from different cloud scenarios in the near infrared. Results are shown for observations in the integrated NIRSpec/G140M wavelength range for tidally locked planets with $T_{\rm global}= 800$~K$\ldots 2600$~K (only equatorial information is used.) Colored stripes indicate the range for cloud scenarios (with cloud mass load, $\rho_g/\rho_d(z)$ for atmospheric layers, where $\langle a \rangle <  0.1~\mu$m is scaled by values between 1 and $10^{-3}$) around different main sequence host stars.
    }
    \label{fig: TDepth_All_NIR_Diff_eq}
\end{figure}

\begin{figure*}[ht]
    \centering

    \includegraphics[width=0.45\textwidth]{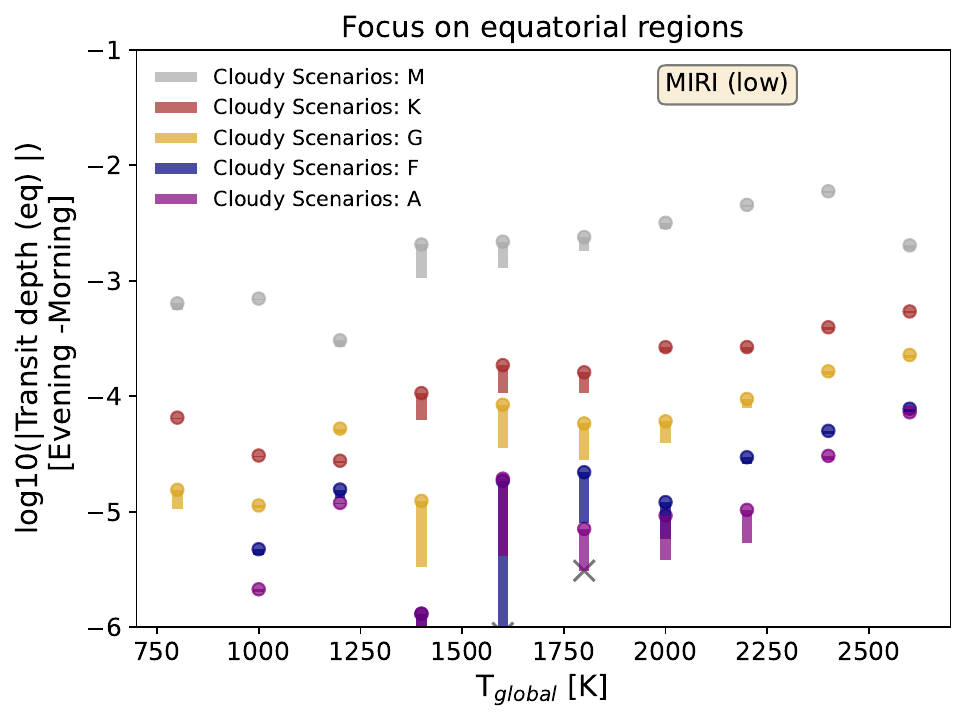}
        \includegraphics[width=0.45\textwidth]{figures/All_Tdepth_diff_All_Both_MIRI_hi.pdf}    
        
    \caption{Equatorial transit asymmetry calculations for planets around diverse host stars from different cloud scenarios for MIRI. Results are shown for observations in the integrated wavelength range between 5 -8~$\mu$m for tidally locked planets with $T_{\rm global}= 800$~K$\ldots 2600$~K. Colored stripes indicate the range for cloud scenarios (with cloud mass load, $\rho_g/\rho_d(z)$ for atmospheric layers, where $\langle a \rangle <  0.1~\mu$m is scaled by values between 1 and $10^{-3}$) around different main sequence host stars.
    }
    \label{fig: TDepth_All_MIRI_Diff_eq}
\end{figure*}

Generally, M stars shows the largest signal to noise ratio but planets orbiting these type of stars are relatively cold (Fig.~\ref{fig: PLATO_targets}). Thus no large horizontal cloud asymmetries are expected. Planets around F and A host stars, are hotter and include ultra-hot Jupiters. However, the comparatively large stellar radii makes transit asymmetry observations challenging (Figs.~\ref{fig: TDepth_All_NIR_Diff_eq}, \ref{fig: TDepth_All_MIRI_Diff_eq}: blue and purple bars for F and A stars, respectively). Potentially, the difference between cloudy and clear atmosphere calculations may be illuminating in the PLATO band. Figure~\ref{fig: TDepth_All_PLATO_Clear_Cloud_eq} shows that once the planets enter the ultra-hot Jupiter regime, the sum of equatorial evening and morning transit depth differences between clear and cloudy atmospheres is not changing strongly with T$_{\rm global}>2000$~K. Conversely, calculations that include all latitudes (Fig.~\ref{fig: TDepth_All_PLATO_Clear_Cloud}) show a stronger decrease of the difference with global temperature for ultra-hot Jupiters. The difference is because in the equatorial case, the evening limb is cloud-free for all ultra-hot Jupiters. When all latitudes are considered, partially cloud higher latitudes are considered that gradually clear up with increasing global temperature. However, revealing such trends would require precise and accurate transit depth measurements that also take into account hot Jupiter radius inflation.

Of all host stars, observations of transit asymmetries in K stars are the most beneficiary and complementary to G stars observations to reveal latitudinal differences. The large relative signal-to-noise ratio would be ideal to reveal particularly large transit asymmetry in the NIR (Fig.~\ref{fig: TDepth_All_NIR_Diff_eq}: orange bars) that may occur if the equatorial evening limb is uniformly hot as assumed in the equatorial transit calculations. For this stellar type, planets with T$_{\rm global}\lesssim1600$~K are observed (Fig.~\ref{fig: PLATO_targets}), which too cold for the ultra-hot Jupiters. Still, the predicted equatorial transit depth asymmetry of $\lesssim 600$~ppm for $\lambda_{\rm Low} =5\,\ldots\,8~\mu$m may allow with accuracies of $\lesssim 200$~ppm to constrain the cloud mass load (Fig.~\ref{fig: TDepth_All_MIRI_Diff} left) for the hottest K planets, T$_{\rm global}=1400\,\ldots\,1600$~K.

\section{Input details for cloud chemistry}

The following table (Tab.~\ref{tab: react}) comprises all reactions solved in the kinetic cloud model.

\begin{table*}
\caption{Chemical surface reactions $r$ assumed to form the solid materials 
   s. }
\label{tab:chemreak}
\resizebox{15.2cm}{!}{
\begin{tabular}{c|c|l|l}
{\bf Index $r$} & {\bf Solid s} & {\bf Surface reaction$^a$} & {\bf Key species} \\
\hline 
1 & TiO$_2$[s]          & TiO$_2$ 
       $\longrightarrow$ TiO$_2$[s]                  & TiO$_2$ \\ 
2 & rutile              & Ti + 2 H$_2$O 
       $\longrightarrow$ TiO$_2$[s] + 2 H$_2$        & Ti     \\
3 & (1)                 & TiO + H$_2$O  
       $\longrightarrow$ TiO$_2$[s] + H$_2$          & TiO     \\ 
4 &                     & TiS + 2 H$_2$O
       $\longrightarrow$ TiO$_2$[s] + H$_2$S + H$_2$ & TiS     \\
\hline 
5 & SiO$_2$[s]          & SiH + 2 H$_2$O
       $\longrightarrow$ SiO$_2$[s] + 2 H$_2$ + H    & SiH \\ 
6 & silica              & SiO + H$_2$O 
       $\longrightarrow$ SiO$_2$[s] + H$_2$          & SiO     \\ 
7 &  (3)                & SiS + 2 H$_2$O 
       $\longrightarrow$ SiO$_2$[s] + H$_2$S + H$_2$ & SiS     \\
\hline 
8 & SiO[s]              & SiO 
       $\longrightarrow$ SiO[s]                  & SiO \\
9 & silicon mono-oxide  & 2 SiH + 2 H$_2$O 
       $\longrightarrow$ 2 SiO[s] + 3 H$_2$      & SiH   \\
10 & (2)                & SiS + H$_2$O 
       $\longrightarrow$ SiO[s] + H$_2$S         & SiS     \\
\hline   
11 & Fe[s]              & Fe 
       $\longrightarrow$ Fe[s]                  & Fe      \\ 
12 & solid iron         & FeO + H$_2$ 
       $\longrightarrow$ Fe[s] + H$_2$O         & FeO     \\
13 & (1)                & FeS + H$_2$ 
       $\longrightarrow$ Fe[s] + H$_2$S         & FeS     \\ 
14 &                    & Fe(OH)$_2$ + H$_2$ 
       $\longrightarrow$ Fe[s] + 2 H$_2$O       & Fe(OH)$_2$ \\ 
15 &                    & 2  FeH
       $\longrightarrow$ 2 Fe[s] + H$_2$         & FeH     \\ 
\hline 
16 & FeO[s]             & FeO 
       $\longrightarrow$ FeO[s]                  & FeO\\
17 & iron\,(II) oxide   & Fe + H$_2$O
       $\longrightarrow$ FeO[s] + H$_2$          & Fe\\
18 & (3)                & FeS + H$_2$O 
       $\longrightarrow$ FeO[s] + H$_2$S         & FeS\\
19 &                    & Fe(OH)$_2$
       $\longrightarrow$ FeO[s] + H$_2$          & Fe(OH)$_2$\\
20 &                    & 2 FeH + 2 H$_2$O
       $\longrightarrow$ 2 FeO[s] + 3 H$_2$      & FeH \\
\hline
21 & FeS[s]             & FeS
       $\longrightarrow$ FeS[s]                       & FeS\\
22 & iron sulphide      & Fe + H$_2$S
       $\longrightarrow$ FeS[s]     + H$_2$           & Fe\\
23 & (3)                & FeO + H$_2$S 
       $\longrightarrow$ FeS[s] + H$_2$O     & $\min\{$FeO, H$_2$S$\}$\\
24 &                    & Fe(OH)$_2$ + H$_2$S     
       $\longrightarrow$ FeS[s] + 2 H$_2$O   & $\min\{$Fe(OH)$_2$, H$_2$S$\}$\\
25 &                    & 2 FeH + 2 H$_2$S
       $\longrightarrow$ 2 FeS[s] + 3 H$_2$  & $\min\{$FeH, H$_2$S$\}$\\
\hline
26 & Fe$_2$O$_3$[s]     & 2 Fe + 3 H$_2$O 
       $\longrightarrow$ Fe$_2$O$_3$[s] + 3 H$_2$        & $\half$Fe\\
27 & iron\,(III) oxide  & 2 FeO + H$_2$O
       $\longrightarrow$ Fe$_2$O$_3$[s] + H$_2$          & $\half$FeO\\
28 & (3)                & 2 FeS + 3 H$_2$O
       $\longrightarrow$ Fe$_2$O$_3$[s] + 2 H$_2$S + H$_2$&$\half$FeS\\
29 &                    & 2 Fe(OH)$_2$ 
       $\longrightarrow$ Fe$_2$O$_3$[s] + H$_2$O + H$_2$ & $\half$Fe(OH)$_2$\\
30 &                    & 2 FeH + 3 H$_2$O
       $\longrightarrow$ Fe$_2$O$_3$[s] + 4 H$_2$ & $\half$FeH\\
\hline
31 & MgO[s]             & Mg + H$_2$O 
      $\longrightarrow$ MgO[s] + H$_2$                & Mg\\
32 & periclase          & 2 MgH + 2 H$_2$O
      $\longrightarrow$ 2 MgO[s] + 3 H$_2$            & $\half$MgH\\ 
33 & (3)                & 2 MgOH
      $\longrightarrow$ 2 MgO[s] + H$_2$              & $\half$MgOH\\
34 &                    & Mg(OH)$_2$
      $\longrightarrow$ MgO[s] + H$_2$O               & Mg(OH)$_2$\\
\hline
35 & MgSiO$_3$[s]     & Mg + SiO + 2 H$_2$O 
     $\longrightarrow$ MgSiO$_3$[s] + H$_2$
                                  & $\min\{$Mg, SiO$\}$\\ 
36 & enstatite        & Mg + SiS + 3 H$_2$O 
     $\longrightarrow$ MgSiO$_3$[s] + H$_2$S + 2 H$_2$ 
                                  & $\min\{$Mg, SiS$\}$\\ 
37 & (3)              & 2 Mg + 2 SiH + 6 H$_2$O 
     $\longrightarrow$ 2 MgSiO$_3$[s] + 7 H$_2$
                                  & $\min\{$Mg, SiH$\}$\\ 
38 &                  & 2 MgOH + 2 SiO + 2 H$_2$O
     $\longrightarrow$ 2 MgSiO$_3$[s] + 3 H$_2$    
                                  & $\min\{\half$MgOH, $\half$SiO$\}$ \\
39 &                  & 2 MgOH + 2 SiS + 4 H$_2$O
     $\longrightarrow$ 2 MgSiO$_3$[s] + 2 H$_2$S + 3 H$_2$ 
                                  & $\min\{\half$MgOH, $\half$SiS$\}$ \\
40 &                  & MgOH + SiH + 2 H$_2$O
     $\longrightarrow$ MgSiO$_3$[s] + 3 H$_2$
                                  & $\min\{\half$MgOH, $\half$SiH$\}$ \\
41 &                  & Mg(OH)$_2$ + SiO 
     $\longrightarrow$ 2 MgSiO$_3$[s] +  H$_2$
                                  & $\min\{$Mg(OH)$_2$, SiO$\}$ \\ 
42 &                  & Mg(OH)$_2$ + SiS + H$_2$O
     $\longrightarrow$ MgSiO$_3$[s] + H$_2$S+ H$_2$
                                  & $\min\{$Mg(OH)$_2$, SiS$\}$ \\
43 &                  & 2 Mg(OH)$_2$ + 2 SiH + 2 H$_2$O
     $\longrightarrow$ 2 MgSiO$_3$[s] + 5 H$_2$
                                  & $\min\{$Mg(OH)$_2$, SiH$\}$ \\
44 &                  & 2 MgH +  2 SiO + 4 H$_2$O
     $\longrightarrow$ 2 MgSiO$_3$[s]+ 5 H$_2$
                                  & $\min\{$MgH, SiO$\}$ \\
45 &                  & 2 MgH +  2 SiS + 6 H$_2$O
     $\longrightarrow$ 2 MgSiO$_3$[s]+ 2 H$_2$S + 5 H$_2$
                                  & $\min\{$MgH, SiS$\}$ \\
46 &                  & MgH + SiH + 3 H$_2$O
     $\longrightarrow$  MgSiO$_3$[s]+ 4 H$_2$
                                  & $\min\{$MgH, SiH$\}$ \\
\hline
47 & Mg$_2$SiO$_4$[s] & 2 Mg + SiO + 3 H$_2$O
     $\longrightarrow$ Mg$_2$SiO$_4$[s] + 3 H$_2$  
                                  & $\min\{\half$Mg, SiO$\}$\\
48 & forsterite       & 2 MgOH + SiO + H$_2$O
     $\longrightarrow$ Mg$_2$SiO$_4$[s] + 2 H$_2$
                                  & $\min\{\half$MgOH, SiO$\}$\\ 
49 & (3)              & 2 Mg(OH)$_2$ + SiO 
     $\longrightarrow$ Mg$_2$SiO$_4$[s] + H$_2$O + H$_2$
                                  & $\min\{\half$Mg(OH)$_2$, SiO$\}$ \\
50 &                  & 2 MgH + SiO + 3 H$_2$O
     $\longrightarrow$ Mg$_2$SiO$_4$[s] + 4 H$_2$
                                  & $\min\{\half$MgH, SiO$\}$ \\
51 &                  & 2 Mg + SiS + 4 H$_2$O            
     $\longrightarrow$ Mg$_2$SiO$_4$[s] + H$_2$S + 3 H$_2$
                                  & $\min\{\half$Mg, SiS\} \\
52 &                  & 2 MgOH + SiS + 2 H$_2$O 
     $\longrightarrow$ Mg$_2$SiO$_4$[s] + H$_2$S + 2 H$_2$
                                  & $\min\{\half$MgOH, SiS\}\\
53 &                  & 2 Mg(OH)$_2$ + SiS 
     $\longrightarrow$ Mg$_2$SiO$_4$[s] + H$_2$ + H$_2$S
                                  & $\min\{\half$Mg(OH)$_2$, SiS\} \\
54 &                  & 2 MgH + SiS + 4 H$_2$O
     $\longrightarrow$ Mg$_2$SiO$_4$[s] + H$_2$S + 4 H$_2$
                                  & $\min\{\half$MgH, SiS$\}$ \\
55 &                  & 4 Mg + 2 SiH + 8 H$_2$O            
     $\longrightarrow$ 2 Mg$_2$SiO$_4$[s] + 9 H$_2$
                                  & $\min\{\half$Mg, SiH\} \\
56 &                  & 4 MgOH + 2 SiH + 4 H$_2$O 
     $\longrightarrow$ 2 Mg$_2$SiO$_4$[s] + 7 H$_2$
                                  & $\min\{\half$MgOH, SiH\}\\
57 &                  & 4 Mg(OH)$_2$ + 2 SiH 
     $\longrightarrow$ 2 Mg$_2$SiO$_4$[s] + 5 H$_2$
                                  & $\min\{\half$Mg(OH)$_2$, SiH\} \\
58 &                  & 4 MgH + 2 SiH + 8 H$_2$O
     $\longrightarrow$ 2 Mg$_2$SiO$_4$[s] + 11 H$_2$
                                  & $\min\{\half$MgH, SiS$\}$ \\
\hline 
59 & Al$_2$O$_3$[s]   & 2 Al + 3 H$_2$O 
     $\longrightarrow$ Al$_2$O$_3$[s] + 3 H$_2$   & $\half$Al\\
60 & aluminia         & 2 AlOH + H$_2$O 
     $\longrightarrow$ Al$_2$O$_3$[s] + 2 H$_2$   & $\half$AlOH \\ 
61 & (3)              &  2 AlH + 3 H$_2$O 
     $\longrightarrow$ Al$_2$O$_3$[s] + 4 H$_2$   & $\half$AlH\\
62 &                  & Al$_2$O + 2 H$_2$O
     $\longrightarrow$ Al$_2$O$_3$[s] + 2 H$_2$   & Al$_2$O\\
63 &                  & 2 AlO$_2$H 
$\longrightarrow$ Al$_2$O$_3$[s] + H$_2$O    & $\half$AlO$_2$H\\
\end{tabular}}
\label{tab: react}
\end{table*}

\begin{table*}
\caption{Table~\ref{tab:chemreak} continued}
\label{tab:chemreak2}
\resizebox{15.2cm}{!}{
\begin{tabular}{c|c|l|l}
{\bf Index $r$} & {\bf Solid s} & {\bf Surface reaction$^a$} & {\bf Key species} \\
\hline
64 & CaTiO$_3$[s]         & Ca + Ti + 3 H$_2$O     
      $\longrightarrow$ CaTiO$_3$[s] + 3 H$_2$       & $\min\{$Ca, Ti$\}$\\  
65 & perovskite           & Ca + TiO + 2 H$_2$O 
      $\longrightarrow$ CaTiO$_3$[s] + 2 H$_2$       & $\min\{$Ca, TiO$\}$\\
66 & (3)                  & Ca + TiO$_2$ + H$_2$O 
      $\longrightarrow$ CaTiO$_3$[s] + H$_2$         & $\min\{$Ca, TiO$_2\}$\\
67 &                      & Ca + TiS + 3 H$_2$O 
      $\longrightarrow$ CaTiO$_3$[s] + H$_2$S + 2 H$_2$  & $\min\{$Ca, TiS$\}$\\
68 &                      & CaO + Ti + 2 H$_2$O 
      $\longrightarrow$ CaTiO$_3$[s] + 2 H$_2$       & $\min\{$CaO, Ti$\}$\\
69 &                      & CaO + TiO + H$_2$O 
      $\longrightarrow$ CaTiO$_3$[s] + H$_2$         & $\min\{$CaO, TiO$\}$\\
70 &                      & CaO + TiO$_2$
      $\longrightarrow$ CaTiO$_3$[s]                 & $\min\{$CaO, TiO$_2\}$\\
71 &                      & CaO + TiS + 2 H$_2$O 
      $\longrightarrow$ CaTiO$_3$[s] + H$_2$S + H$_2$ & $\min\{$CaO, TiO$\}$\\
72 &                      & CaS + Ti + 3 H$_2$O 
      $\longrightarrow$ CaTiO$_3$[s] + H$_2$S + H$_2$ & $\min\{$CaS, Ti$\}$\\
73 &                      & CaS + TiO + 2 H$_2$O 
      $\longrightarrow$ CaTiO$_3$[s] + H$_2$S + 2 H$_2$ &$\min\{$CaS, TiO$\}$\\
74 &                      & CaS + TiO$_2$ + H$_2$O 
      $\longrightarrow$ CaTiO$_3$[s] + H$_2$S        & $\min\{$CaS, TiO$_2\}$\\
75 &                      & CaS + TiS + 3 H$_2$O 
      $\longrightarrow$ CaTiO$_3$[s] + 2 H$_2$S + H$_2$ &$\min\{$CaS, TiO$\}$\\
76 &                      & Ca(OH)$_2$ + Ti + H$_2$O 
      $\longrightarrow$ CaTiO$_3$[s] + 2 H$_2$  & $\min\{$Ca(OH)$_2$, Ti$\}$\\
77 &                      & Ca(OH)$_2$ + TiO 
      $\longrightarrow$ CaTiO$_3$[s] + H$_2$    & $\min\{$Ca(OH)$_2$, TiO$\}$\\
78 &                      & Ca(OH)$_2$ + TiO$_2$ 
      $\longrightarrow$ CaTiO$_3$[s] + H$_2$O   &$\min\rm\{Ca(OH)_2,TiO_2\}$\\
79 &                      & Ca(OH)$_2$ + TiS + H$_2$O
      $\longrightarrow$ CaTiO$_3$[s] + H$_2$S + H$_2$   & $\min\{$Ca(OH)$_2$, TiO$\}$\\
80 &                      & 2 CaH + 2 Ti + 6 H$_2$O
      $\longrightarrow$ 2 CaTiO$_3$[s] + 7 H$_2$   &   $\min\{$CaH, Ti$\}$\\
81 &                      & 2 CaH + 2 TiO + 4 H$_2$O
      $\longrightarrow$ 2 CaTiO$_3$[s] + 5 H$_2$   &   $\min\{$CaH, TiO$\}$\\
82 &                      & 2 CaH + 2 TiO$_2$ + 2 H$_2$O
      $\longrightarrow$ 2 CaTiO$_3$[s] + 3 H$_2$   &   $\min\{$CaH, TiO$_2$ $\}$\\
83 &                      & 2 CaH + 2 TiS + 6 H$_2$O
      $\longrightarrow$ 2 CaTiO$_3$[s] + 2 H$_2$S +5 H$_2$   &   $\min\{$CaH, TiS$\}$\\
84 &                      & 2 CaOH + 2 Ti + 4 H$_2$O
      $\longrightarrow$ 2 CaTiO$_3$[s] + 5 H$_2$   &   $\min\{$CaOH, Ti$\}$\\
85 &                      & 2 CaOH + 2 TiO + 2 H$_2$O
      $\longrightarrow$ 2 CaTiO$_3$[s] + 3 H$_2$   &   $\min\{$CaOH, TiO$\}$\\
86 &                      & 2 CaOH + 2 TiO$_2$
      $\longrightarrow$ 2 CaTiO$_3$[s] + H$_2$   &   $\min\{$CaOH, TiO$_2$ $\}$\\
87 &                      & 2 CaOH + 2 TiS + 4 H$_2$O
      $\longrightarrow$ 2 CaTiO$_3$[s] + 2 H$_2$S + 3 H$_2$   &   $\min\{$CaOH, TiS$\}$\\
\hline
88 & CaSiO$_3$[s]         & Ca + SiO + 2 H$_2$O
      $\longrightarrow$ CaSiO$_3$[s] + 2 H$_2$   &   $\min\{$Ca, SiO$\}$\\
89 & Wollastonite         & Ca + SiS + 3 H$_2$O
      $\longrightarrow$ CaSiO$_3$[s] + H$_2$S + 2 H$_2$   &   $\min\{$Ca, SiS$\}$\\
90 & (4)                  & 2 Ca + 2 SiH + 6 H$_2$O
      $\longrightarrow$ 2 CaSiO$_3$[s] + 7 H$_2$   &   $\min\{$Ca, SiH$\}$\\
91 &                      & CaO + SiO + 1 H$_2$O
      $\longrightarrow$ CaSiO$_3$[s] + H$_2$   &   $\min\{$CaO, SiO$\}$\\
92 &                      & CaO + SiS + 2 H$_2$O
      $\longrightarrow$ CaSiO$_3$[s] + H$_2$S + H$_2$   &   $\min\{$CaO, SiS$\}$\\
93 &                      & 2 CaO + 2 SiH + 4 H$_2$O
      $\longrightarrow$ 2 CaSiO$_3$[s] + 5 H$_2$   &   $\min\{$CaO, SiH$\}$\\
94 &                      & CaS + SiO + 2 H$_2$O
      $\longrightarrow$ CaSiO$_3$[s] + H$_2$S + H$_2$  &  $\min\{$CaS, SiO$\}$\\
95 &                      & CaS + SiS + 3 H$_2$O
      $\longrightarrow$ CaSiO$_3$[s] + 2 H$_2$S + H$_2$  &  $\min\{$CaS, SiS$\}$\\
96 &                      & 2 CaS + 2 SiH + 6 H$_2$O
      $\longrightarrow$ 2 CaSiO$_3$[s] + 2 H$_2$S + 5 H$_2$  &  $\min\{$CaS, SiH$\}$\\
97 &                      & 2 CaOH + 2 SiO + 2 H$_2$O
      $\longrightarrow$ 2 CaSiO$_3$[s] + 5 H$_2$   &  $\min\{$CaOH, SiO$\}$\\
98 &                      & 2 CaOH + 2 SiS + 4 H$_2$O
      $\longrightarrow$ 2 CaSiO$_3$[s] + 2 H$_2$S + 3 H$_2$   &  $\min\{$CaOH, SiS$\}$\\
99 &                      &  CaOH +  SiH + 2 H$_2$O
      $\longrightarrow$  CaSiO$_3$[s] + 3 H$_2$   &  $\min\{$CaOH, SiH$\}$\\
100 &                      &  Ca(OH)$_2$ +  SiO
      $\longrightarrow$  CaSiO$_3$[s] + H$_2$   &  $\min\{$Ca(OH)$_2$, SiO$\}$\\
101 &                      &  Ca(OH)$_2$ +  SiS + H$_2$O
      $\longrightarrow$  CaSiO$_3$[s] + H$_2$S + H$_2$   &  $\min\{$Ca(OH)$_2$, SiS$\}$\\
102 &                      & 2 Ca(OH)$_2$ + 2 SiH + 2 H$_2$O
      $\longrightarrow$  2 CaSiO$_3$[s] + 5 H$_2$   &  $\min\{$Ca(OH)$_2$, SiH$\}$\\
103 &                      & 2 CaH + 2 SiO + 4 H$_2$O
      $\longrightarrow$  2 CaSiO$_3$[s] + 5 H$_2$   &  $\min\{$CaH, SiO$\}$\\
104 &                      & 2 CaH + 2 SiS + 6 H$_2$O
      $\longrightarrow$  2 CaSiO$_3$[s] + 2 H$_2$S + 5 H$_2$   &  $\min\{$CaH, SiS$\}$\\
105 &                      & CaH + SiH + 3 H$_2$O
      $\longrightarrow$  CaSiO$_3$[s] + 4 H$_2$   &  $\min\{$CaH, SiH$\}$\\
\hline
106 & Fe$_2$SiO$_4$[s]      & 2 Fe + SiO + 3 H$_2$O
      $\longrightarrow$  Fe$_2$SiO$_4$[s] + 3 H$_2$  &  $\min\{$ $\half$Fe, SiO$\}$\\
107 & Fayalite              & 2 Fe + SiS + 4 H$_2$O
      $\longrightarrow$  Fe$_2$SiO$_4$[s] + H$_2$S + 3 H$_2$  &  $\min\{$ $\half$Fe, SiS$\}$\\
108 & (4)                   & 4 Fe + 2 SiH + 8 H$_2$O
      $\longrightarrow$  2 Fe$_2$SiO$_4$[s] + 9 H$_2$  &  $\min\{$ $\half$Fe, SiH$\}$\\
109 &                       & 2 FeO + SiO + H$_2$O
      $\longrightarrow$  Fe$_2$SiO$_4$[s] + H$_2$  &  $\min\{$ $\half$FeO, SiO$\}$\\
110 &                       & 2 FeO + SiS + 2 H$_2$O
      $\longrightarrow$  Fe$_2$SiO$_4$[s] + H$_2$S + H$_2$  &  $\min\{$ $\half$FeO, SiS$\}$\\
111 &                       & 4 FeO + 2 SiH + 4 H$_2$O
      $\longrightarrow$  2 Fe$_2$SiO$_4$[s] + 5 H$_2$  &  $\min\{$ $\half$FeO, SiH$\}$\\
112 &                       & 2 FeS + SiO + 3 H$_2$O
      $\longrightarrow$  Fe$_2$SiO$_4$[s] + 2 H$_2$S + H$_2$  &  $\min\{$ $\half$FeS, SiO$\}$\\
113 &                       & 2 FeS + SiS + 4 H$_2$O
      $\longrightarrow$  Fe$_2$SiO$_4$[s] + 3 H$_2$S + H$_2$  &  $\min\{$ $\half$FeS, SiS$\}$\\
114 &                       & 4 FeS + 2 SiH + 8 H$_2$O
      $\longrightarrow$  2 Fe$_2$SiO$_4$[s] + 4 H$_2$S + 5 H$_2$  &  $\min\{$ $\half$FeS, SiH$\}$\\
115 &                       & 2 Fe(OH)$_2$ + SiO
      $\longrightarrow$  Fe$_2$SiO$_4$[s] + H$_2$O + H$_2$  &  $\min\{$ $\half$Fe(OH)$_2$, SiO$\}$\\
116 &                       & 2 Fe(OH)$_2$ + SiS
      $\longrightarrow$  Fe$_2$SiO$_4$[s] + H$_2$S + H$_2$  &  $\min\{$ $\half$Fe(OH)$_2$, SiS$\}$\\
117 &                       & 4 Fe(OH)$_2$ + 2 SiH
      $\longrightarrow$  2 Fe$_2$SiO$_4$[s] + 5 H$_2$  &  $\min\{$ $\half$Fe(OH)$_2$, SiH$\}$\\
118 &                       & 2 FeH + SiO + 3 H$_2$O
      $\longrightarrow$  Fe$_2$SiO$_4$[s] + 4 H$_2$  &  $\min\{$ $\half$FeH, SiO$\}$\\
119 &                       & 2 FeH + SiS + 4 H$_2$O
      $\longrightarrow$  Fe$_2$SiO$_4$[s] + H$_2$S + 4 H$_2$  &  $\min\{$ $\half$Fe(OH)$_2$, SiS$\}$\\
120 &                       & 4 FeH + 2 SiH + 8 H$_2$O
      $\longrightarrow$  2 Fe$_2$SiO$_4$[s] + 11 H$_2$  &  $\min\{$ $\half$Fe(OH)$_2$, SiH$\}$ \\
\hline
121 & KCl[s]                  & K + Cl
      $\longrightarrow$  KCl[s]      & $\min$\{K, Cl\} \\
122 & Sylvite                & KCl
      $\longrightarrow$  KCl[s]      & KCl \\
123 & (7)                   & 2 K + 2 HCl
      $\longrightarrow$  2 KCl[s] + H$_2$      & $\min\{\half$ K, $\half$ KCl\} \\
124 &                       & (KCl)$_2$
      $\longrightarrow$   2 KCl[s]      & (KCl)$_2$ \\
125 &                       & KOH + HCl
      $\longrightarrow$   KCl[s] + H$_2$O      & $\min\{$ KOH, HCl\} \\
126 &                       & KH + HCl
      $\longrightarrow$   KCl[s] + H$_2$      & $\min\{$KH, HCl\} \\

\end{tabular}}
\end{table*}

\begin{table*}
\caption{Table~\ref{tab:chemreak} continued}
\label{tab:chemreak3}
\resizebox{11.2cm}{!}{
\begin{tabular}{c|c|l|l}
{\bf Index $r$} & {\bf Solid s} & {\bf Surface reaction$^a$} & {\bf Key species} \\
\hline
127 & NaCl[s]               & 2 Na + 2 HCL
        $\longrightarrow$ 2 NaCl[s] + H$_2$   & $\min\{$Na, HCl\} \\
128 & Halite                & NaCl
        $\longrightarrow$   NaCl[s]            & NaCl \\
129 & (7)                   & (NaCl)$_2$
        $\longrightarrow$ 2 NaCl[s]            & (NaCl)$_2$ \\
130 &                       & NaOH + HCl
        $\longrightarrow$ 2 NaCl[s] + H$_2$O   & $\min\{$NaOH, NaCl\} \\
131 &                       & NaH + HCl
        $\longrightarrow$ 2 NaCl[s] + H$_2$   & $\min\{$NaH, HCl\} \\
132 &                       & (NaOH)$_2$ + 2 HCl
        $\longrightarrow$ 2 NaCl[s] + 2 H$_2$O   & $\min\{$(NaOH)$_2$, $\half$HCl\} \\        
\end{tabular}}
\newline
a: The efficiency of the reaction is limited by
   the collision rate of the key species, which has the lowest
   abundance among the reactants. The notation $\half$ in the
   r.h.s.~column means that only every second collision (and sticking)
   event initiates one reaction. Data sources for the
   supersaturation ratios (and saturation vapor pressures): (1)
   Helling \& Woitke (2006); (2) Nuth \& Ferguson (2006); (3) Sharp \&
   Huebner (1990); (4) Woitke et al. (2017)
\end{table*}

\newpage

\end{appendix}

\end{document}